\documentclass[12pt]{article}
\usepackage[dvips]{graphics}  
\usepackage{epsfig}           
\usepackage{glas}

\newcommand{\eVdist}{\kern-0.06667em}

\newcommand{\stru}[2]{%
   \relax\ifmmode\hbox{\vrule height#1 depth#2 width0pt}%
   \else\vrule height#1 depth#2 width0pt\fi}

\newcommand{\Ronum}[1]{\uppercase\expandafter{\romannumeral#1}}
\newcommand{\ronum}[1]{\expandafter{\romannumeral#1}}

\DeclareMathAlphabet{\mathbf}{OT1}{cmr}{bx}{sl}

\newcommand{\slashfrac}[2]{%
  \raisebox{0.5ex}{\ensuremath #1}\kern-0.12em/\kern-0.08em
  \raisebox{-.8ex}{\ensuremath #2}}

\newcommand{\sqr}[3]{%
    {\vcenter{\hrule height.#3ex\hbox{\vrule width.#2ex height#1ex
     \kern#1ex\vrule width.#3ex}\hrule height.#2ex}}}

\catcode`\@=11 
\newcommand{\parenbar}{\mathpalette\p@renb@r}
\def\p@renb@r#1#2{\vbox{%
  \ifx#1\scriptscriptstyle \dimen@.7em\dimen@ii.2em\else
  \ifx#1\scriptstyle \dimen@.8em\dimen@ii.25em\else
  \dimen@1em\dimen@ii.4em\fi\fi \offinterlineskip
  \ialign{\hfill##\hfill\cr
    \vbox{\hrule width\dimen@ii}\cr
    \noalign{\vskip-.3ex}%
    \hbox to\dimen@{$\mathchar300\hfil\mathchar301$}\cr
    \noalign{\vskip-.3ex}%
    $#1#2$\cr}}}
\catcode`\@=12 

\newcommand{\IP}{{\rm I$\kern-0.01667em$P}\xspace}

\mathchardef\qsm=63
\mathchardef\pls=43
\mathchardef\mns=512
\mathchardef\plm=518
\mathchardef\eql=61
\mathchardef\smallleft=300
\mathchardef\smallright=301
\mathchardef\les=316
\mathchardef\gre=318
\mathchardef\leq=532
\mathchardef\grq=533

\catcode`\@=11 
\newcounter{pict@width}
\newcounter{pict@height}
\newlength{\pict@scale}
\newcommand{\psfigadd}[4]{%
\setcounter{pict@width}{1*\ratio{#2+\pict@scale/2}{\pict@scale}}
\setcounter{pict@height}{1*\ratio{#3+\pict@scale/2}{\pict@scale}}
\setlength{\unitlength}{\pict@scale}
\hbox to #2{\hspace{-\fill}\begin{picture}(\thepict@width,\thepict@height)
\put(0,0){\psfig{figure=#1,width=#2,height=#3,clip=}}
\SetScale{0.283466457}
\SetWidth{1.763889}
{#4}
\end{picture}}
}
\newcounter{pict@widthfst}
\newcounter{pict@widthscd}
\newcounter{pict@widthtot}
\newcommand{\psfigaddtwo}[7]{%
\setcounter{pict@widthfst}{1*\ratio{#2+\pict@scale/2}{\pict@scale}}
\setcounter{pict@widthscd}{1*\ratio{#2+#4+\pict@scale/2}{\pict@scale}}
\setcounter{pict@widthtot}{1*\ratio{#2+#4+#6+\pict@scale/2}{\pict@scale}}
\setcounter{pict@height}{1*\ratio{#3+\pict@scale/2}{\pict@scale}}
\setlength{\unitlength}{\pict@scale}
\hbox{\hspace{-\fill}\begin{picture}(\thepict@widthtot,\thepict@height)
\put(0,0){\psfig{figure=#1,width=#2,height=#3,clip=}}
\put(\thepict@widthscd,0){\psfig{figure=#5,width=#6,height=#3,clip=}}
\SetScale{0.283466457}
\SetWidth{1.763889}
{#7}
\end{picture}}
}
\newcommand{\psfigror}[4]{%
\setcounter{pict@width}{1*\ratio{#2+\pict@scale/2}{\pict@scale}}
\setcounter{pict@height}{1*\ratio{#3+\pict@scale/2}{\pict@scale}}
\setlength{\unitlength}{\pict@scale}
\hbox{\begin{picture}(\thepict@width,\thepict@height)
\put(0,\thepict@height){\psfig{figure=#1,width=#3,height=#2,clip=,angle=270}}
\SetScale{0.283466457}
\SetWidth{1.763889}
{#4}
\end{picture}}
}
\newcommand{\psfigrol}[4]{%
\setcounter{pict@width}{1*\ratio{#2+\pict@scale/2}{\pict@scale}}
\setcounter{pict@height}{1*\ratio{#3+\pict@scale/2}{\pict@scale}}
\setlength{\unitlength}{\pict@scale}
\hbox{\begin{picture}(\thepict@width,\thepict@height)
\put(0,0){\psfig{figure=#1,width=#3,height=#2,clip=,angle=90}}
\SetScale{0.283466457}
\SetWidth{1.763889}
{#4}
\end{picture}}
}
\catcode`\@=12 
\newlength\listtextwidth

\catcode`\@=11 
\newlength{\@tabfninsert}
\newlength{\@tabfnwidth}
\newcommand{\tabfootnote}[2]{%
  \setlength{\@tabfninsert}{0.8em}
  \setlength{\@tabfnwidth}{\textwidth}
  \addtolength{\@tabfnwidth}{-\@tabfninsert}
  \addtolength{\@tabfnwidth}{-0.4em}
  \noindent\makebox[\@tabfninsert][r]{\footnotesize$^{#1}$\hfil}\hfill%
  \parbox[t]{\@tabfnwidth}{\footnotesize #2\hfill}}
\catcode`\@=12 

\def\st{\scriptstyle}

\def\be{\begin{equation}}
\def\ee{\end{equation}}
\def\bea{\begin{eqnarray}}
\def\eea{\end{eqnarray}}

\def\b2hh{$B^0_{(s)} \to h^+h^{'-}$}
\def\bpipi{$B^0 \to \pi^{+}\pi^{-}$}
\def\VeloR{\texttt{VeloR}}
\def\VeloSpace{\texttt{VeloSpace}}
\def\Forward{\texttt{Forward}}
\def\Matching{\texttt{Matching}}

\begin{document}
\begin{titlepage}{GLAS-PPE/2008-10}{6$^{\underline{\rm{th}}}$ August 2008}
\title{Impact of misalignments on the analysis of B decays}%
\author{M.~Gersabeck$^1$, E.~Rodrigues$^1$, J. Nardulli$^2$\\
\\
$^1$ University of Glasgow, Glasgow G12 8QQ, Scotland\\
$^2$ Rutherford Appleton Laboratory, Didcot}
\vspace*{1.0cm}
\begin{abstract}
This note investigates the effects of a misaligned tracking system
on the analysis of $B$ decays.
Misalignment effects of both the vertex locator and the inner and
outer T-stations have been studied.
$z$-scaling effects of the vertex locator are also considered.
It is proven that misalignments of the order of the detector
single-hit resolutions have little or negligible effects on the
quality of the reconstruction and of the analysis of B decays.
The studies were performed with a sample of \b2hh decays,
but the impact of misalignments on the performance of the pattern
recognition algorithms and on the primary vertex resolutions,
assessed for the first time, are rather general and not restricted
to \b2hh decays.
\end{abstract}
\vspace*{1.0cm}
\begin{center}
\textit{LHCb Public Note, LHCb-2008-012}
\end{center}
\newpage
\end{titlepage}


\tableofcontents 
\newpage  

\section{Introduction and motivation}
\label{sec:int}
An accurate and efficient tracking system is of crucial importance to 
the success of the LHCb experiment \cite{LHCb,optimTDR}.
The alignment of the tracking system is of great importance, as
misalignments potentially cause losses in both tracking and physics
performances.

\begin{figure}[htbp]
\vspace*{0.3cm}
\begin{center}
\scalebox{0.6}{\includegraphics{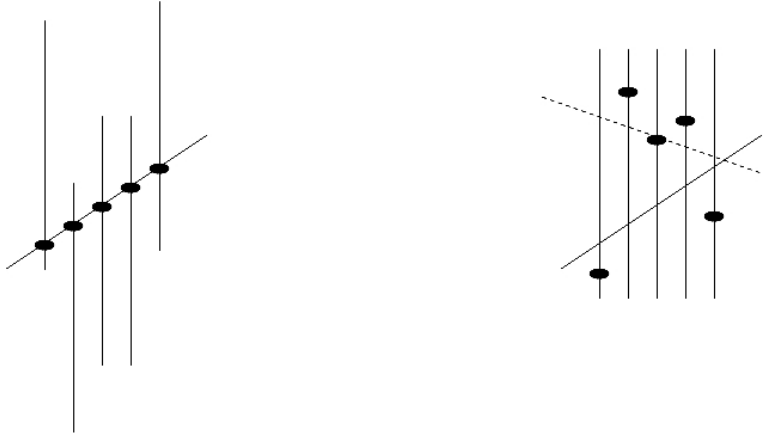}}
\end{center}
\caption{The basic alignment problem:
         a particle passes through a misaligned detector (left) but is
         fitted using the uncorrected geometry (right).
         Both the ``true'' track (full line) and the ``mis-reconstructed''
         track (dashed line) are visible.}
\label{fig:event}
\end{figure}

Understanding the alignment of the tracking detector planes is of paramount
importance in track reconstruction. This can be schematically demonstrated
with the simple example shown in Figure~\ref{fig:event}.
Here, a particle passes through a misaligned detector (left panel), but is
fitted assuming the uncorrected geometry (right panel).
As a consequence, wrong hit positions are assigned to the track and the
tracking performance deteriorates.

First studies of the deterioration of the LHCb tracking and (software) trigger
performance due to residual misalignments in the Vertex Locator (VELO) were
discussed in~\cite{petrie}~\footnote{Note that these studies relate to
a rather old and obsolete version of the trigger.}.
A study of the consequences of a misaligned Outer Tracker on the
signal and background separation of \b2hh decays
can be found in~\cite{jacopothesis}.

Here, the effects of misalignments of both the
Vertex Locator and the tracking T-stations on the analysis of the
\b2hh decays are investigated in detail.
Not only the effects on the pattern recognition, but also on the
event selection and reconstruction performance are described.

It is important to emphasise that the systematic study presented in this
note aims at assessing the effect of misalignments purely based on their size.

The next section details the implementation of misalignments and the
data samples used for the study. Section~\ref{sec:b2hh} presents the
impact of misalignments on the analysis of \b2hh decays separately for
several ``classes'' of misalignments of the tracking detectors VELO,
Inner Tracker (IT) and Outer Tracker (OT).
The conclusion and areas of future work are discussed in Section~\ref{sec:conclusions}.

\section{Implementation of misalignments}
\label{sec:misalignments}

\subsection{Misalignment scales}
\label{sec:misalignments_scales}
As mentioned in the previous section, here the effects
of misalignments based solely on their magnitude are considered.
In particular, no assumptions are made based on the quality of the metrology
or the expected performance of the alignment algorithms.

The misalignment effects are looked at as a function of a
``misalignment scale''.
The scales were chosen to be roughly $~1/3$ of the detector single-hit
resolution -- called ``1$\sigma$''.
Misalignments were then applied to each VELO module and sensor,
each IT box and OT layer following a Gaussian distribution with a sigma
corresponding to the 1$\sigma$ values.
The list of these misalignment 1$\sigma$ scales is summarised in
Table~\ref{tab:scales}.
Figure~\ref{fig:db} graphically confirms the Gaussian distributions
with some examples.

\begin{table}
\vspace{0.5cm}
\begin{center}
\begin{tabular}{|c||c|c|c||c|c|c|}
\hline
Detector & \multicolumn{3}{|c||}{Translations ($\mathrm{\mu m}$)} & \multicolumn{3}{|c|}{Rotations ($\mathrm{mrad}$)}\\
 & $\Delta_x$ & $\Delta_y$ & $\Delta_z$ & $R_x$ & $R_y$ & $R_z$\\
\hline\hline
VELO modules &  3 &  3 &  10 & 1.00 & 1.00 & 0.20 \\
VELO sensors &  3 &  3 &  10 & 1.00 & 1.00 & 0.20 \\
IT boxes    & 15 & 15 &  50 & 0.10 & 0.10 & 0.10 \\
OT layers    & 50 &  0 & 100 & 0.05 & 0.05 & 0.05 \\
\hline
\end{tabular}
\caption{Misalignment ``1$\sigma$'' scales for the VELO modules and sensors, the IT boxes and OT layers.}
\label{tab:scales}
\end{center}
\vspace{0.5cm}
\end{table}

For each sub-detector 10 sets of such 1$\sigma$ misalignments were generated,
to avoid any potentially ``friendly'' or ``catastrophic'' set of misalignments.
Likewise, this procedure was repeated with the creation of 10 similar sets
for each VELO module and sensor and each IT box and OT layer with
misalignment scales increased by factors of 3 (3$\sigma$) and 5 (5$\sigma$).

Each of these 10 misalignment sets were implemented and stored in dedicated
(conditions) databases.
In total 9 databases were produced, corresponding to the 1$\sigma$, 3$\sigma$
and 5$\sigma$ misalignments for the VELO, IT and OT detectors.

\subsection{Data samples}
\label{sec:misalignments_samples}
The study was performed with a 20$k$ sample of \bpipi\hspace{1.0mm}
events~\footnote{For the sake of simplicity only one of the \b2hh family
of decays was considered, as their different final states and
$B$-mother are not relevant in the present study.} for each scenario:
\begin{itemize}
\item perfect alignment (denoted 0$\sigma$ in the rest of the note);
\item 1$\sigma$, 3$\sigma$ and 5$\sigma$ misalignments for the
      following cases:\\
      VELO misalignments, IT and OT misalignments, and misalignments of
      VELO, IT and OT.
\end{itemize}
Each 20$k$ sample consists in reality of 10 sub-samples of 2$k$ events,
each of which was processed with a different one of the 10 sets of a
particular misalignment scenario.
In addition, the effects of a systematic change in the VELO $z$-scale
have also been studied.

\subsection{Event processing}
\label{sec:misalignments_processing}
All the events were generated and digitized with a perfect geometry
(\texttt{Gauss} generation program version v25r8 and
\texttt{Boole} digitization program version v12r10).
Starting always from the same digitized data samples,
the misalignments were only introduced at reconstruction level, where pattern
recognition, track fitting, primary vertexing and particle identification
are performed. The version \texttt{v32r2} of the \texttt{Brunel}
reconstruction software was used for this task.
Physics analysis was later performed with the \texttt{DaVinci} program
version \texttt{v19r9}.

\section{Impact of misalignments on the analysis\\of \b2hh decays}
\label{sec:b2hh}
In this section the direct effect of misalignments on the selection of the
\b2hh decays (as extensively described
in~\cite{dc04b2hh_selection}) is analysed.
First, the misalignments of the VELO are considered
(in Section~\ref{sec:b2hh_velo}), then the ones of the IT and OT tracking
stations (in Section~\ref{sec:b2hh_t}). Next the combined effects
of both the VELO and the tracking stations were considered
(in Section~\ref{sec:b2hh_all}). 
The last subsection, \ref{sec:b2hh_z-scale}, is devoted to the analysis
of effects of $z$-scaling of the VELO detector.

In all four studies, first the effect of misalignments on the pattern
recognition algorithms efficiencies is described, subsequently the effect
on the event selection is considered and every single cut is described
in detail.
Finally, the variations of momentum and $B^0$ mass, vertex and proper time
resolutions are shown as a function of the misalignments.

In Table~\ref{tab:shh} the selection cuts applied to the \b2hh
channels are shown. A more detailed explanation of all cuts can be
found in~\cite{dc04b2hh_selection}.

\begin{table}[hbtp]
\vspace{0.5cm}
\begin{center}
\begin{tabular}{|l|c|}
\hline
\b2hh selection parameter & cut value \\ 
\hline \hline
smallest $p_t$($\mathrm{GeV}$) of the daughters & $>$ 1.0\\
largest $p_t$($\mathrm{GeV}$) of the daughters  & $>$ 3.0 \\
$B^0_{(s)}$ $p_t$($\mathrm{GeV}$)               & $>$ 1.2\\
\hline
smallest $IP/\sigma_{IP}$  of the daughters     & $>$ 6 \\
largest $IP/\sigma_{IP}$ of the daughters       & $>$ 12\\
$B^0_{(s)}$  $IP/\sigma_{IP}$                   & $<$ 2.5\\
\hline
$B^0_{(s)}$ vertex fit $\chi^2$                 & $<$ 5\\
($L$)$/\sigma_L$                                & $>$ 18\\
$|\Delta$m$|$($\mathrm{MeV}$)                   & $<$ 50\\
\hline
\end{tabular}
\caption{Selection cuts applied to the \b2hh channels.}
\label{tab:shh}
\end{center}
\vspace{0.5cm}
\end{table}


\subsection{Misalignments in the VELO}
\label{sec:b2hh_velo}

\subsubsection{Effect on the pattern recognition}
Once the misalignments are introduced at the reconstruction level, as explained
in Section~\ref{sec:misalignments}, their effects need to be studied both on
the pattern recognition (track finding efficiencies) and on the event
selection (efficiency for finding the correct decay).

The pattern recognition algorithms~\footnote{For more details
about the definitions of the pattern recognition efficiencies
see~\cite{tracking}.} considered are the ones that find:
\begin{itemize}
\item tracks in the VELO detector in $r$-$z$ and 3D-space. The algorithms are
      hereafter denoted by \VeloR\ and \VeloSpace, respectively;
\item tracks that traverse the whole LHCb detector (called ``long tracks'').
      The two existing long tracking algorithms are hereafter denoted
      \Forward\ and \Matching.
\end{itemize}

In Table~\ref{tab:pr_velo} the efficiencies for the \VeloR, \VeloSpace,
\Forward, and \Matching\ pattern recognition algorithms are shown for the
0$\sigma$, 1$\sigma$, 3$\sigma$ and 5$\sigma$ scenarios.
All efficiencies are quoted for all long tracks in the event with
no momentum cut applied.

A clear relative degradation of $6.1\%$ for the \VeloSpace\ track finding
efficiency is observed in the $5\sigma$ scenario.
For the $1\sigma$ scenario there is hardly any effect.
The deterioration in the \Forward\ and \Matching\ long tracking efficiencies
can be fully attributed to the worsening in the VELO part.
Hence, these algorithms are not directly affected by VELO misalignments,
as expected.

All pattern recognition efficiencies quoted throughout the note refer to
averages obtained from the 10 sub-samples explained in
Section~\ref{sec:misalignments}. Further details are discussed in
Appendix~\ref{sec:subsamples}.

\begin{table}[hbtp]
\vspace{0.5cm}
\begin{center}
\begin{tabular}{|c||c|c|c|c|}
\hline
Misalignment & \VeloR          & \VeloSpace    & \Forward         & \Matching \\
scenario     & efficiency (\%) & efficiency (\%) & efficiency (\%) & efficiency (\%) \\
\hline\hline
0$\sigma$ & $98.0 \pm 0.1$  & $97.0 \pm 0.1$  & $85.9 \pm 0.2$  & $81.1 \pm 0.2$\\
1$\sigma$ & $98.0 \pm 0.1$  & $96.7 \pm 0.1$  & $85.6 \pm 0.2$  & $80.9 \pm 0.2$\\
3$\sigma$ & $98.0 \pm 0.1$  & $93.9 \pm 0.8$  & $83.1 \pm 0.6$  & $78.3 \pm 0.6$\\
5$\sigma$ & $97.7 \pm 0.2$  & $91.1 \pm 1.7$  & $80.1 \pm 1.6$  & $75.5 \pm 1.5$\\
\hline
\end{tabular}
\caption{\VeloR, \VeloSpace, \Forward\ and \Matching\ pattern recognition
         efficiencies for a perfectly aligned VELO and various VELO
         misalignment scenarios.}
\label{tab:pr_velo}
\end{center}
\vspace{0.5cm}
\end{table}

\subsubsection{Effect on the event selection}
\label{sec:b2hh_velo_sel}
In this section the effect of misalignments of the VELO on the
various discriminating variables shown in Table~\ref{tab:shh} is analysed.

In this instance it is important to realise that the selection cuts on these
discriminating variables were optimized for a perfect alignment. It may be
that the deteriorations reported hereafter can be reduced to some extent
with a cuts re-optimization for the non-perfectly aligned scenarios.
This same comment holds for all the studies summarised in the remainder of
the note.

In Figures~\ref{fig:sel_velo} and~\ref{fig:sel_velo_2} the distributions of the
discriminating variables are shown for the 0$\sigma$, 1$\sigma$, 3$\sigma$
and 5$\sigma$ cases.
In each plot the full line represents the $0\sigma$ case; the dashed line
represents the $1\sigma$ case; the dotted line represents the $3\sigma$ case
and the dot-dashed line represents the $5\sigma$ case.
Note that all plots are obtained after applying the
full selection on all the variables but the plotted one.
The cut value is indicated by a vertical line. 
In case of the $p_{T}$ and impact parameter significance cuts on the pions,
where one threshold is applied to both pions and another has to be exceeded
by at least one of them, the common threshold is depicted by a solid line,
while the additional threshold is shown as a dashed line.
In addition, normalised integrals are shown to give a direct comparison of
the acceptances for different misalignments~\footnote{In case of the $p_{T}$
and impact parameter significance cuts, which have been switched off
simultaneously, the acceptance related to the second cut can not be directly
read off the acceptance functions.}.

The most sensitive variable to VELO misalignments is the impact parameter
significance of the $B^0$ (Figure~\ref{fig:sel_velo}(a)).
The acceptance of this cut drops by a factor two for misalignments as large
as in the $5\sigma$ case.
Other discriminating variables that are affected by VELO misalignments are
the $B^0$ decay vertex $\chi^2$ (relative loss $\approx{}30\%$) and the
daughters' impact parameter significance (relative loss $<10\%$).
The overview of the number of selected events is given in
Table~\ref{tab:sel_velo}.

\begin{table}[hbtp]
\vspace{0.5cm}
\begin{center}
\begin{tabular}{|c|c|}
\hline
Misalignment & Number of \\
scenario     & selected events \\
\hline\hline
0$\sigma$         & 4141  (100\%) \\
1$\sigma$         & 3829  (92.5\%)\\
3$\sigma$         & 2194  (53.0\%)\\
5$\sigma$         & 1082  (26.1\%)\\
\hline
\end{tabular}
\caption{Number of selected events after running the \b2hh selection
for the different VELO misalignment scenarios.}
\label{tab:sel_velo}
\end{center}
\vspace{0.5cm}
\end{table}

\subsubsection{Effect on resolutions}
\label{sec:b2hh_velo_res}
After having studied the effect of misalignments on the event selection it is
equally important to evaluate their impact on physics analysis observables;
therefore, momentum and mass resolution have been studied as well as the
resolutions on the primary and secondary ($B^0$ decay) vertices and on the
proper time. These resolutions are shown in Figures~\ref{fig:res_velo}
and~\ref{fig:res_velo_2} while their values
(the sigmas of single-Gaussian fits) are summarised in
Tables~\ref{tab:resolution_velo} and~\ref{tab:resolution_velo_2}.

The momentum and mass resolutions are only affected on the
percent level by VELO misalignments, while the vertex related quantities are
strongly affected.
For the latter the resolutions worsen by roughly a factor two for the $5\sigma$
case compared to the fully aligned scenario.
Note that, in particular, the results on the precision of the reconstruction
of primary vertices are fully general and not restricted to \b2hh decays.

In addition to the ``bare'' resolutions, the effect of misalignments on the
assigned proper time error has been studied. Figure~\ref{fig:tau_velo} shows
the distribution of the proper time error for the various misalignment
scenarios as well as the respective pull distributions.
It is clearly visible and confirmed by the numbers in Table~\ref{tab:tau_velo}
that even in the aligned case there is a bias in the estimation of the proper
time and that the errors are under-estimated.
With misalignments both the bias and the error estimation worsen significantly.
Note that the errors used in the pull distributions do not account for  
the change in the effective single-hit resolution due to misalignments.
Hence, these figures should only be seen as another way  
of illustrating the degrading precision as a function of misalignments.

\begin{table}
\vspace{0.3cm}
\begin{center}
\begin{tabular}{|c||c|c|c|}
\hline
                               & Momentum   & Mass             & Proper time \\
\raisebox{1.5ex}{Misalignment} & resolution & resolution       & resolution \\
\raisebox{1.75ex}{scenario}    & (\%)       & ($\mathrm{MeV}$) & ($\mathrm{fs}$) \\
\hline\hline
0$\sigma$         & 0.49 & 22.5 & 37.7\\
1$\sigma$         & 0.50 & 22.5 & 39.4 \\
3$\sigma$         & 0.52 & 22.2 & 58.1 \\
5$\sigma$         & 0.52 & 23.5 & 82.0 \\
\hline
\end{tabular}
\caption{Values of the resolutions on the daughters' momentum, the $B^0$ mass
and the $B^0$ proper time for the different VELO misalignment scenarios.
The resolutions correspond to the sigmas of single-Gaussian fits.
The errors on all resolutions are around 1-1.5 \%.}
\label{tab:resolution_velo}
\end{center}
\vspace{0.3cm}
\end{table}

\begin{table}
\vspace{0.3cm}
\begin{center}
\begin{tabular}{|c||c|c|c||c|c|c|}
\hline
Misalignment & \multicolumn{3}{|c||}{Primary vertex} & \multicolumn{3}{|c|}{$B^0$ vertex}\\
scenario  &  \multicolumn{3}{|c||}{resolutions ($\mathrm{\mu m}$)} &
\multicolumn{3}{|c|}{resolutions ($\mathrm{\mu m}$)} \\
& $x$ & $y$ & $z$ & $x$ & $y$ & $z$\\
\hline\hline
0$\sigma$ &  9 & 9  & 41  & 14 & 14 & 147 \\
1$\sigma$ & 10 & 10 & 48  & 15 & 15 & 155 \\
3$\sigma$ & 16 & 16 & 81  & 21 & 21 & 226 \\
5$\sigma$ & 23 & 27 & 147 & 28 & 29 & 262 \\
\hline
\end{tabular}
\caption{Values of the position resolutions on the primary and the $B^0$ decay
vertices for the different VELO misalignment scenarios.
The resolutions correspond to the sigmas of single-Gaussian fits.
The errors on all resolutions are around 1-2 \%.}
\label{tab:resolution_velo_2}
\end{center}
\vspace{0.3cm}
\end{table}

\begin{table}
\vspace{0.3cm}
\begin{center}
\begin{tabular}{|c||c|c|}
\hline
Misalignment scenario & Mean & Sigma\\
\hline\hline
0$\sigma$ & $0.06\pm{}0.02$ & $1.14\pm{}0.01$\\
1$\sigma$ & $0.09\pm{}0.02$ & $1.20\pm{}0.01$\\
3$\sigma$ & $0.09\pm{}0.03$ & $1.62\pm{}0.02$\\
5$\sigma$ & $0.18\pm{}0.04$ & $2.08\pm{}0.03$\\
\hline
\end{tabular}
\caption{Values for the mean and sigma of the proper time pulls
for the different VELO misalignment scenarios.}
\label{tab:tau_velo}
\end{center}
\vspace{0.3cm}
\end{table}

\mbox{}
\vspace*{2.0cm}
\subsection{Misalignments in the T stations}
\label{sec:b2hh_t}

\subsubsection{Effect on the pattern recognition}
%
In Table~\ref{tab:pr_ot} both the \Matching\ and the \Forward\ pattern
recognition efficiencies are shown for the 0$\sigma$, 1$\sigma$, 3$\sigma$ and
5$\sigma$ scenarios. Again, all efficiencies are quoted for all long tracks
in the event with no momentum cut applied.
It can be seen that for the set of misalignments considered there is a relative
loss, for the 5$\sigma$ case, of 0.6\% for the \Forward\ efficiency and of
5\% for the \Matching\ efficiency.

\begin{table}[hbtp]
\vspace{0.5cm}
\begin{center}
\begin{tabular}{|c|c|c|}
\hline
Misalignment & \Forward        & \Matching \\ 
scenario     & efficiency (\%) & efficiency (\%) \\
\hline\hline
0$\sigma$    & $85.9 \pm 0.2$   & $81.1 \pm 0.2$\\
1$\sigma$    & $85.8 \pm 0.2$   & $81.0 \pm 0.2$\\
3$\sigma$    & $85.6 \pm 0.2$   & $79.9 \pm 0.4$\\
5$\sigma$    & $85.4 \pm 0.3$   & $77.2 \pm 1.3$\\
\hline
\end{tabular}
\caption{\Matching\ and \Forward\ pattern recognition efficiencies
for various misalignment scenarios of the T-stations.}
\label{tab:pr_ot}
\end{center}
\vspace{0.5cm}
\end{table}

\subsubsection{Effect on the event selection}
In this section the effect of misalignments of the T-stations on the
various discriminating variables shown in Table~\ref{tab:shh} is analysed.

In Figures~\ref{fig:sel_ot} and \ref{fig:sel_ot_2} the distributions of the
discriminating variables are shown for the 0$\sigma$, 1$\sigma$, 3$\sigma$
and 5$\sigma$ cases (refer to Section~\ref{sec:b2hh_velo_sel} for an
explanation of the distributions, lines and cuts applied).

None of the discriminating variables is strongly affected
by the misalignments considered. This explains the relatively small
-- compared to the VELO case -- loss in number of selected events, shown in
Table~\ref{tab:sel_ot}, which amounts to 4.2\% in the worst-case scenario.

\subsubsection{Effect on resolutions}
In Figures~\ref{fig:resolution_ot} and~\ref{fig:resolution_ot_2} the
$B^0$ daughters' momentum, $B^0$ mass and proper time and primary vertex
and $B^0$ vertex resolutions are shown for the 0$\sigma$, 1$\sigma$,
3$\sigma$ and 5$\sigma$ cases. The values of the resolutions (the sigmas of
single-Gaussian fits) are summarised in
Tables~\ref{tab:resolution_ot} and~\ref{tab:resolution_ot_2}.

\begin{table}[hbtp]
\vspace{0.3cm}
\begin{center}
\begin{tabular}{|c|c|}
\hline
Misalignment & Number of \\
scenario     & selected events \\
\hline\hline
0$\sigma$    & 4141  (100\%) \\
1$\sigma$    & 4131  (99.8\%)\\
3$\sigma$    & 4095  (98.9\%)\\
5$\sigma$    & 3969  (95.8\%)\\
\hline
\end{tabular}
\caption{Number of selected events after running the \b2hh selection
for the different T-stations misalignment scenarios.}
\label{tab:sel_ot}
\end{center}
\vspace{0.3cm}
\end{table}

\begin{table}[hbtp]
\vspace{0.3cm}
\begin{center}
\begin{tabular}{|c||c|c|c|}
\hline
                               & Momentum   & Mass             & Proper time \\
\raisebox{1.5ex}{Misalignment} & resolution & resolution       & resolution \\
\raisebox{1.75ex}{scenario}    & (\%)       & ($\mathrm{MeV}$) & ($\mathrm{fs}$) \\
\hline\hline
0$\sigma$         & 0.49 & 22.5 & 37.7\\
1$\sigma$         & 0.50 & 22.6 & 37.4 \\
3$\sigma$         & 0.54 & 23.4 & 37.4 \\
5$\sigma$         & 0.59 & 25.8 & 38.8 \\
\hline
\end{tabular}
\caption{Values of the resolutions on the daughters' momentum, the $B^0$ mass
and the $B^0$ proper time for the different T-stations misalignment scenarios.
The resolutions correspond to the sigmas of single-Gaussian fits.
The errors on all resolutions are around 1-1.5 \%.}
\label{tab:resolution_ot}
\end{center}
\vspace{0.3cm}
\end{table}

\begin{table}[hbtp]
\vspace{0.3cm}
\begin{center}
\begin{tabular}{|c||c|c|c||c|c|c|}
\hline
Misalignment & \multicolumn{3}{|c||}{Primary vertex} & \multicolumn{3}{|c|}{$B^0$ vertex}\\
scenario  &  \multicolumn{3}{|c||}{resolutions ($\mathrm{\mu m}$)} &
\multicolumn{3}{|c|}{resolutions ($\mathrm{\mu m}$)} \\
& $x$ & $y$ & $z$ & $x$ & $y$ & $z$\\
\hline\hline
0$\sigma$ &  9 &  9 & 41 & 14 & 14 & 147 \\
1$\sigma$ &  9 &  9 & 42 & 14 & 14 & 146 \\
3$\sigma$ &  9 &  9 & 42 & 14 & 14 & 145 \\
5$\sigma$ &  9 &  9 & 42 & 14 & 14 & 142 \\
\hline
\end{tabular}
\caption{Values of the position resolutions on the primary and the $B^0$ decay
vertices for the different T-stations misalignment scenarios.
The errors on all resolutions are around 1-2 \%.}
\label{tab:resolution_ot_2}
\end{center}
\vspace{0.3cm}
\end{table}

Hardly any effect is visible on either the primary and $B^0$ decay vertex
resolutions or on the $B^0$ proper time resolution. This is expected because
the resolution on these quantities is dominated by the VELO resolutions
and alignment. On the contrary, a 17\% effect can be seen on the momentum
resolution of the daughters and a 15\% effect on the $B^0$ invariant mass.
Being a two-body decay, the $B^0$ invariant mass resolution is dominated by
the momentum resolution of the daughters, which explains the effect on
the $B^0$ mass resolution.

The effect of misalignments on the proper time error has also been studied.
Figure~\ref{fig:tau_ot} shows the distribution of the proper time error for
the various misalignment scenarios as well as the respective pull
distributions. As already shown in the VELO case,
Section~\ref{sec:b2hh_velo_res}, it can be observed that even in the
aligned case there is a bias in the estimation of the proper time
and that the errors are under-estimated (see Table~\ref{tab:tau_t}).
No significant worsening is observed in case of misalignments.

\begin{table}[hbtp]
\vspace{0.5cm}
\begin{center}
\begin{tabular}{|c||c|c|}
\hline
Misalignment scenario & Mean & Sigma\\
\hline\hline
0$\sigma$ & $0.06\pm{}0.02$ & $1.14\pm{}0.01$\\
1$\sigma$ & $0.06\pm{}0.02$ & $1.14\pm{}0.01$\\
3$\sigma$ & $0.06\pm{}0.02$ & $1.15\pm{}0.01$\\
5$\sigma$ & $0.05\pm{}0.02$ & $1.17\pm{}0.01$\\
\hline
\end{tabular}
\caption{Values for the mean and sigma of the proper time pulls for the different T-stations misalignment scenarios.}
\label{tab:tau_t}
\end{center}
\vspace{0.5cm}
\end{table}

\subsection{Misalignments in the VELO and T stations}
\label{sec:b2hh_all}

\subsubsection{Effect on the pattern recognition}
In Table~\ref{tab:pr_all} the \VeloR, \VeloSpace, \Forward\ and \Matching\
pattern recognition efficiencies for all long tracks in the event with
no momentum cut applied are shown for the
0$\sigma$, 1$\sigma$, 3$\sigma$ and the 5$\sigma$ scenarios.
For the set of misalignments considered there is
a relative loss of 8.6\% for the \Forward\ efficiency and of 12.9\% for
the \Matching\ efficiency. These numbers roughly correspond to the combined
losses due to the misalignments applied independently in the VELO
and in the T-stations, shown in the previous subsections. This indicates that
the two effects are largely uncorrelated.

\begin{table}[hbtp]
\vspace{0.5cm}
\begin{center}
\begin{tabular}{|c|c|c|c|c|}
\hline
Misalignment & \VeloR          & \VeloSpace      & \Forward        & \Matching\\ 
scenario     & efficiency (\%) & efficiency (\%) & efficiency (\%) & efficiency (\%)\\
\hline\hline
0$\sigma$    & $98.0 \pm 0.1$  & $97.0 \pm 0.1$  & $85.9 \pm 0.2$  & $81.1 \pm 0.2$\\
1$\sigma$    & $98.0 \pm 0.1$  & $96.8 \pm 0.1$  & $85.6 \pm 0.2$  & $80.8 \pm 0.2$\\
3$\sigma$    & $98.0 \pm 0.1$  & $94.3 \pm 0.4$  & $83.3 \pm 0.5$  & $77.3 \pm 0.7$\\
5$\sigma$    & $97.8 \pm 0.2$  & $90.1 \pm 1.7$  & $78.5 \pm 1.8$  & $70.6 \pm 1.9$\\
\hline
\end{tabular}
\caption{\VeloR, \VeloSpace, \Forward\ and \Matching\ pattern recognition
efficiencies for various misalignment scenarios of both the VELO
and the T-stations.}
\label{tab:pr_all}
\end{center}
\vspace{0.5cm}
\end{table}

\subsubsection{Effect on the event selection}
In Table~\ref{tab:selall} the number of selected events is shown for the
different misalignment scenarios of both the VELO and the T-stations.

As already shown, if only the T-stations misalignments are considered,
the loss in the number of selected events amounts to 4.2\%, while in the
VELO case, the loss in number of selected events amounts to 73.9\%.
It can be concluded that the 75.6\% loss in number of selected events,
here seen in the worst-case scenario, is mostly due to losses induced by
misalignments in the VELO.
Therefore the distributions of the single-cut variables of the selection
have not been studied again.

It should be kept in mind that though the effects of misalignments on the
performance of the particle identification have not been studied here,
the latter is expected to be influenced mainly by T-stations misalignments.

\begin{table}[hbtp]
\vspace{0.5cm}
\begin{center}
\begin{tabular}{|c|c|}
\hline
Misalignment & Number of \\
scenario     & selected events \\
\hline\hline
0$\sigma$    & 4141  (100\%) \\
1$\sigma$    & 3807  (91.9\%)\\
3$\sigma$    & 2041  (49.3\%)\\
5$\sigma$    & 1009  (24.4\%)\\
\hline
\end{tabular}
\caption{Number of selected events after running the \b2hh selection
for the different misalignment scenarios of both the VELO
and the T-stations considered.}
\label{tab:selall}
\end{center}
\vspace{0.5cm}
\end{table}

\subsubsection{Effect on resolutions}
In Figures~\ref{fig:resolution_all} and \ref{fig:resolution_all_2} the
$B^0$ daughters' momentum, $B^0$ mass and proper time and primary vertex
and $B^0$ vertex resolutions are shown for the 0$\sigma$, 1$\sigma$,
3$\sigma$ and 5$\sigma$ cases. The values of the resolutions (the sigmas of
single-Gaussian fits) are summarised in
Tables~\ref{tab:resolution_all} and \ref{tab:resolution_all_2}.

\begin{table}[hbtp]
\vspace{0.5cm}
\begin{center}
\begin{tabular}{|c||c|c||c|c|}
\hline
                               & Momentum   & Mass             & Proper time \\
\raisebox{1.5ex}{Misalignment} & resolution & resolution       & resolution \\
\raisebox{1.75ex}{scenario}    & (\%)       & ($\mathrm{MeV}$) & ($\mathrm{fs}$) \\
\hline\hline
0$\sigma$         & 0.49 & 22.5 & 37.7\\
1$\sigma$         & 0.50 & 22.3 & 40.9 \\
3$\sigma$         & 0.56 & 25.1 & 58.0 \\
5$\sigma$         & 0.63 & 25.5 & 78.6 \\
\hline
\end{tabular}
\caption{Values of the resolutions on the daughters' momentum, the $B^0$ mass
and the $B^0$ proper time for the different misalignment scenarios of both the
VELO and the T-stations.
The resolutions correspond to the sigmas of single-Gaussian fits.
The errors on all resolutions are around 1-1.5 \%.}
\label{tab:resolution_all}
\end{center}
\vspace{0.5cm}
\end{table}

Comparing these results with the ones previously shown for independent 
misalignments of the VELO and of the T-stations, it can be seen that while
VELO misalignments strongly influence the primary and the $B^0$ vertex
resolutions, and consequently the proper time resolution, T-stations
misalignments have an effect on the daughters' momentum resolution and
therefore on the $B^0$ mass resolution. Both misalignments have complementary
effects.

\begin{table}[hbtp]
\vspace{0.5cm}
\begin{center}
\begin{tabular}{|c||c|c|c||c|c|c|}
\hline
Misalignment & \multicolumn{3}{|c||}{Primary vertex} & \multicolumn{3}{|c|}{$B^0$ vertex}\\
scenario  &  \multicolumn{3}{|c||}{( resolutions $\mathrm{\mu m}$)} &
\multicolumn{3}{|c|}{resolutions ($\mathrm{\mu m}$)} \\
& $x$ & $y$ & $z$ & $x$ & $y$ & $z$\\
\hline\hline
0$\sigma$ &  9 & 9  & 41  & 14 & 14 & 147 \\
1$\sigma$ & 10 & 10 & 48  & 15 & 15 & 159 \\
3$\sigma$ & 14 & 17 & 84  & 20 & 21 & 214 \\
5$\sigma$ & 23 & 27 & 153 & 26 & 31 & 260 \\
\hline
\end{tabular}
\caption{Values of the position resolutions on the primary and the $B^0$ decay
vertices for the different misalignment scenarios of both the VELO
and the T-stations.
The errors on all resolutions are around 1-2 \%.}
\label{tab:resolution_all_2}
\end{center}
\vspace{0.5cm}
\end{table}

Finally, the effect of misalignments on the proper time error has been studied
(see Table~\ref{tab:tau_all}).
Figure~\ref{fig:tau_all} shows the distribution of the proper time error for
the various misalignment scenarios as well as the respective pull
distributions. Again, a bias is observed in the estimation of the proper
time, and the proper time errors are under-estimated.
Comparing these results with the ones previously shown in
Tables~\ref{tab:tau_velo} and Table~\ref{tab:tau_t},
it can be concluded that these results strongly resemble the ones in
Table~\ref{tab:tau_velo}, indicating that the worsening seen here is
originated by the misalignments in the VELO.

\begin{table}[hbtp]
\vspace{0.5cm}
\begin{center}
\begin{tabular}{|c||c|c|}
\hline
Misalignment scenario & Mean & Sigma\\
\hline\hline
0$\sigma$ & $0.06\pm{}0.02$ & $1.14\pm{}0.01$\\
1$\sigma$ & $0.05\pm{}0.02$ & $1.22\pm{}0.02$\\
3$\sigma$ & $0.11\pm{}0.04$ & $1.63\pm{}0.03$\\
5$\sigma$ & $0.15\pm{}0.07$ & $2.10\pm{}0.06$\\
\hline
\end{tabular}
\caption{Values for the mean and sigma of the proper time pulls for the
different  misalignment scenarios of both the VELO and the T-stations.}
\label{tab:tau_all}
\end{center}
\vspace{0.5cm}
\end{table}

\subsection{Effects of changes in the VELO $z$-scale}
\label{sec:b2hh_z-scale}
In addition to studying the effects of random misalignments, the change
of the VELO $z$-scale has been examined.
This is of particular interest to lifetime measurements as it potentially
directly introduces a bias in the measured proper time.

A $z$-scaling effect could be expected from an expansion due to temperature
variations of the VELO components, particularly the Aluminium base plate
onto which the individual modules are screwed.
However, the base plate is kept constant at $20^{\circ}\mathrm{C}$ by
additional local heating.
In addition, the scaling should be limited by the carbon-fibre constraint
system that keeps the modules in place with a precision of
$100~\mathrm{\mu{}m}$ and which is less prone to temperature-induced expansion
given its material\footnote{A conservative estimate using a temperature change
of $10$ K yields a scaling in the $z$-direction of $2\times{}10^{-5}$. The $10$ K are estimated as a maximal change in the temperature of the constraint system as it has a large area contact to the base plate at $20^{\circ}$C and only a small cross-section with the VELO modules at about $-5^{\circ}$C.}.

To assess the influence of an incorrect knowledge of the VELO $z$-scale,
four scenarios with different $z$-scales have been simulated and studied.
For each scenario the $z$-position of each module has been changed according
to the equation
\begin{equation}
z_{module}\rightarrow{}z_{module}\cdot{}(1+scale),
\end{equation}
where $scale$ takes the four values $\frac{1}{3}10^{-4}$, $10^{-4}$,
$\frac{1}{3}10^{-3}$, and $10^{-3}$ for the four scenarios, respectively.

\subsubsection{Effect on the pattern recognition}
%
The first quantities to be studied with a changed VELO $z$-scale were the
pattern recognition efficiencies.
As shown in Table~\ref{tab:pr_z_scaling} no deterioration has been observed
up to a change in the $z$-scale of $1/3\times{}10^{-3}$.
This is expected for the VELO-based pattern recognitions as a $z$-scaling
effectively only changes the track slopes.
For the largest $z$-scaling under study small losses in the VELO-based
pattern recognition efficiencies are observed.
These also propagate to the \Forward\ and \Matching\ efficiencies.

\begin{table}[hbtp]
\vspace{0.5cm}
\begin{center}
\begin{tabular}{|c|c|c|c|c|}
\hline
$z$-scale & \VeloR          & \VeloSpace      & \Forward        & \Matching\\
          & efficiency (\%) & efficiency (\%) & efficiency (\%) & efficiency (\%)\\
\hline\hline
 $1.00000$ & 98.0  & 97.0 & 85.9 & 81.1\\
 $1.00003$ & 98.0  & 97.0 & 85.9 & 81.2\\
 $1.00010$ & 98.0  & 97.0 & 85.9 & 81.2\\
 $1.00033$ & 98.0  & 96.8 & 85.7 & 81.0\\
 $1.00100$ & 96.5  & 94.3 & 83.8 & 79.0\\
\hline
\end{tabular}
\caption{\VeloR, \VeloSpace, \Forward\ and \Matching\ pattern recognition
efficiencies for the various VELO $z$-scaling misalignment scenarios.}
\label{tab:pr_z_scaling}
\end{center}
\vspace{0.5cm}
\end{table}

\subsubsection{Effect on the event selection}
%
When studying the influence of various $z$-scales on the event selection
the situation observed for the pattern recognition performances repeats itself.
The overview of the number of selected events is given in
Table~\ref{tab:sel_z}.
The first four scales under study show only a minor loss in the number of
selected events, while a relative loss of about $20\%$ is observed for the
largest $z$-scale.
As for the studies in the previous chapters, this is due to a
worsening in the resolution of the various cut parameters, where particularly
the VELO-related quantities have shown great sensitivity.

\begin{table}[hbtp]
\vspace{0.5cm}
\begin{center}
\begin{tabular}{|c|c|}
\hline
$z$-scale & Number of \\
    & selected events \\
\hline\hline
$1.00000$         & 4141  (100.0\%)\\
$1.00003$         & 4137  (99.9\%)\\
$1.00010$         & 4142  (100.0\%)\\
$1.00033$         & 4063  (98.1\%)\\
$1.00100$         & 3273  (79.0\%)\\
\hline
\end{tabular}
\caption{Number of selected events after running the \b2hh selection
for the various VELO $z$-scaling misalignment scenarios.}
\label{tab:sel_z}
\end{center}
\vspace{0.5cm}
\end{table}

\subsubsection{Effect on resolutions}
The effect of an incorrectly known VELO $z$-scale on the resolutions of
various physics quantities is summarised in Tables~\ref{tab:resolution_z}
and~\ref{tab:resolutionz_2}.
The relevant resolution distributions are pictured in Figures~\ref{fig:res_z}
and~\ref{fig:res_z_2}.

For the first three $z$-scaling scenarios the observed
changes in the resolutions are minimal.
Only for the two largest $z$-scaling cases one observes a sizeable deterioration
in particular of the proper time and vertex resolutions.

\begin{table}[hbtp]
\vspace{0.5cm}
\begin{center}
\begin{tabular}{|c||c|c|c|}
\hline
          & Momentum   & Mass             & Proper time \\
$z$-scale & resolution & resolution       & resolution \\
          & (\%)       & ($\mathrm{MeV}$) & ($\mathrm{fs}$) \\
\hline\hline
$1.00000$         & 0.49 & 22.5 & 37.7 \\
$1.00003$         & 0.49 & 22.2 & 37.7 \\
$1.00010$         & 0.49 & 22.1 & 37.7 \\
$1.00033$         & 0.49 & 22.0 & 38.5 \\
$1.00100$         & 0.50 & 22.0 & 46.8 \\
\hline
\end{tabular}
\caption{Values of the resolutions on the daughters' momentum, the $B^0$ mass
and the $B^0$ proper time for the various VELO $z$-scaling misalignment
scenarios.
The resolutions correspond to the sigmas of single-Gaussian fits.
The errors on all resolutions are around 1-1.5 \%.}
\label{tab:resolution_z}
\end{center}
\vspace{0.5cm}
\end{table}

\begin{table}[hbtp]
\vspace{0.5cm}
\begin{center}
\begin{tabular}{|c||c|c|c||c|c|c|}
\hline
$z$-scale & \multicolumn{3}{|c||}{Primary vertex} & \multicolumn{3}{|c|}{$B^0$ vertex}\\
          &  \multicolumn{3}{|c||}{resolutions ($\mathrm{\mu m}$)}
                                                  & \multicolumn{3}{|c|}{resolutions ($\mathrm{\mu m}$)} \\
    & $x$ & $y$ & $z$ & $x$ & $y$ & $z$\\
\hline\hline
$1.00000$ &  9 &  9 & 41  & 14 & 14 & 147 \\
$1.00003$ &  9 &  9 & 42  & 14 & 14 & 147 \\
$1.00010$ &  9 &  9 & 42  & 14 & 14 & 145 \\
$1.00033$ &  9 &  9 & 46  & 14 & 14 & 149 \\
$1.00100$ & 11 & 11 & 72  & 16 & 15 & 184 \\
\hline
\end{tabular}
\caption{Values of the resolutions of the primary and the $B^0$ decay vertices
for the various VELO $z$-scaling scenarios.
The errors on all resolutions are around 1-1.5 \%.}
\label{tab:resolutionz_2}
\end{center}
\vspace{0.5cm}
\end{table}

Looking at the pull distributions for the reconstructed proper time shown in
Figure~\ref{fig:tau_z} and their summary in Table~\ref{tab:tau_z},
it appears that there is no significant change in the proper time bias due to
a change in the $z$ scale. 
This is expected as, even for the largest $z$-scale under study, the estimated effect on the pull mean is of the order of its uncertainty.

\begin{table}[hbtp]
\vspace{0.5cm}
\begin{center}
\begin{tabular}{|c||c|c|}
\hline
$z$-scale & Mean & Sigma\\
\hline\hline
$1.00000$ & $0.06\pm{}0.02$ & $1.14\pm{}0.01$\\
$1.00003$ & $0.06\pm{}0.02$ & $1.15\pm{}0.01$\\
$1.00010$ & $0.07\pm{}0.02$ & $1.15\pm{}0.02$\\
$1.00033$ & $0.07\pm{}0.02$ & $1.15\pm{}0.01$\\
$1.00100$ & $0.05\pm{}0.02$ & $1.35\pm{}0.02$\\
\hline
\end{tabular}
\caption{Values for the mean and sigma of the proper time pulls
for the various VELO $z$-scaling scenarios.}
\label{tab:tau_z}
\end{center}
\vspace{0.5cm}
\end{table}

\clearpage
\section{Conclusions}
\label{sec:conclusions}
In this note an extensive study of the effects of misalignments of the
tracking stations on the analysis of B decays has been presented,
illustrated by the example channel of \b2hh.

The effects of the VELO and of the downstream tracking
stations IT and OT are rather decoupled. A summary for various quantities
related to the $B^0$ candidates is given in Table~\ref{tab:summary}.

\begin{table}[hbtp]
\vspace{0.5cm}
\begin{center}
\begin{tabular}{|l||c|c|}
\hline
 & Affected by & Affected by\\
 & VELO misalignments & T misalignments\\
\hline\hline
$B^0$ daughters momentum &  no & yes\\
$B^0$ mass               &  no & yes\\
$B^0$ vertex             & yes & no\\
$B^0$ impact parameter   & yes & no\\
$B^0$ proper time        & yes & no\\
\hline
\end{tabular}
\caption{Summary of the combined effect of misalignments in the VELO and
the T-stations on various physics quantities involved in the selection of
\b2hh events.}
\label{tab:summary}
\end{center}
\vspace{0.5cm}
\end{table}

This study has shown that misalignments of the order of one third of the detector
single-hit resolutions (our ``1$\sigma$'' scales) have little or negligible
effects on the quality of the reconstruction
and of the analysis of \b2hh decays.

The impact of misalignments on the performance of the pattern recognition
algorithms and on the primary vertex resolutions has also been assessed
for the first time. 
It is important to realise that these quantitative results obtained are
rather general and not restricted to \b2hh decays -- they are independent
of the actual B decay.

A natural follow-up study would involve the assessment of residual effects
after application of the several dedicated alignment algorithms being developed
at present.
Preliminary studies with the latest versions of the algorithms seem to
indicate a performance leading to residual misalignments of the order of
our ``1$\sigma$'' scales. A detailed discussion is left for a future analysis.

\appendix
\section{Comments on misalignments sub-samples}
\label{sec:subsamples}

As explained in Section~\ref{sec:misalignments}, all 20$k$ samples of
each scenario considered have been produced as 10 sub-samples of 2$k$ events,
each of which using a different set of the 10 sets of a particular
misalignment scenario.
This procedure was chosen to avoid any potentially ``friendly'' or
``catastrophic'' set of misalignments.

\begin{figure}[p]
\vfill
\begin{center}
\setlength{\unitlength}{1.0cm}
\begin{picture}(10.,14.)
\put(0.0,7.){\scalebox{0.60}{\includegraphics{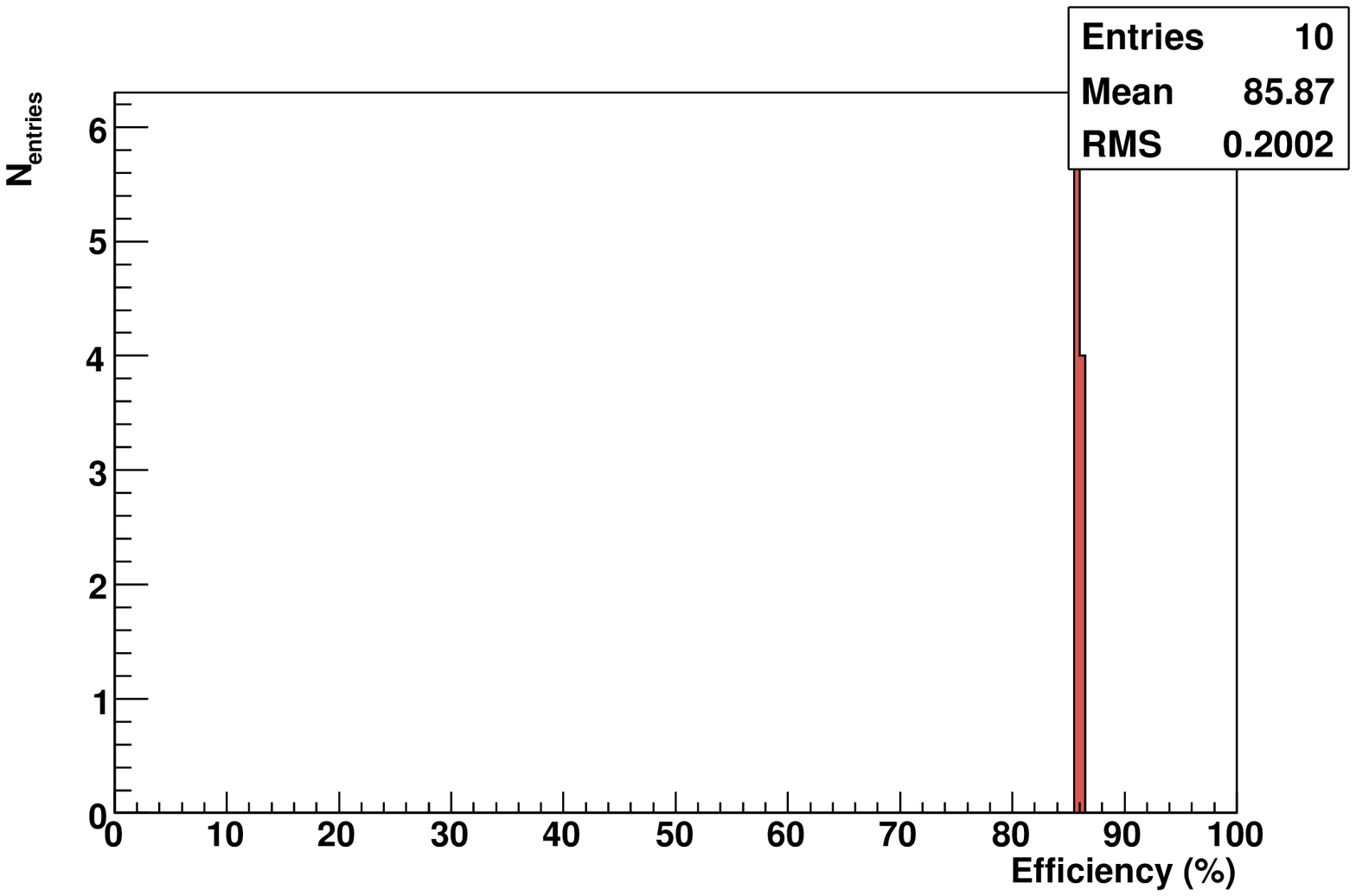}}}
\put(0.0,0.){\scalebox{0.60}{\includegraphics{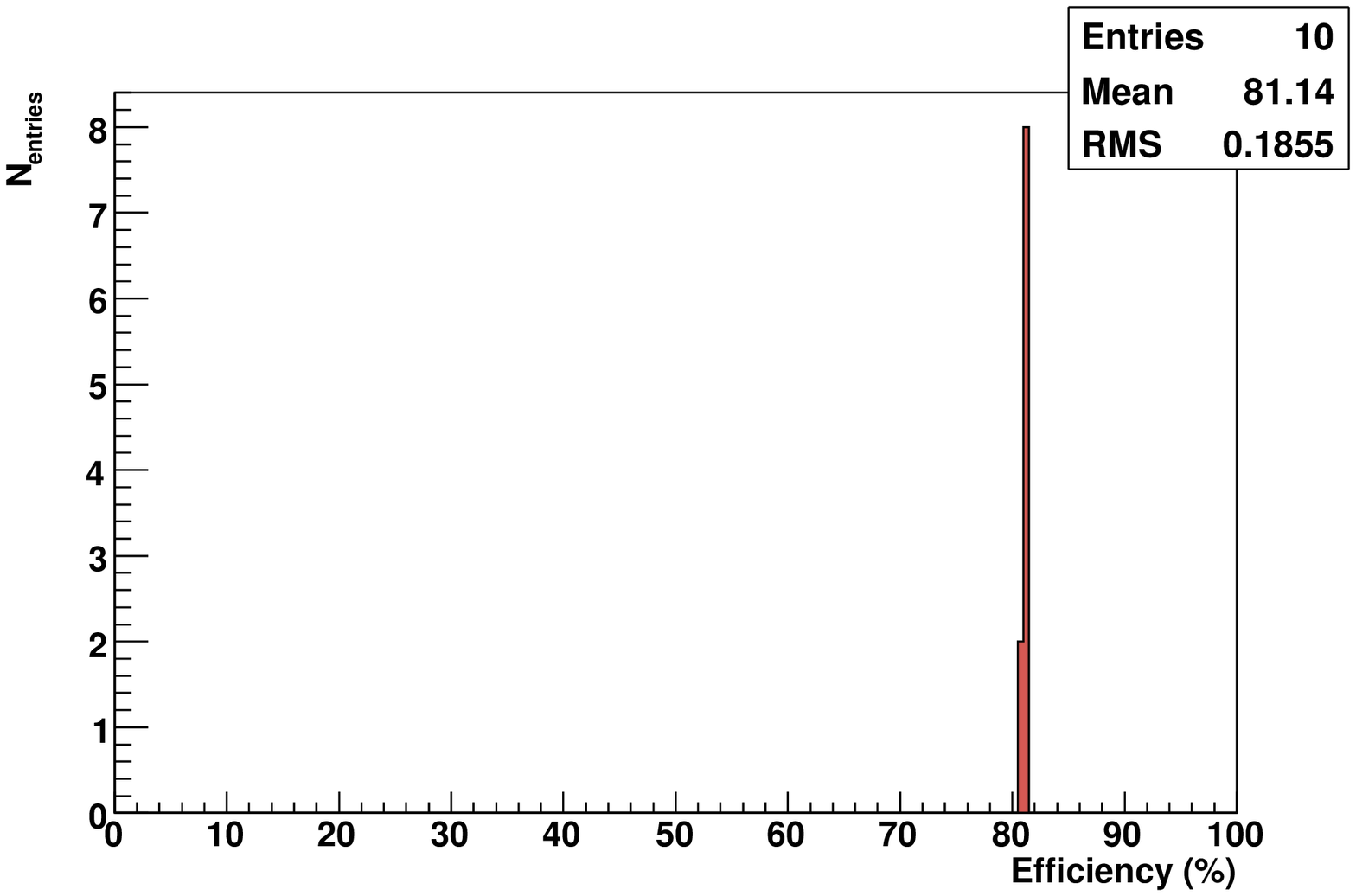}}}
\put(2.0,11.5){\small (a)}
\put(2.0,4.5){\small (b)}
\end{picture}
\end{center}
\caption{Pattern recognition efficiencies with a perfect alignment for
(a) the \Forward\ and (b) the \Matching\ algorithms.}
\label{fig:pr_0sigma}
\vfill
\end{figure}

\begin{figure}[p]
\vfill
\begin{center}
\setlength{\unitlength}{1.0cm}
\begin{picture}(10.,14.)
\put(0.0,7.){\scalebox{0.60}{\includegraphics{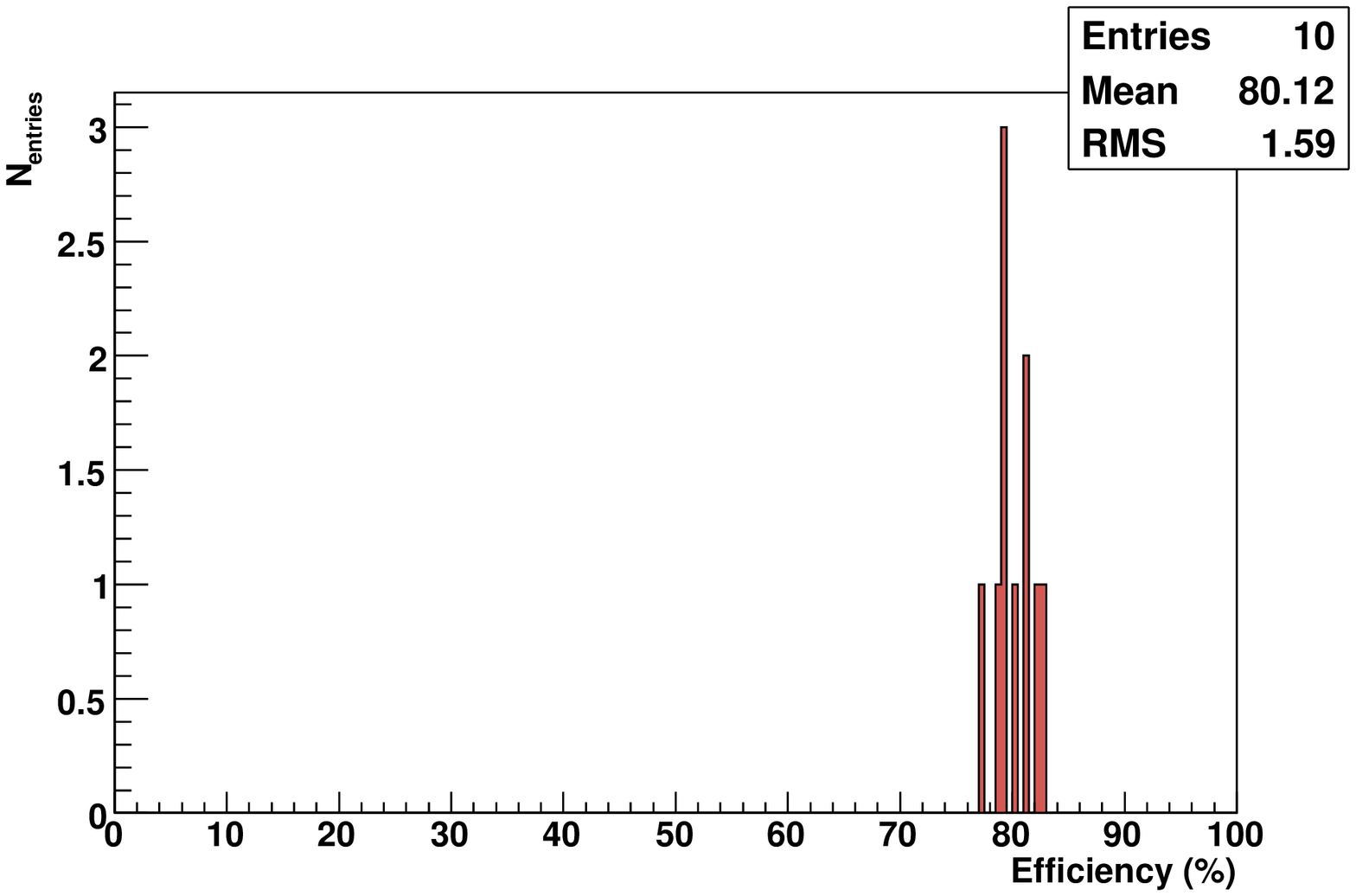}}}
\put(0.0,0.){\scalebox{0.60}{\includegraphics{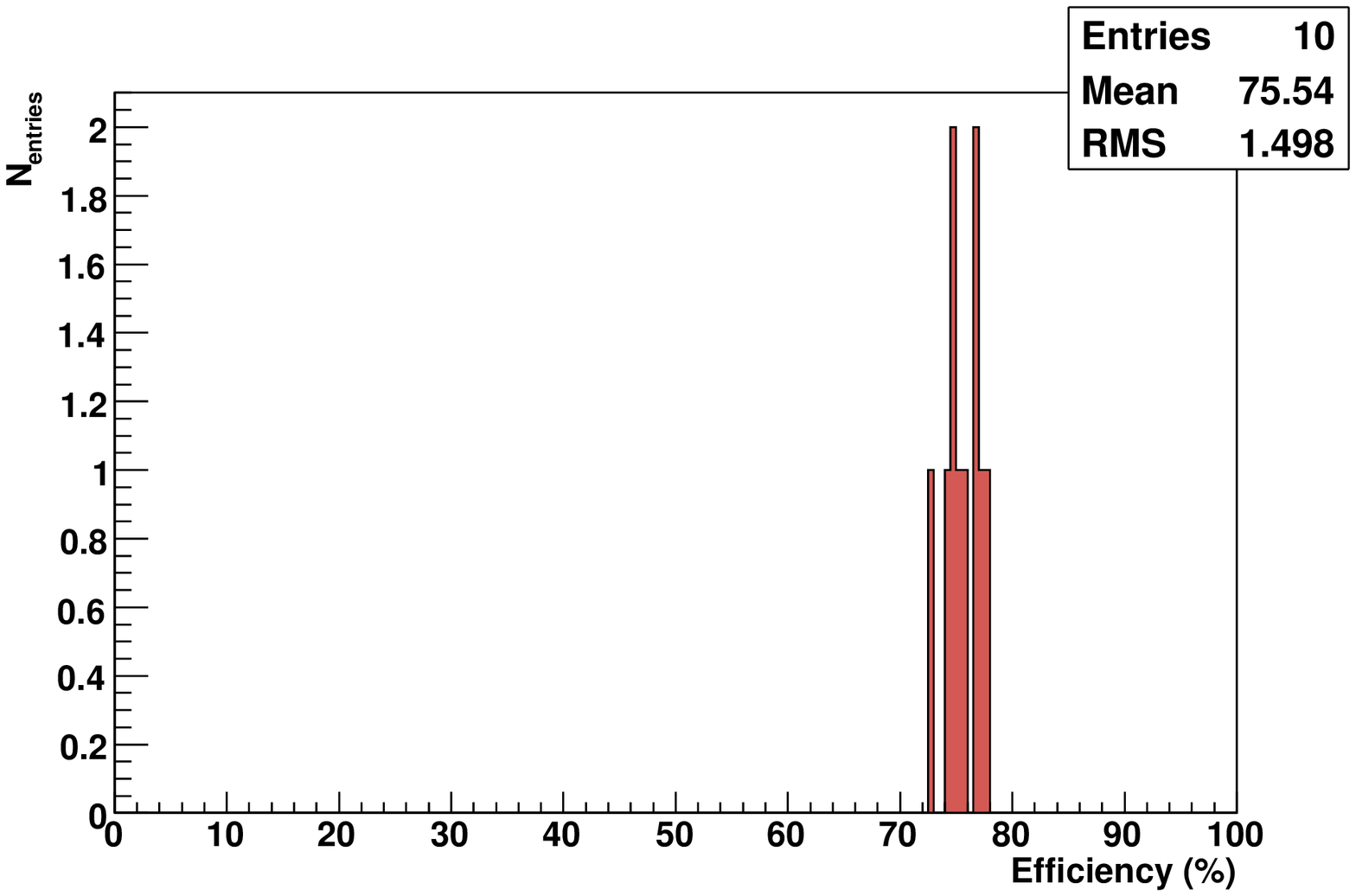}}}
\put(2.0,11.5){\small (a)}
\put(2.0,4.5){\small (b)}
\end{picture}
\end{center}
\caption{Pattern recognition efficiencies with 5$\sigma$ misalignments
of the VELO for (a) the \Forward\ and (b) the \Matching\ algorithms.}
\label{fig:pr_velo}
\vfill
\end{figure}

\begin{figure}[p]
\vfill
\begin{center}
\setlength{\unitlength}{1.0cm}
\begin{picture}(10.,14.)
\put(0.0,7.){\scalebox{0.60}{\includegraphics{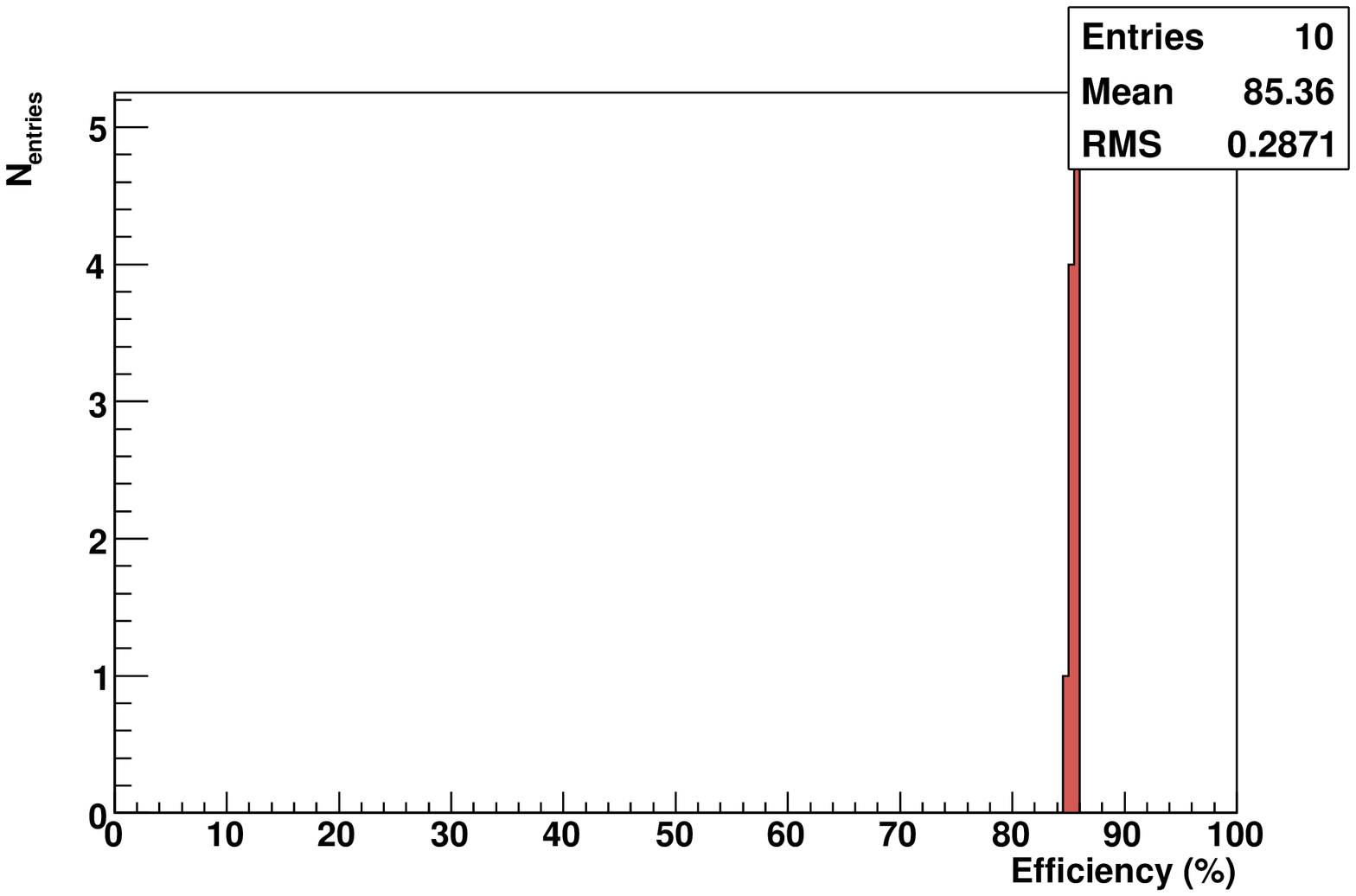}}}
\put(0.0,0.){\scalebox{0.60}{\includegraphics{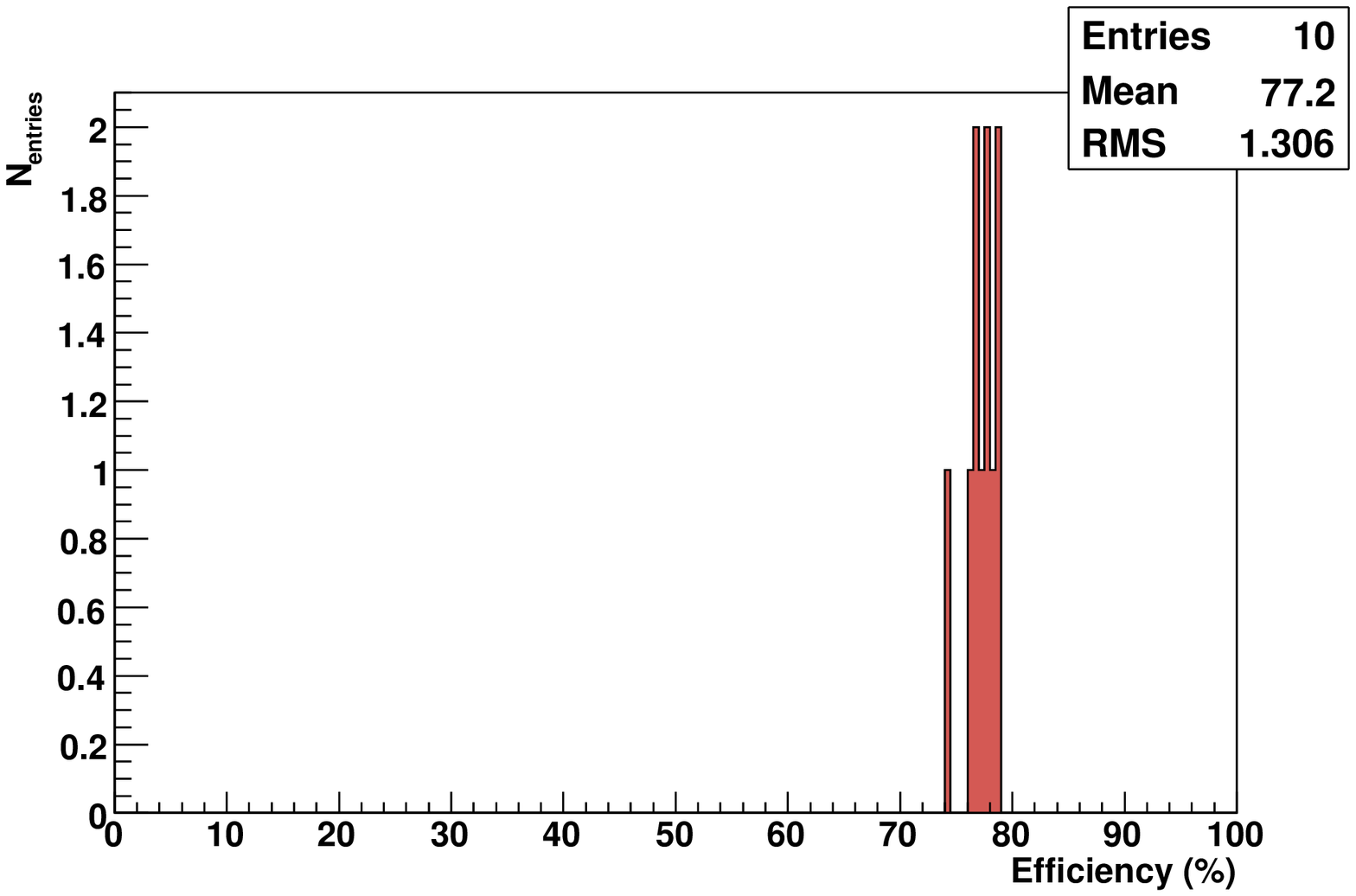}}}
\put(2.0,11.5){\small (a)}
\put(2.0,4.5){\small (b)}
\end{picture}
\end{center}
\caption{Pattern recognition efficiencies with 5$\sigma$ misalignments
of the T-stations for (a) the \Forward\ and (b) the \Matching\ algorithms.}
\label{fig:pr_ot}
\vfill
\end{figure}

\begin{figure}[p]
\vfill
\begin{center}
\setlength{\unitlength}{1.0cm}
\begin{picture}(10.,14.)
\put(0.0,7.){\scalebox{0.60}{\includegraphics{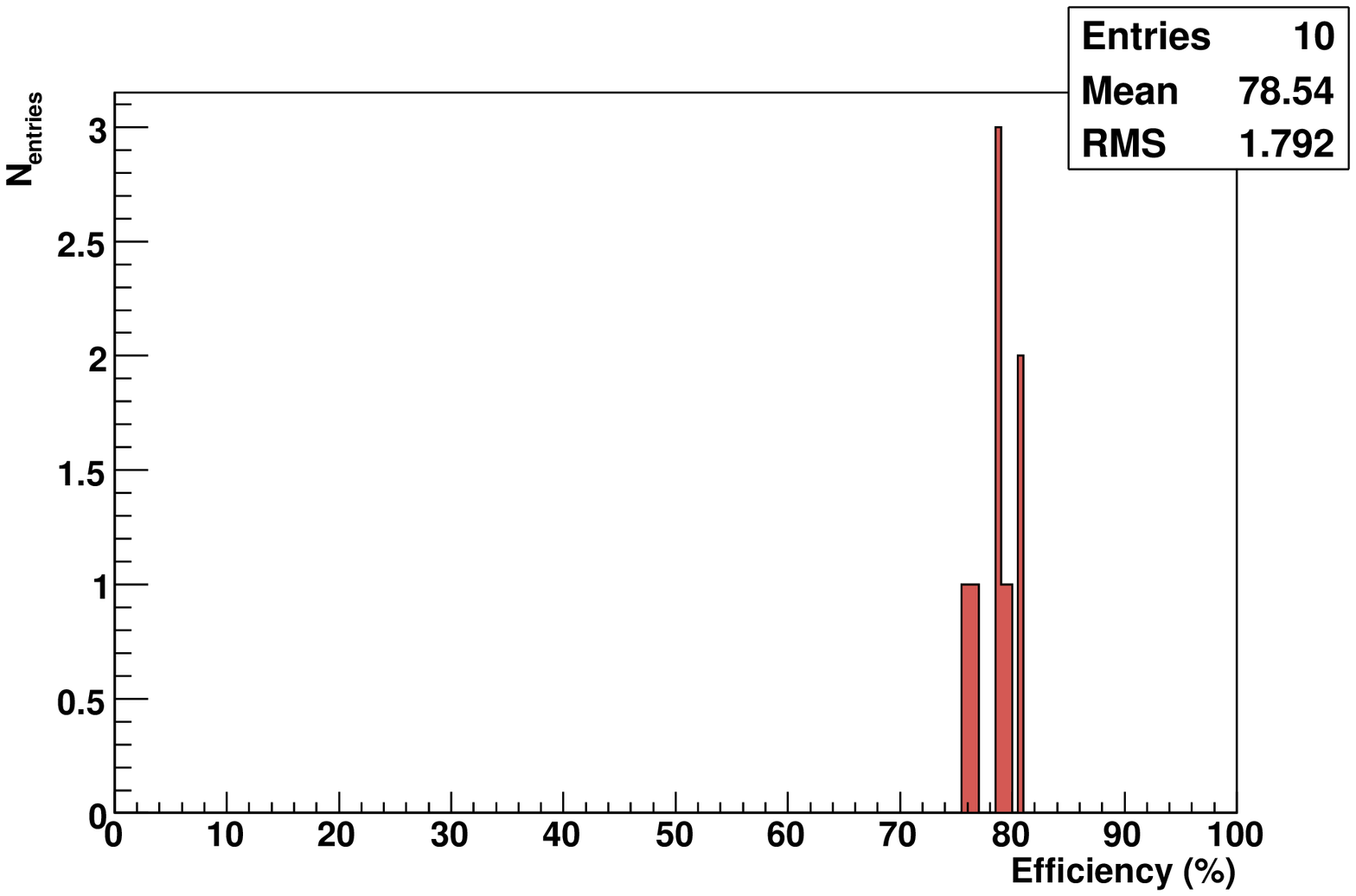}}}
\put(0.0,0.){\scalebox{0.60}{\includegraphics{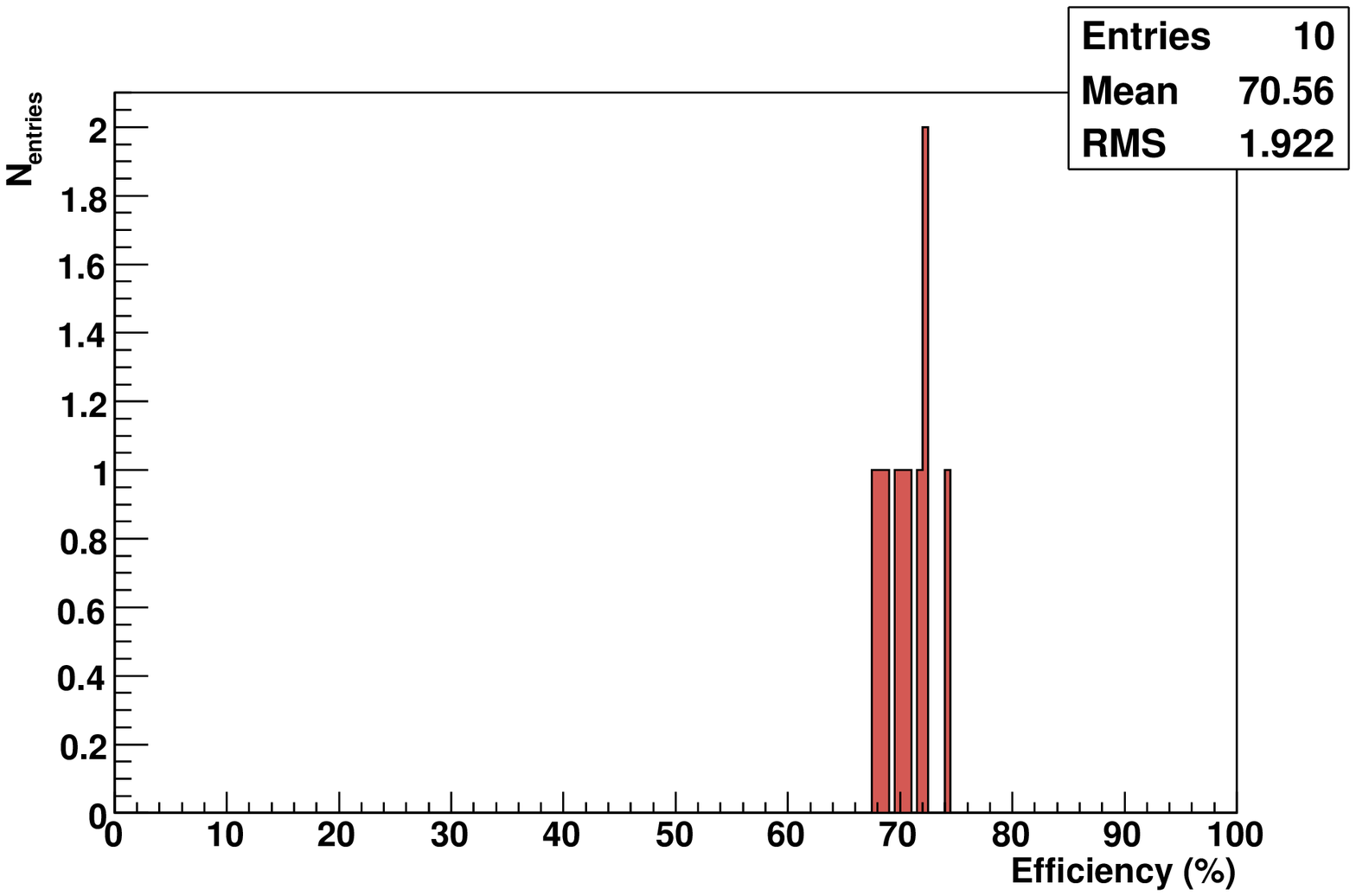}}}
\put(2.0,11.5){\small (a)}
\put(2.0,4.5){\small (b)}
\end{picture}
\end{center}
\caption{Pattern recognition efficiencies with 5$\sigma$ misalignments
of the VELO and the T-stations for (a) the \Forward\ and
(b) the \Matching\ algorithms.}
\label{fig:pr_all}
\vfill
\end{figure}

Figures~\ref{fig:pr_0sigma},~\ref{fig:pr_velo},~\ref{fig:pr_ot}
and~\ref{fig:pr_all} exemplify the distributions (over the 10 sets) of
pattern recognition efficiencies for the \Matching and the \Forward
algorithms respectively for the 0$\sigma$ case and for the $5\sigma$
misalignments cases of the VELO, the T-stations, and the VELO and
T-stations.
Pattern recognition efficiency variations of the order of a few percent are
observed on a per sub-sample basis.

Furthermore, regarding the measurement of resolutions it can be argued that when averaging over 10 samples of different misalignments one measures not only the average resolution but also a contribution coming from the spread of a potential bias in the single samples.
To assess the size of this effect the resolutions and biases have been measured on each of the 10 samples.
Thereafter, the average resolution was compared to the spread of the observed biases by taking a ratio of these quantities.
In the ideal case of a negligible bias this ratio would take values larger than 1.
Over all the resolutions under study this quantity varied between 14 and 36 in the $0\sigma$ case and between 3 and 15 in the $5\sigma$ case.
This means that even in the worst case the contribution of the variation in the bias is a factor 3 smaller than the average resolution.
When combining these numbers, by naively adding them in quadrature, one arrives at the conclusion that a variation in the bias of the single misalignment samples should account for at most $10\%$ of the measured resolution.
Hence the method of averaging over 10 samples of different misalignments is valid and does not lead to significantly wrong results.



\begin{figure}[p]
\begin{center}
\setlength{\unitlength}{1.0cm}
\begin{picture}(14.,14.)
\put(0.0,7.5){\scalebox{0.35}{\includegraphics{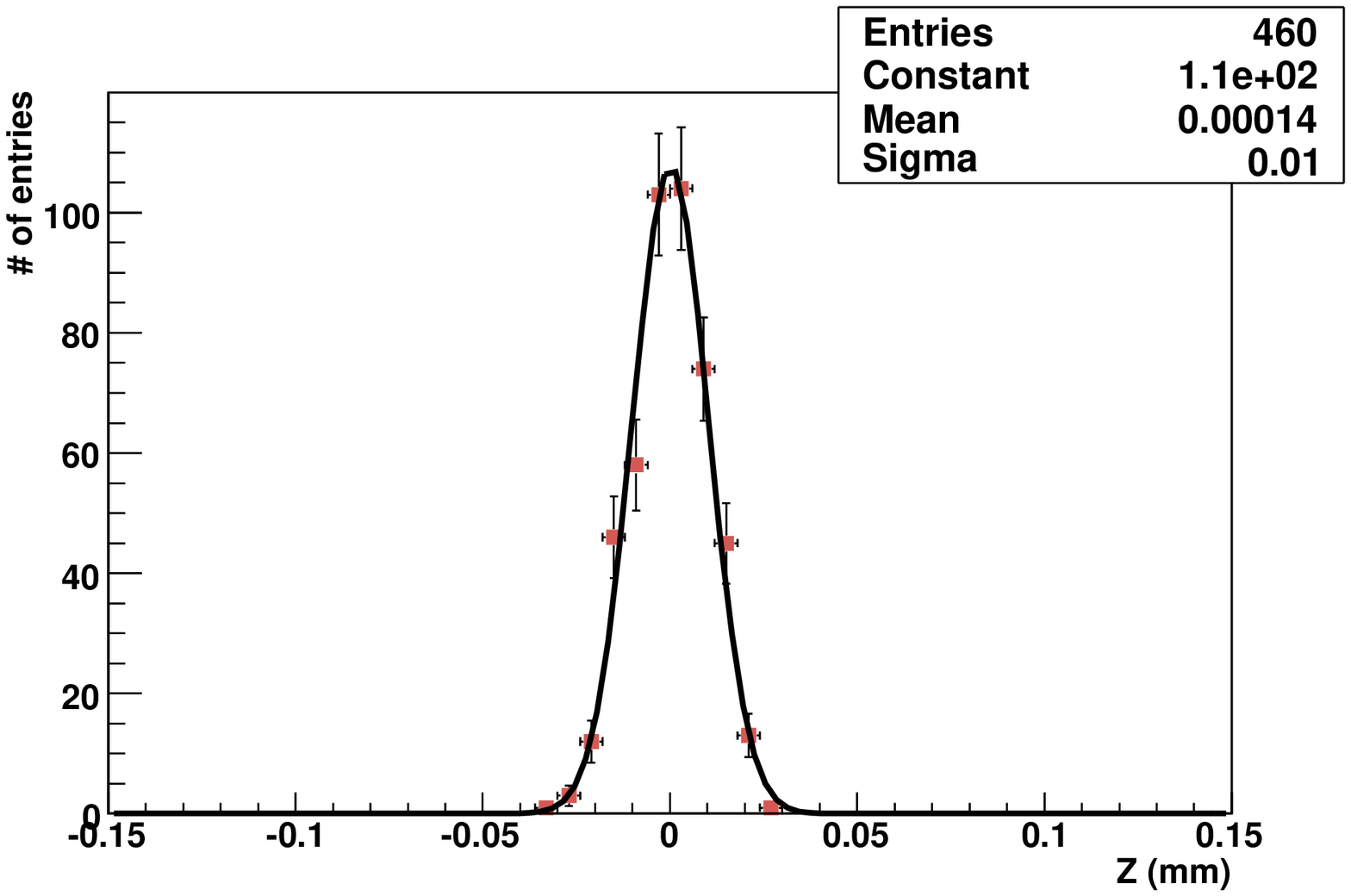}}}
\put(7.5,7.5){\scalebox{0.35}{\includegraphics{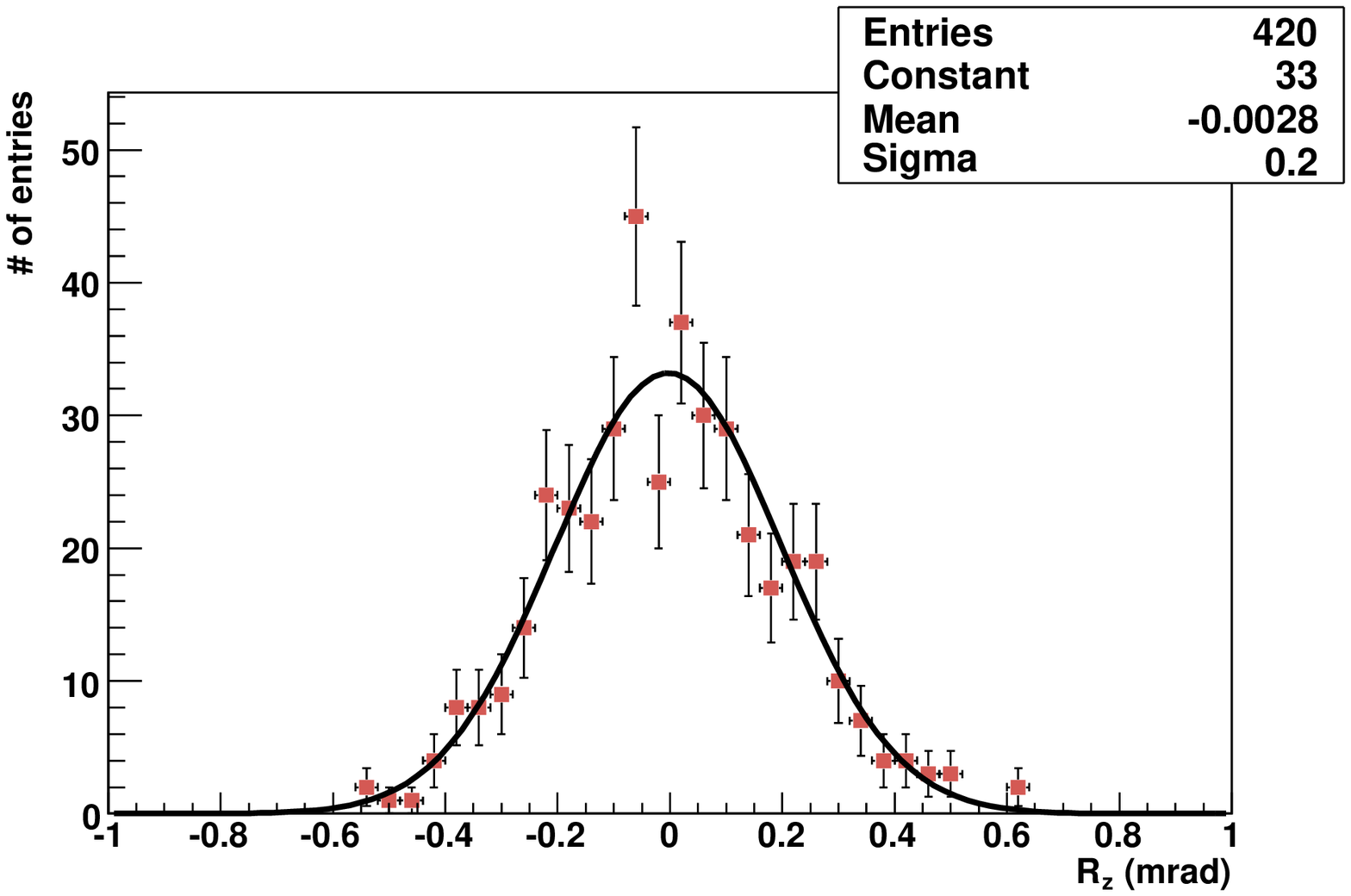}}}
\put(0.0,0.0){\scalebox{0.35}{\includegraphics{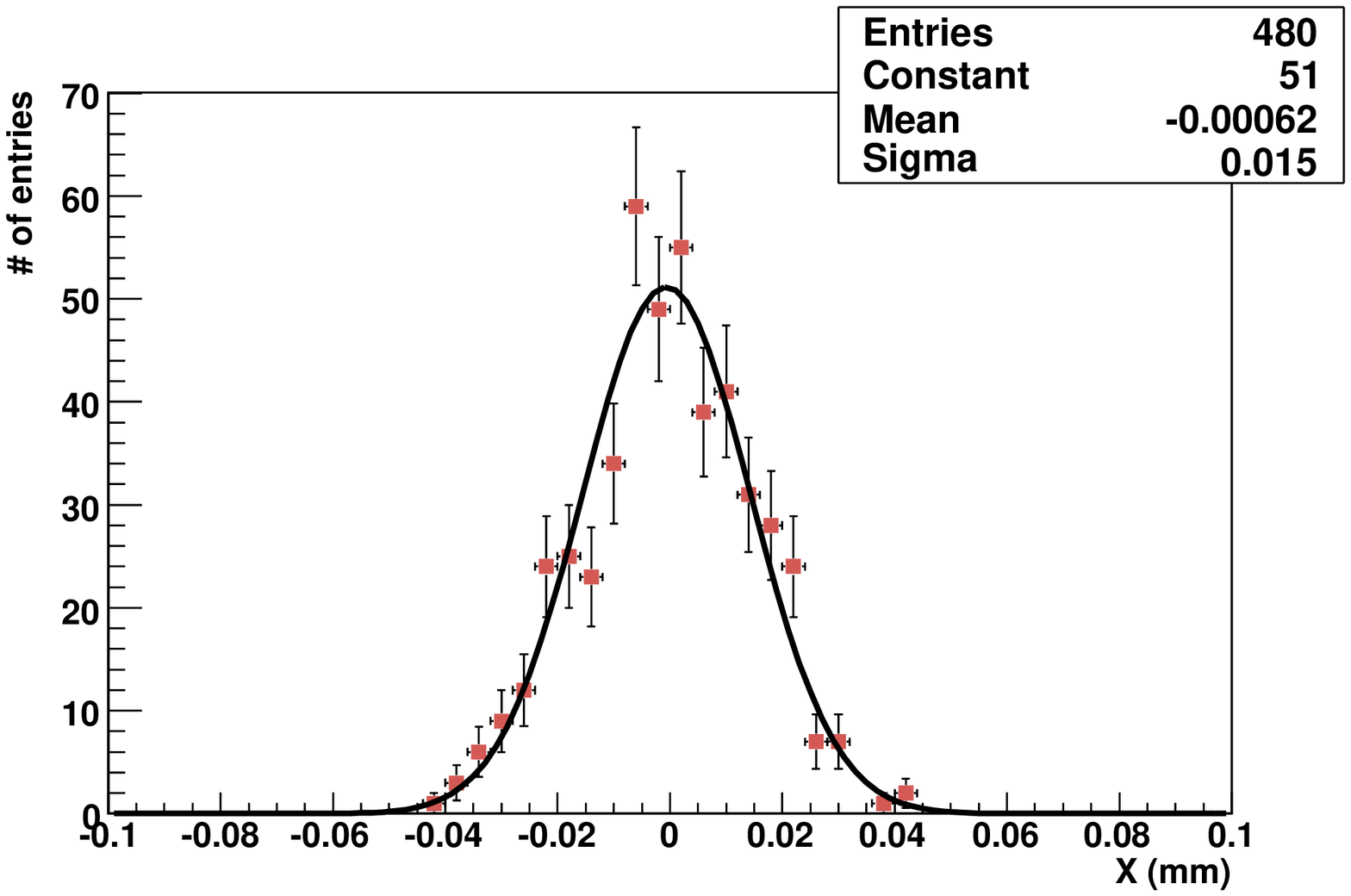}}}
\put(7.5,0.0){\scalebox{0.35}{\includegraphics{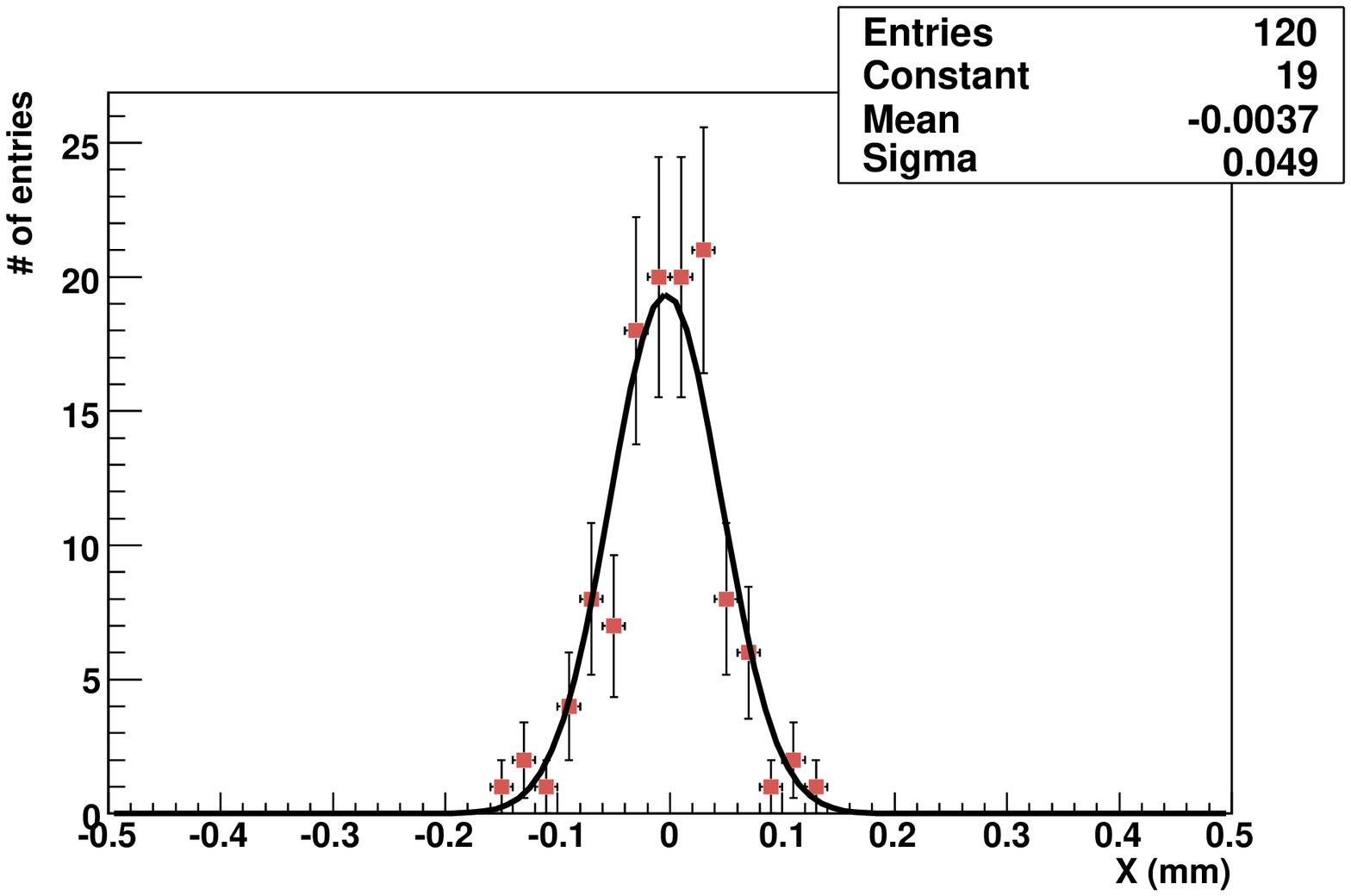}}}
\put(2.0,12.5){\small (a)}
\put(9.0,12.5){\small (b)}
\put(2.0,5.0){\small (c)}
\put(9.0,5.0){\small (d)}
\end{picture}
\end{center}
\caption{Gaussian distributions for the 1$\sigma$ misalignment scales
introduced in the conditions databases for (a) VELO modules $z$ translations,
(b) VELO sensors $z$ rotations, (c) IT boxes $x$ translations and
(d) OT layers $x$ translations.}
\label{fig:db}
\vfill
\end{figure}

\begin{figure}[p]
\vfill
\begin{center}
\setlength{\unitlength}{1.0cm}
\begin{picture}(14.,18.5)
\put(0.0,12.6){\scalebox{0.32}{\includegraphics{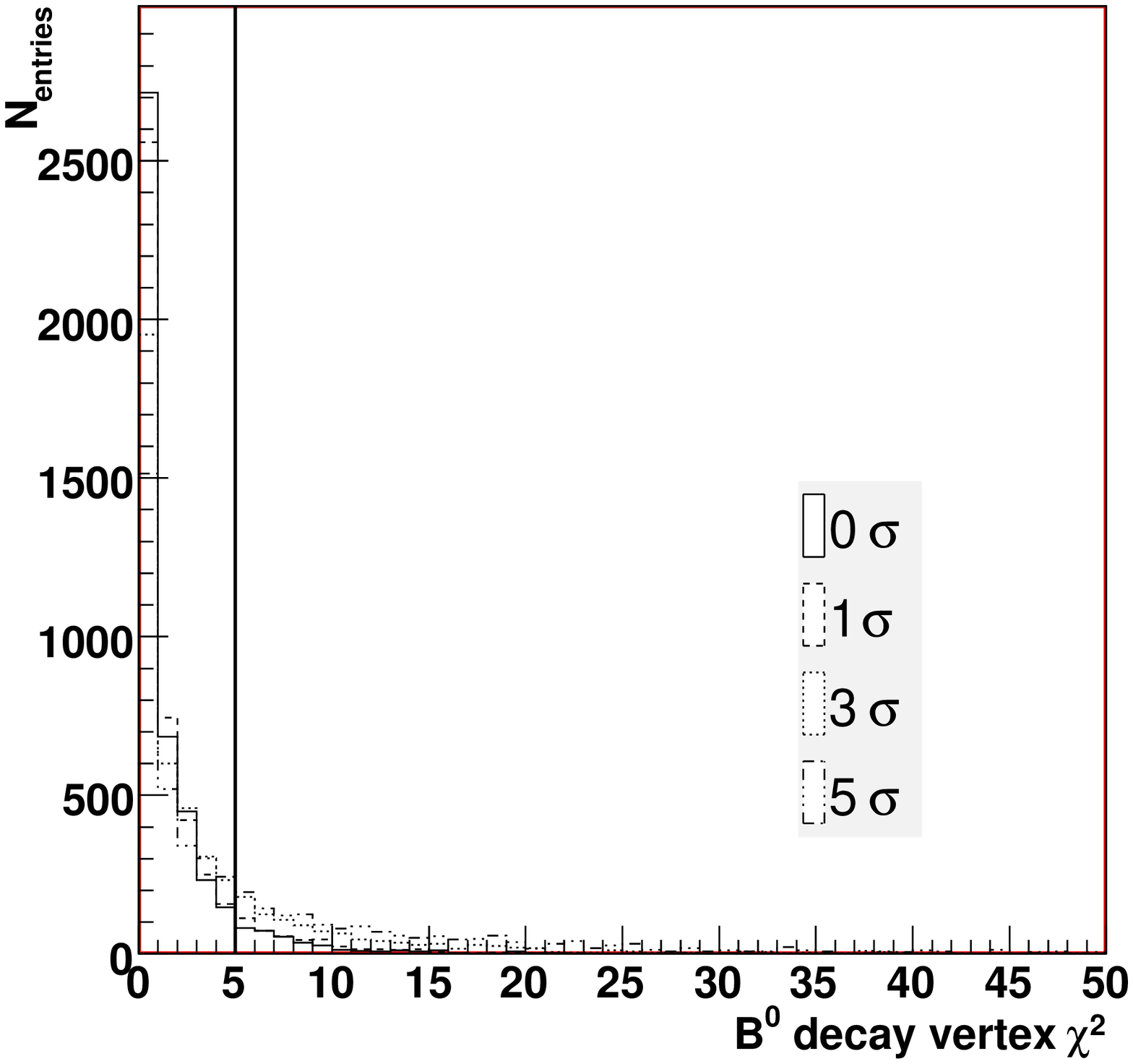}}}
\put(7.0,12.6){\scalebox{0.32}{\includegraphics{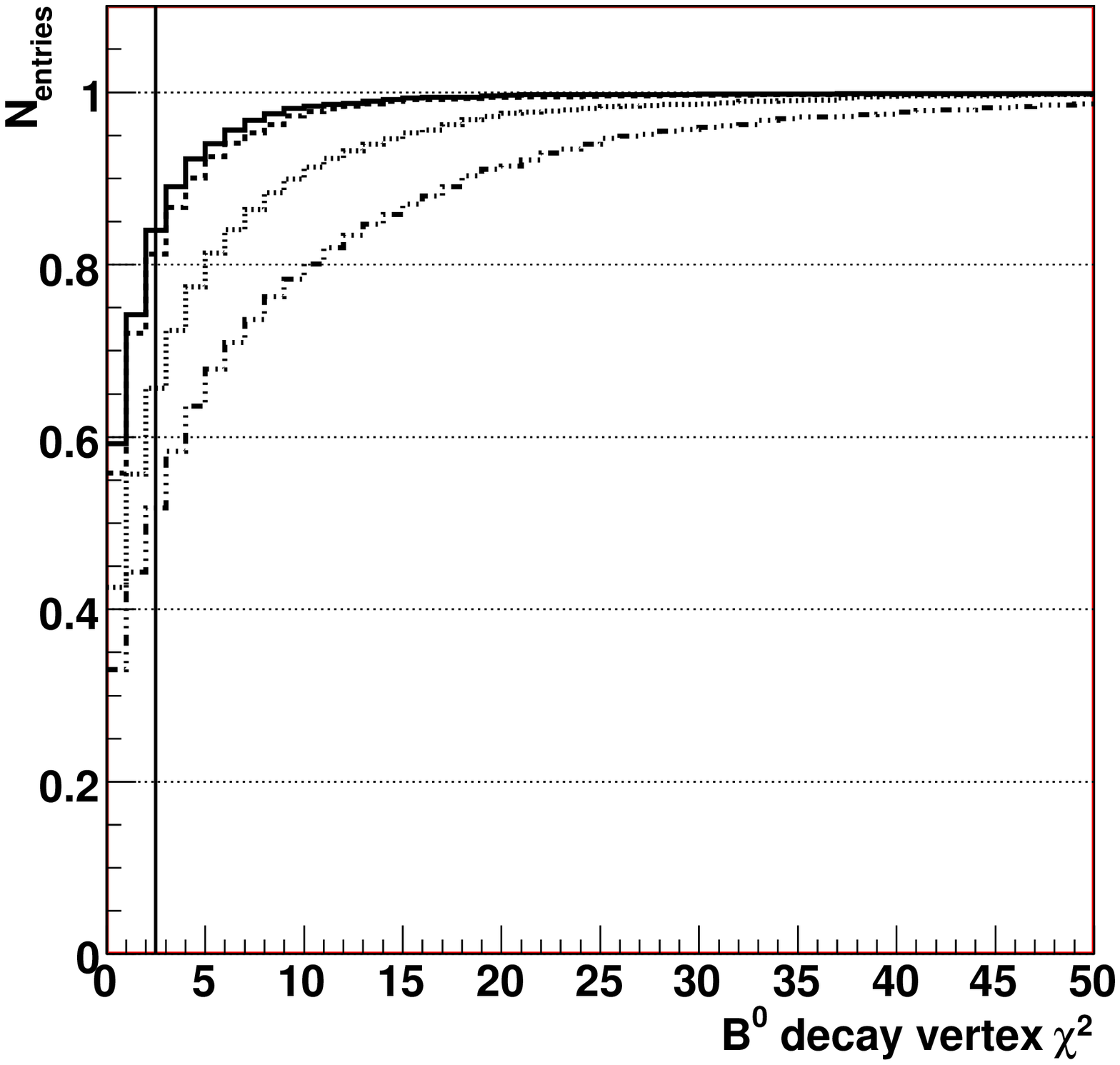}}}
\put(0.,6.3){\scalebox{0.32}{\includegraphics{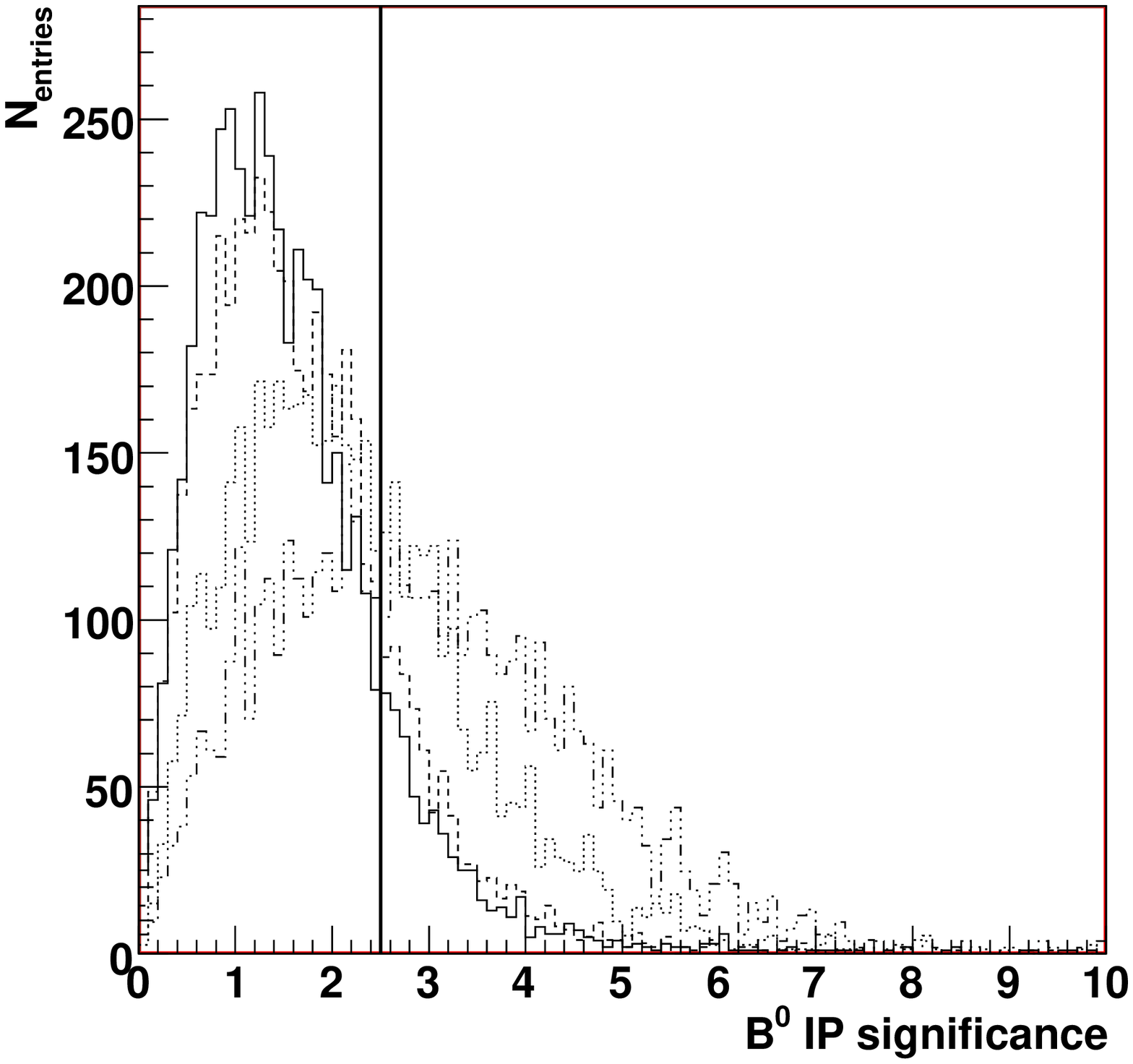}}}
\put(7.0,6.3){\scalebox{0.32}{\includegraphics{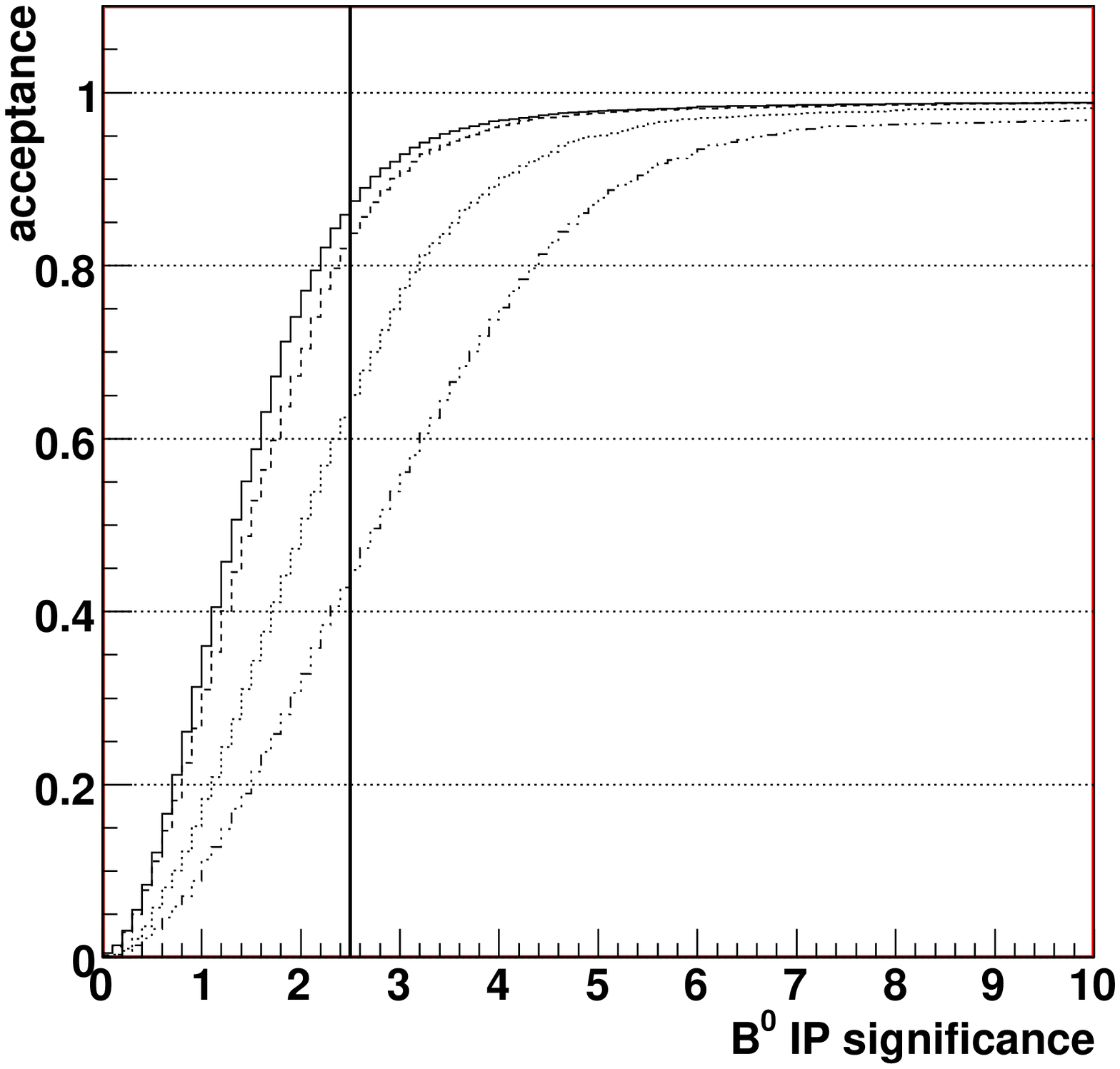}}}
\put(0.0,0.){\scalebox{0.32}{\includegraphics{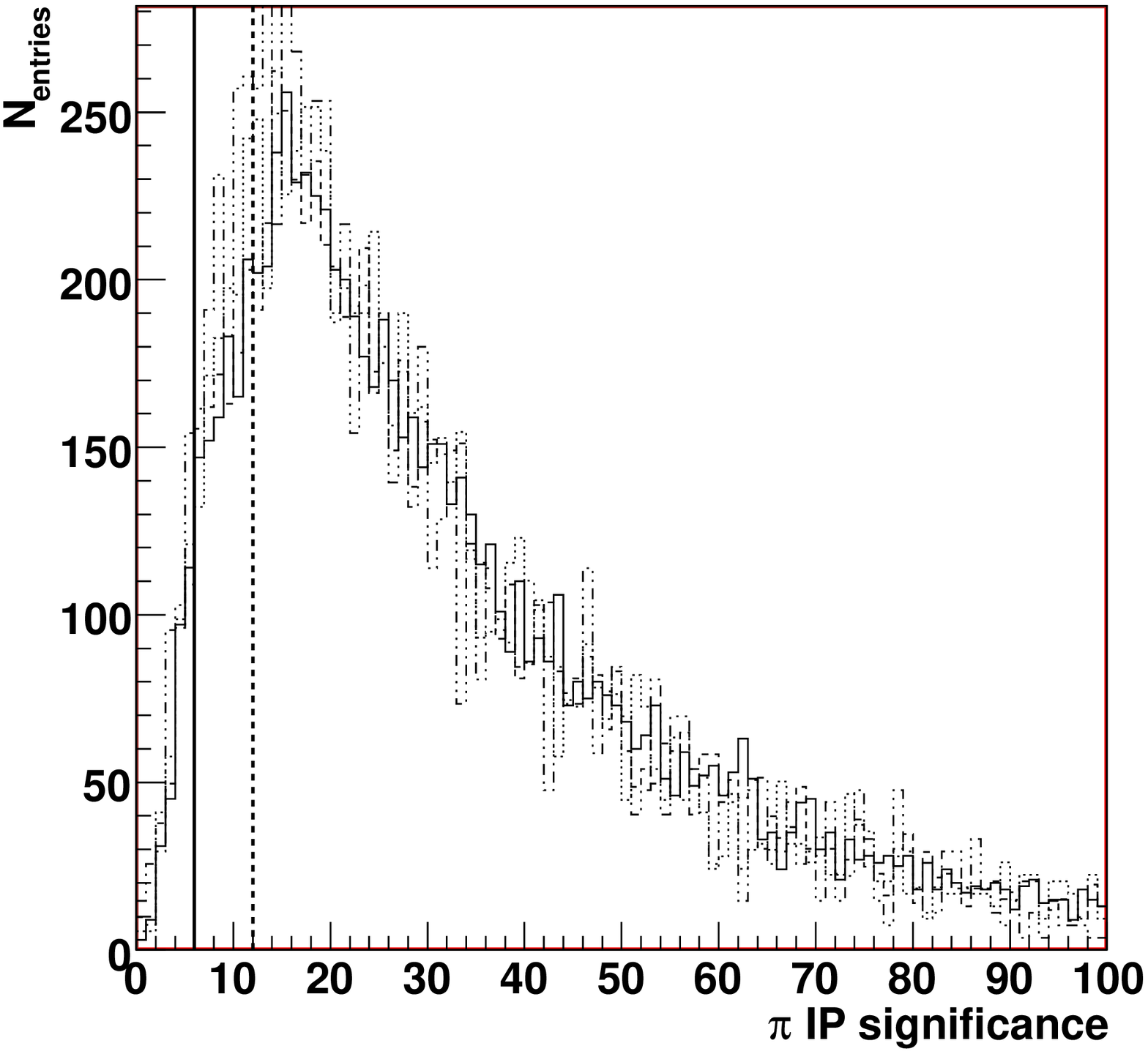}}}
\put(7.0,0.){\scalebox{0.32}{\includegraphics{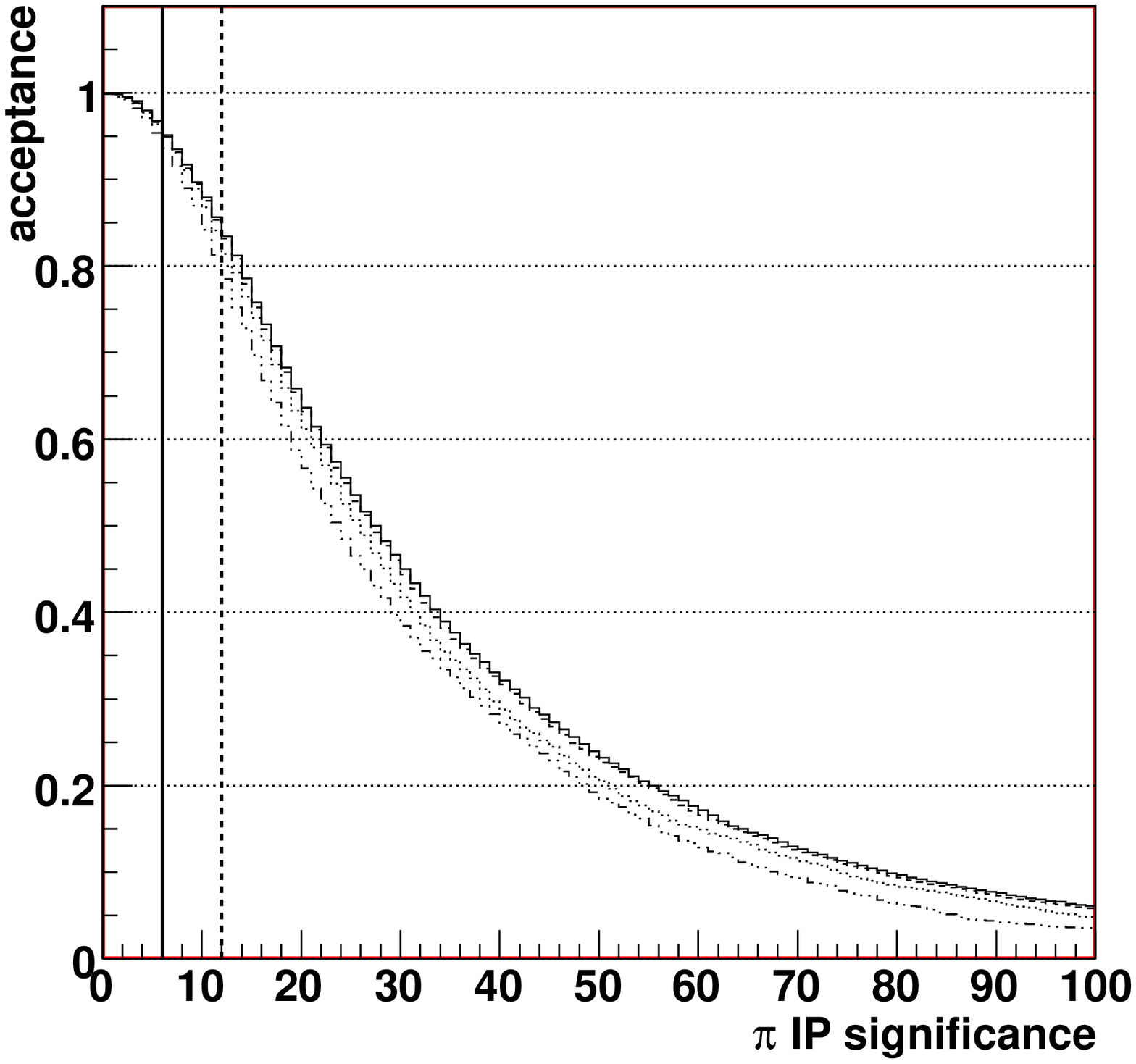}}}
\put(4.0,16.5){\small (a)}
\put(11.0,16.5){\small (b)}
\put(4.0,10.5){\small (c)}
\put(11.0,10.5){\small (d)}
\put(4.0,4.5){\small (e)}
\put(11.0,4.5){\small (f)}
\end{picture}
\end{center}
\caption{Effect of VELO misalignments on the $B^0$ decay vertex $\chi^2$
and impact parameter significances for the $B^0$ candidate and its
daughter pions.
The right-hand-side distributions correspond to the integrated left-hand-side
distributions.
The full line represents the $0\sigma$ misalignment scenario; the dashed line
represents the $1\sigma$ scenario; the dotted line represents the $3\sigma$
scenario and the dot-dashed line represents the $5\sigma$ scenario.
The vertical cut lines are detailed in Section~\ref{sec:b2hh_velo_sel}.}
\label{fig:sel_velo}
\vfill
\end{figure}

\clearpage
\begin{figure}[p]
\vfill
\begin{center}
\setlength{\unitlength}{1.0cm}
\begin{picture}(14.,18.5)
\put(0.,12.6){\scalebox{0.32}{\includegraphics{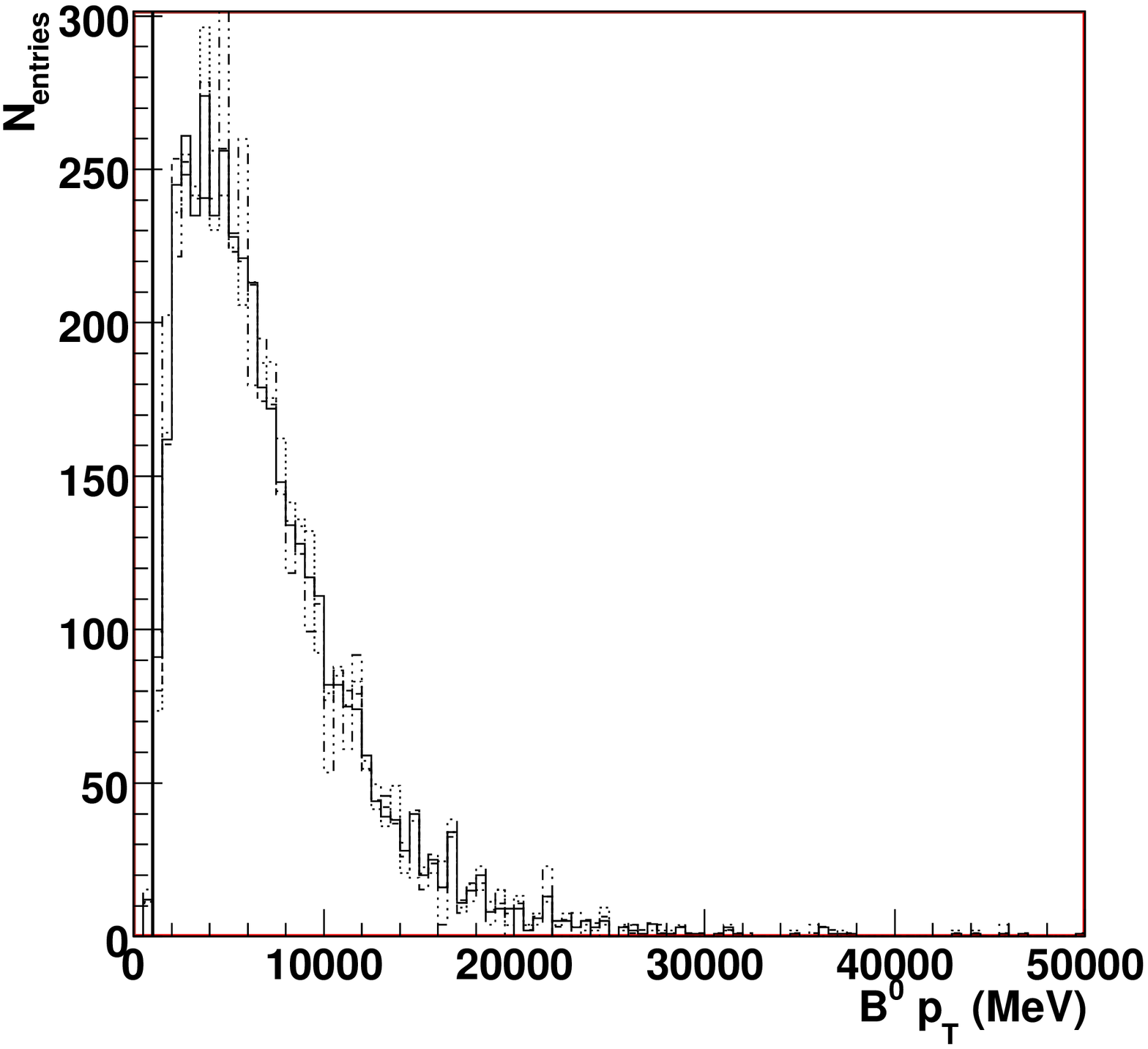}}}
\put(7.0,12.6){\scalebox{0.32}{\includegraphics{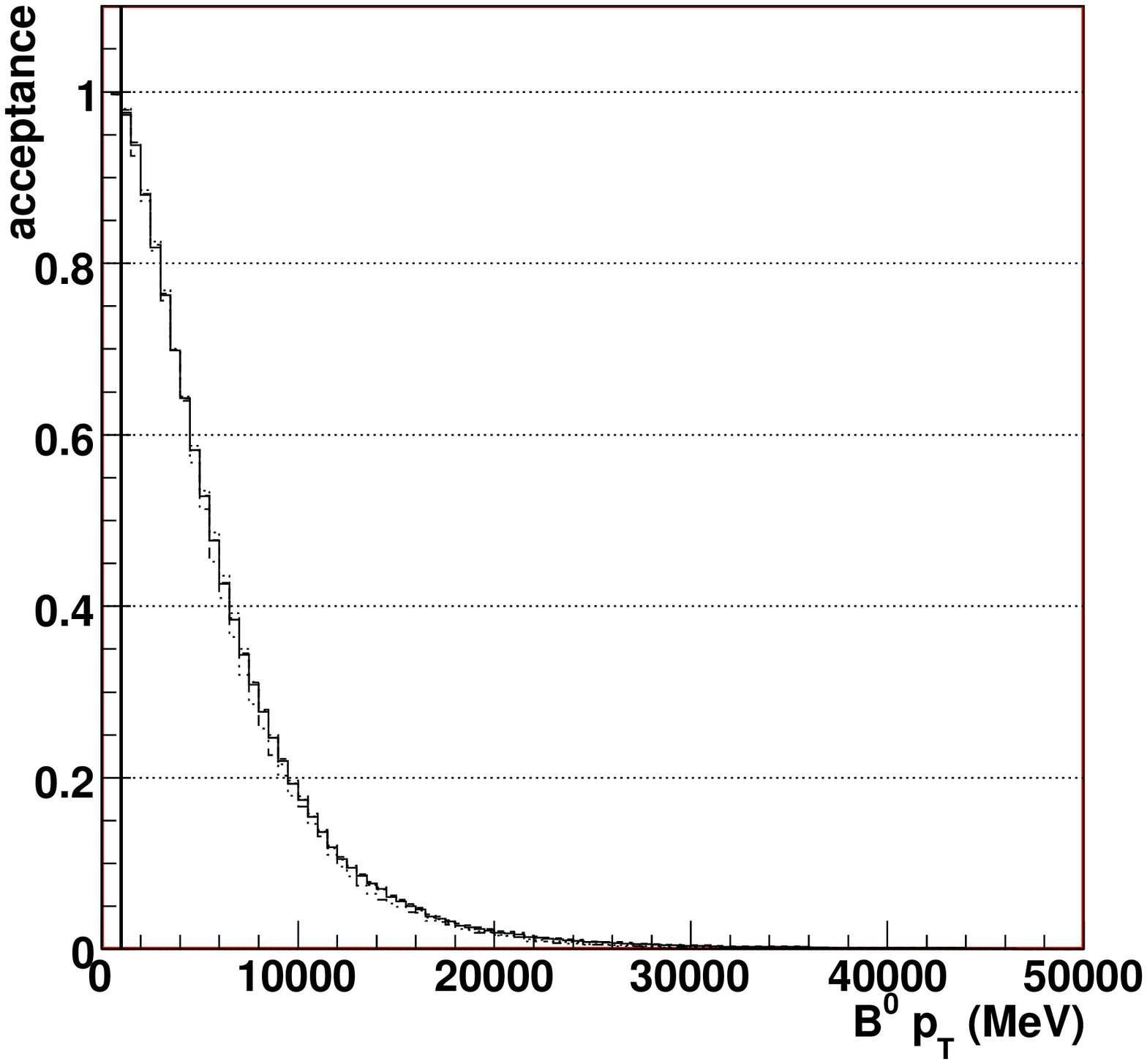}}}
\put(0.0,6.3){\scalebox{0.32}{\includegraphics{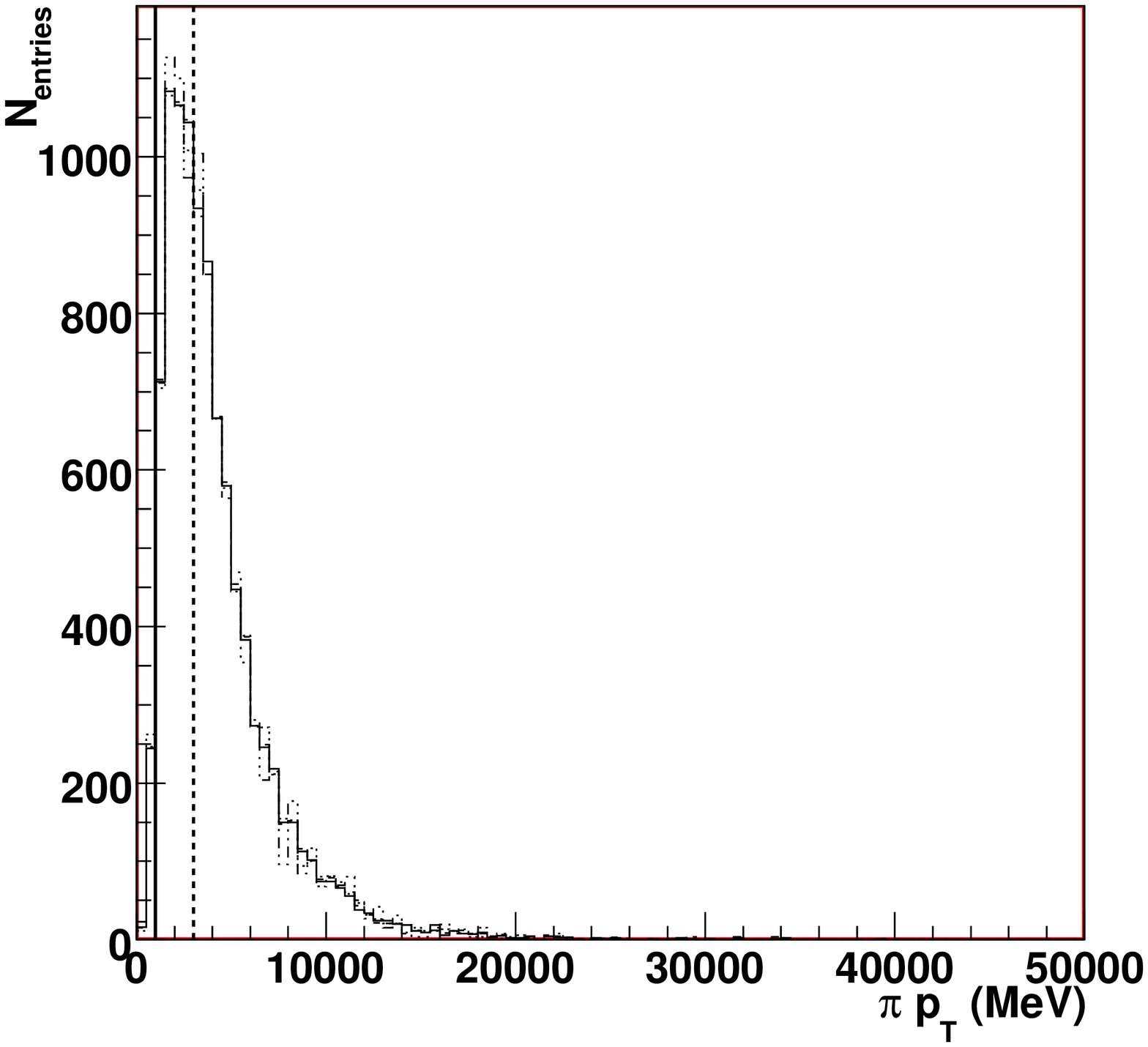}}}
\put(7.0,6.3){\scalebox{0.32}{\includegraphics{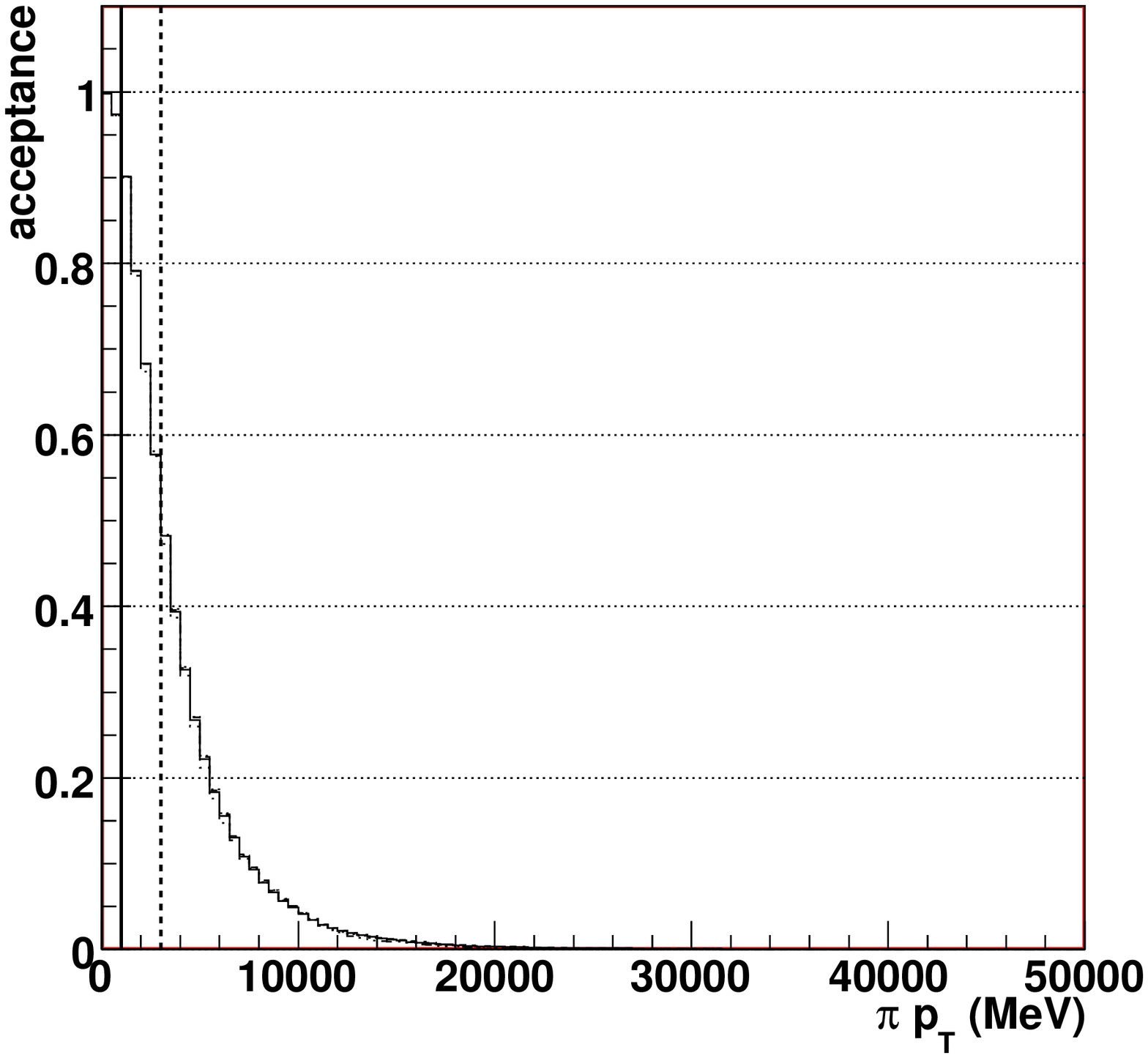}}}
\put(0.0,0.){\scalebox{0.32}{\includegraphics{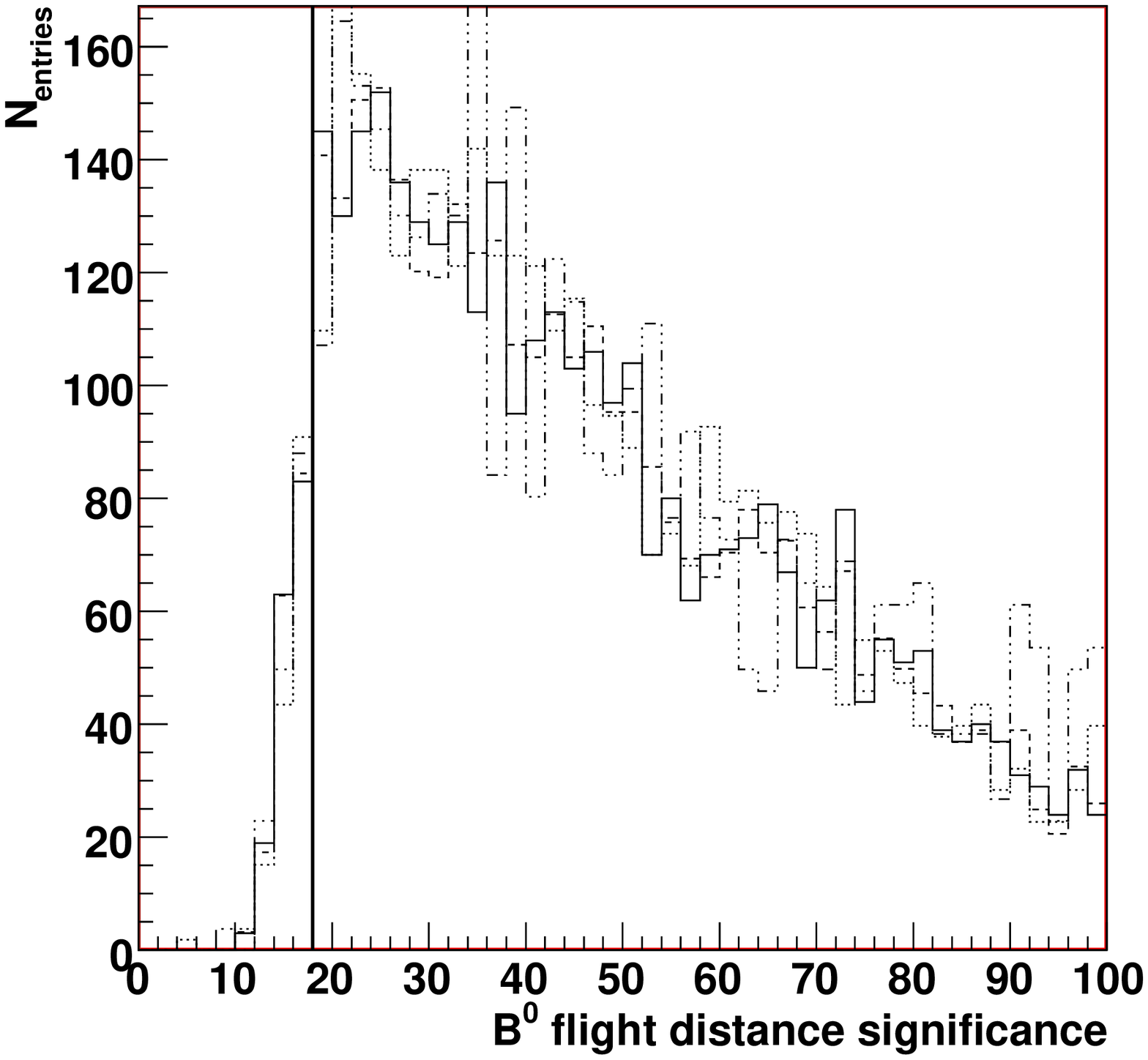}}}
\put(7.0,0.){\scalebox{0.32}{\includegraphics{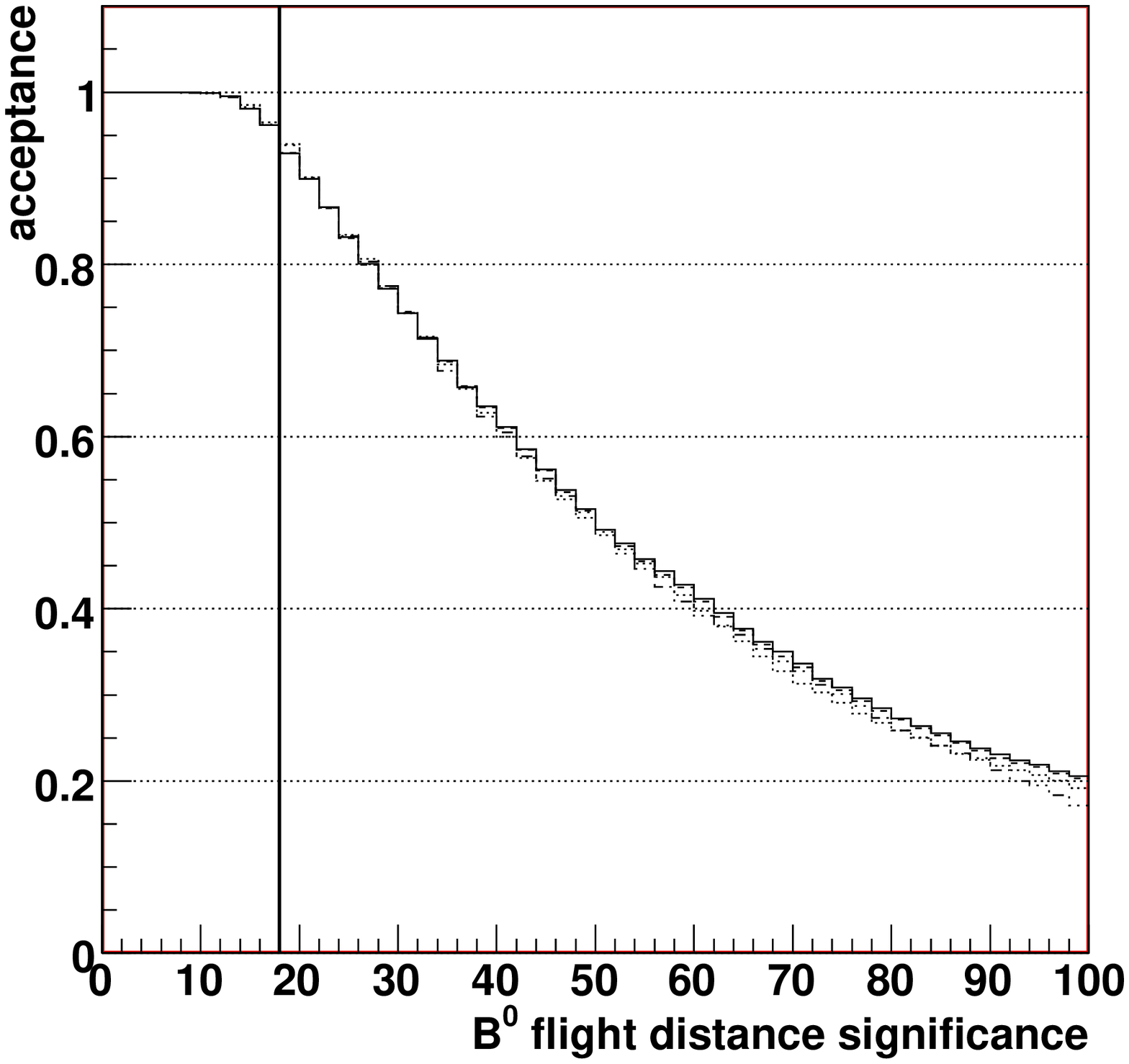}}}
\put(4.0,16.5){\small (a)}
\put(11.0,16.5){\small (b)}
\put(4.0,10.5){\small (c)}
\put(11.0,10.5){\small (d)}
\put(4.0,4.5){\small (e)}
\put(11.0,4.5){\small (f)}
\end{picture}
\end{center}
\caption{Effect of VELO misalignments on the transverse momentum of the
$B^0$ and daughter pions, and $B^0$ flight distance significance.
The right-hand-side distributions correspond to the integrated left-hand-side
distributions.
The various line styles are as explained in Figure~\ref{fig:sel_velo}.
The vertical cut lines are detailed in Section~\ref{sec:b2hh_velo_sel}.}
\label{fig:sel_velo_2}
\vfill
\end{figure}

\clearpage
\begin{figure}[p]
\vfill
\begin{center}
\setlength{\unitlength}{1.0cm}
\begin{picture}(14.,13.)
\put(0.,6.5){\scalebox{0.32}{\includegraphics{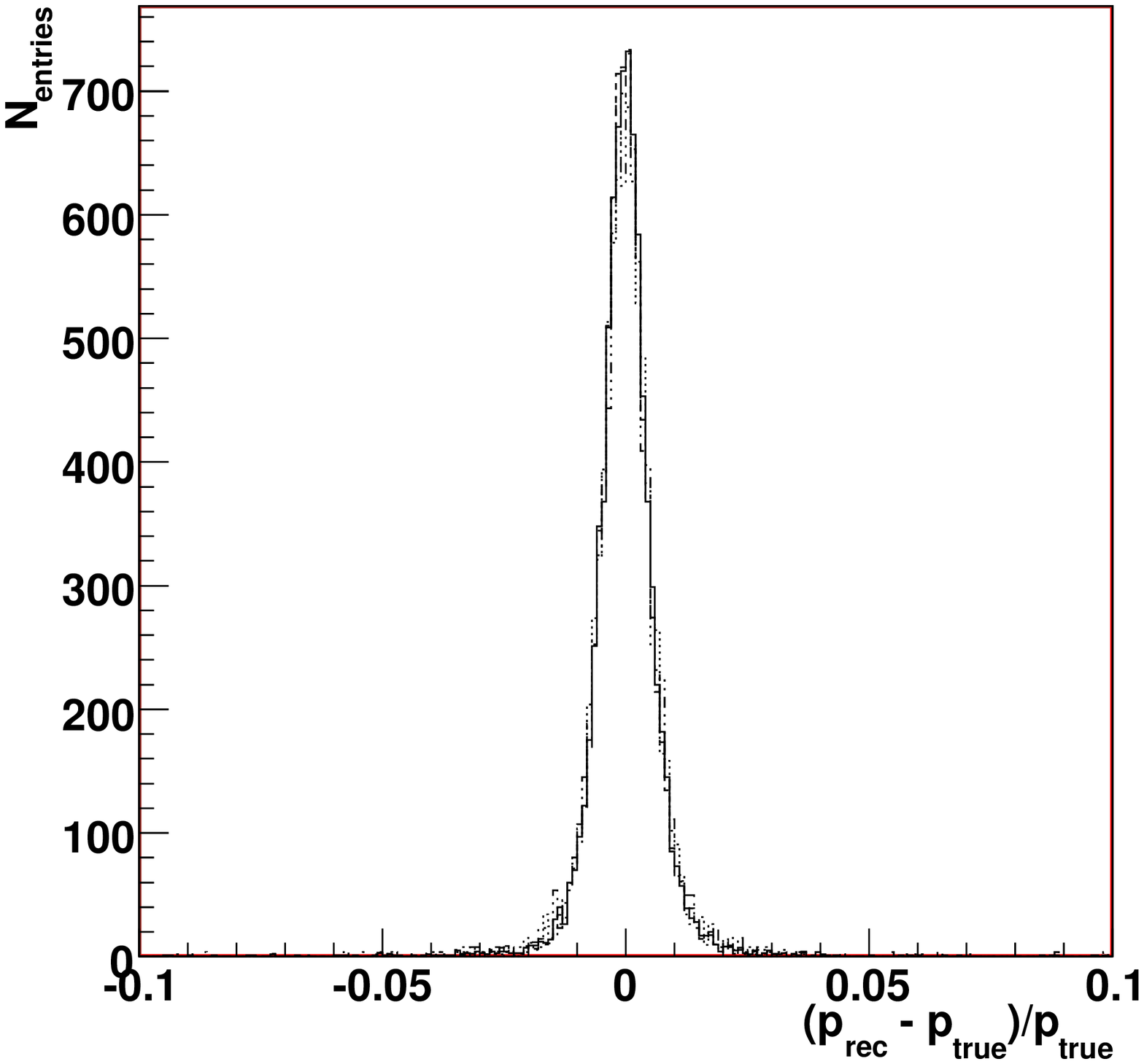}}}
\put(7.,6.5){\scalebox{0.32}{\includegraphics{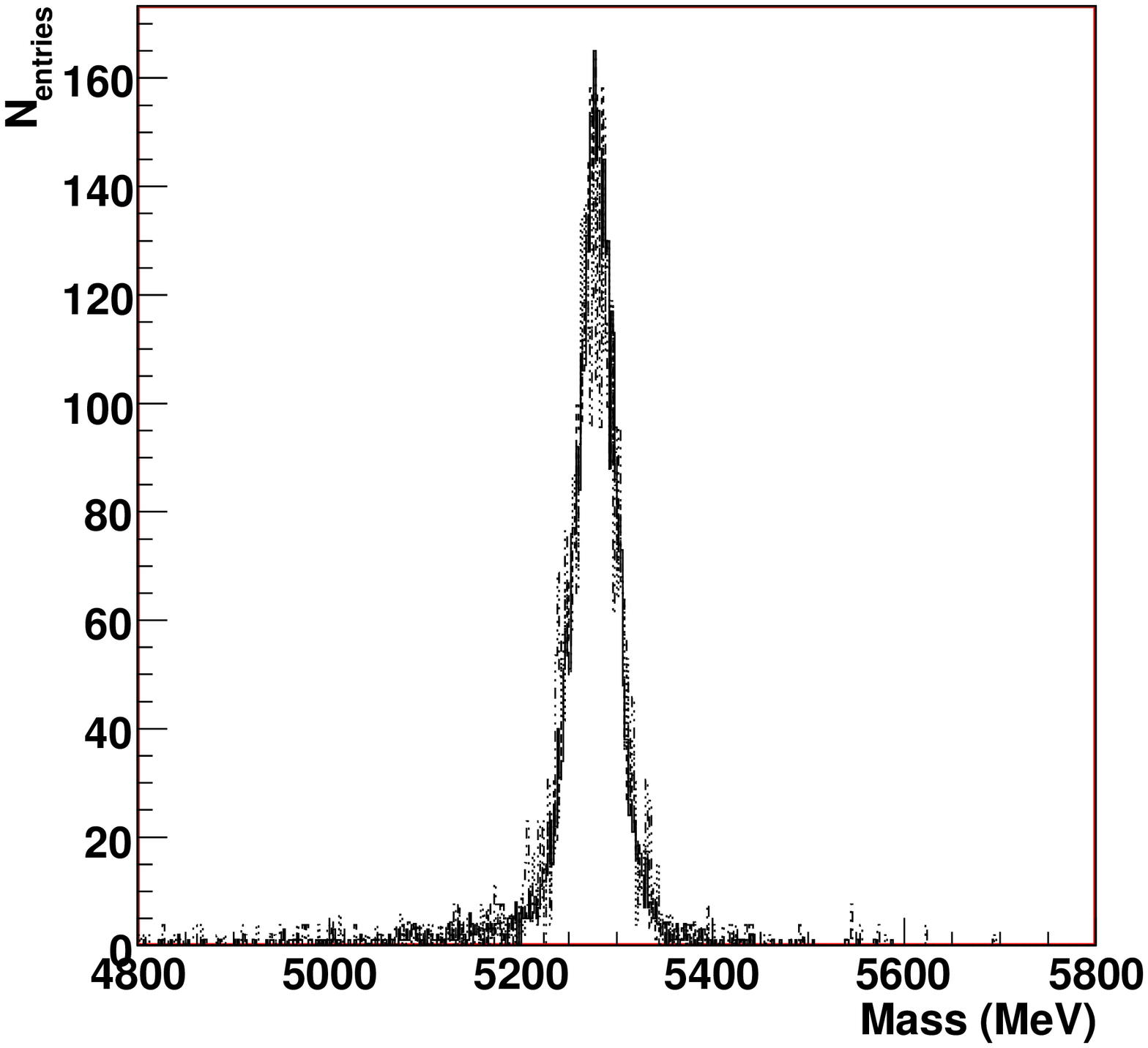}}}
\put(0.,0.){\scalebox{0.32}{\includegraphics{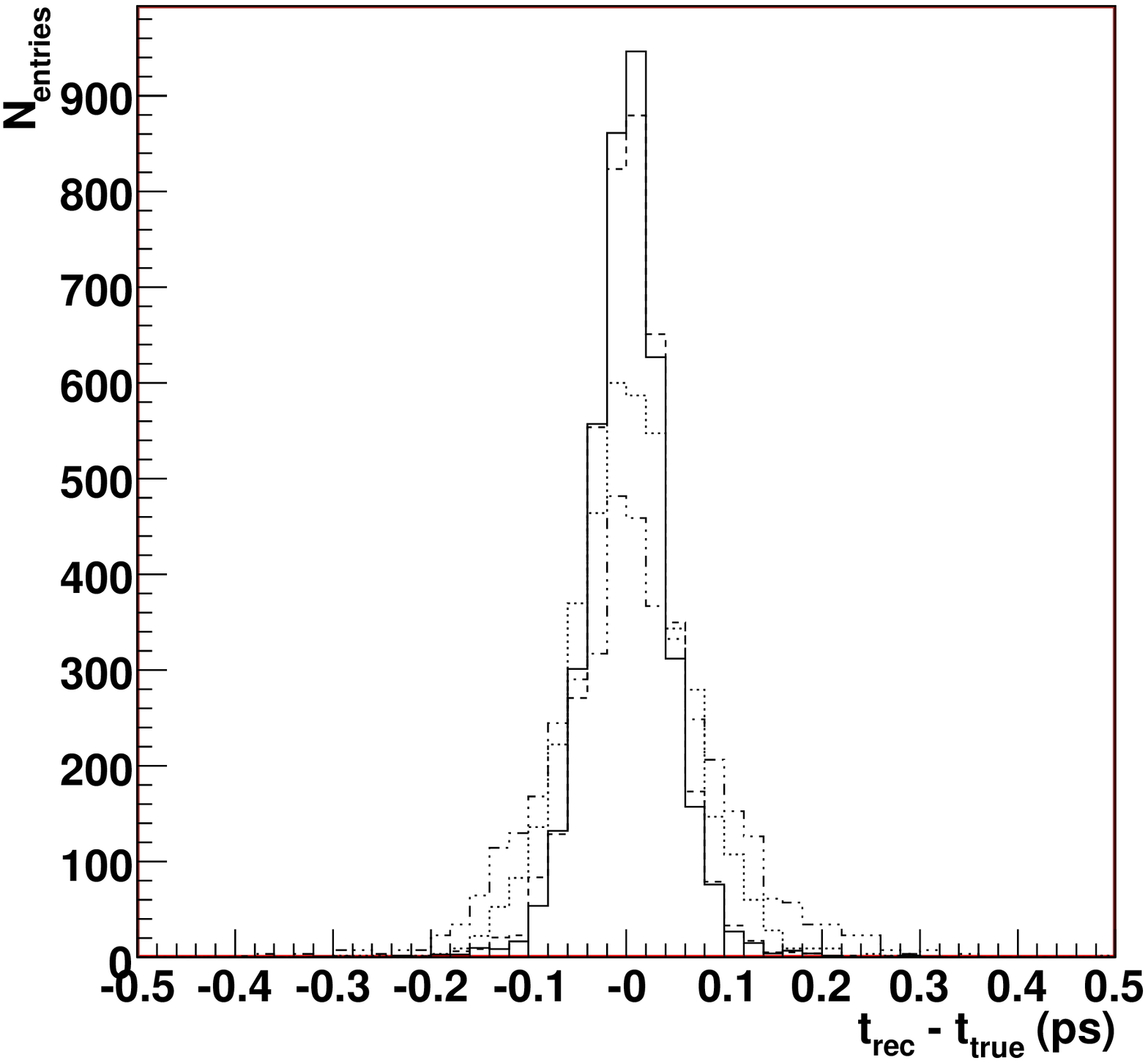}}}
\put(2.0,10.5){\small (a)}
\put(9.0,10.5){\small (b)}
\put(2.0,4.5){\small (c)}
\end{picture}
\end{center}
\caption{Effect of VELO misalignments on the resolutions in (a) momentum of
the daughter pions, in (b) $B^0$ invariant mass and in (c) $B^0$ proper time.
The various line styles are as explained in Figure~\ref{fig:sel_velo}.}
\label{fig:res_velo}
\vfill
\end{figure}

\clearpage
\begin{figure}[p]
\vfill
\begin{center}
\setlength{\unitlength}{1.0cm}
\begin{picture}(14.,18.)
\put(0.0,0.){\scalebox{0.9}{\includegraphics[angle=90]{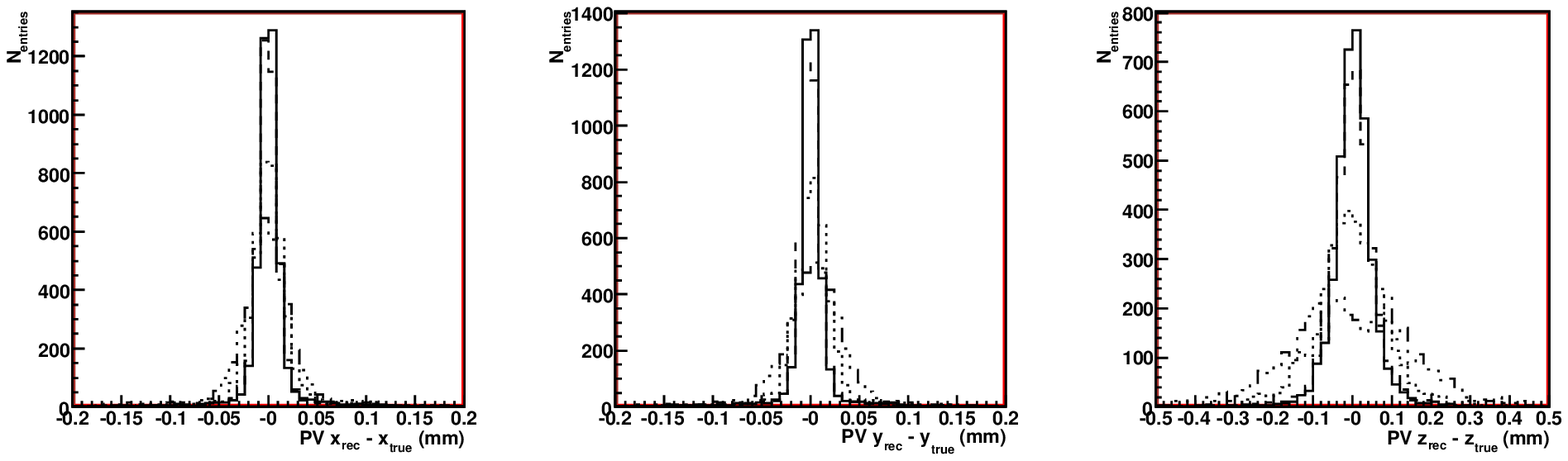}}}
\put(7.0,0.){\scalebox{0.9}{\includegraphics[angle=90]{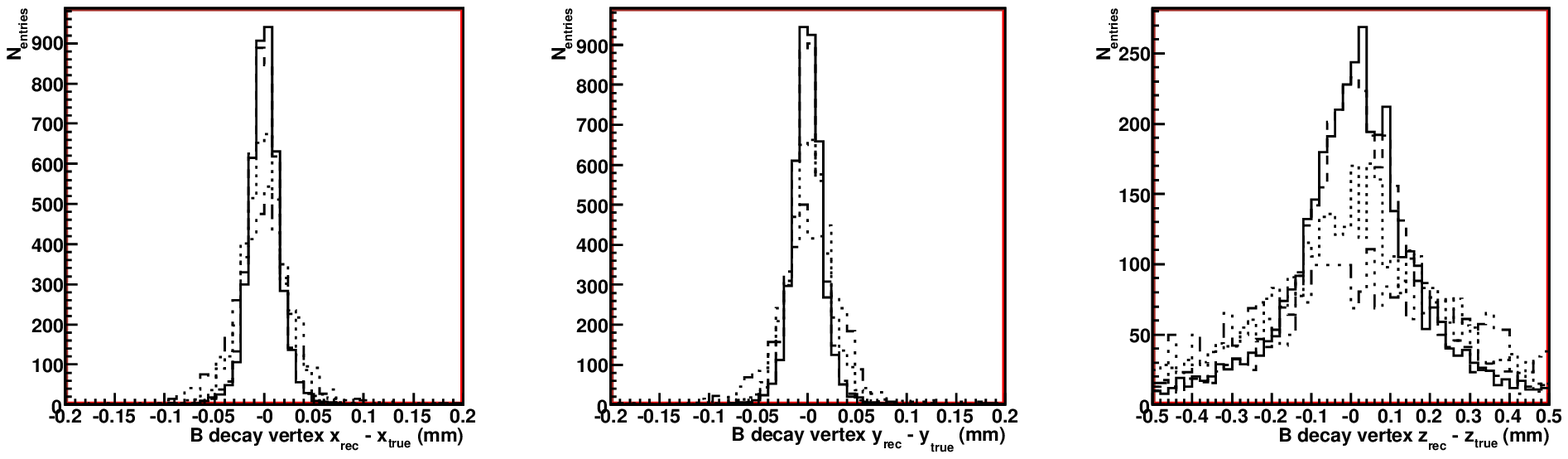}}}
\put(2.0,4.5){\small (a)}
\put(9.0,4.5){\small (b)}
\end{picture}
\end{center}
\caption{Effect of VELO misalignments on the resolutions of the
(a) primary vertex and (b) the $B^0$ vertex.
The various line styles are as explained in Figure~\ref{fig:sel_velo}.}
\label{fig:res_velo_2}
\vfill
\end{figure}

\begin{figure}[p]
\vfill
\begin{center}
\setlength{\unitlength}{1.0cm}
\begin{picture}(14.,6.)
\put(0.,0.){\scalebox{0.32}{\includegraphics{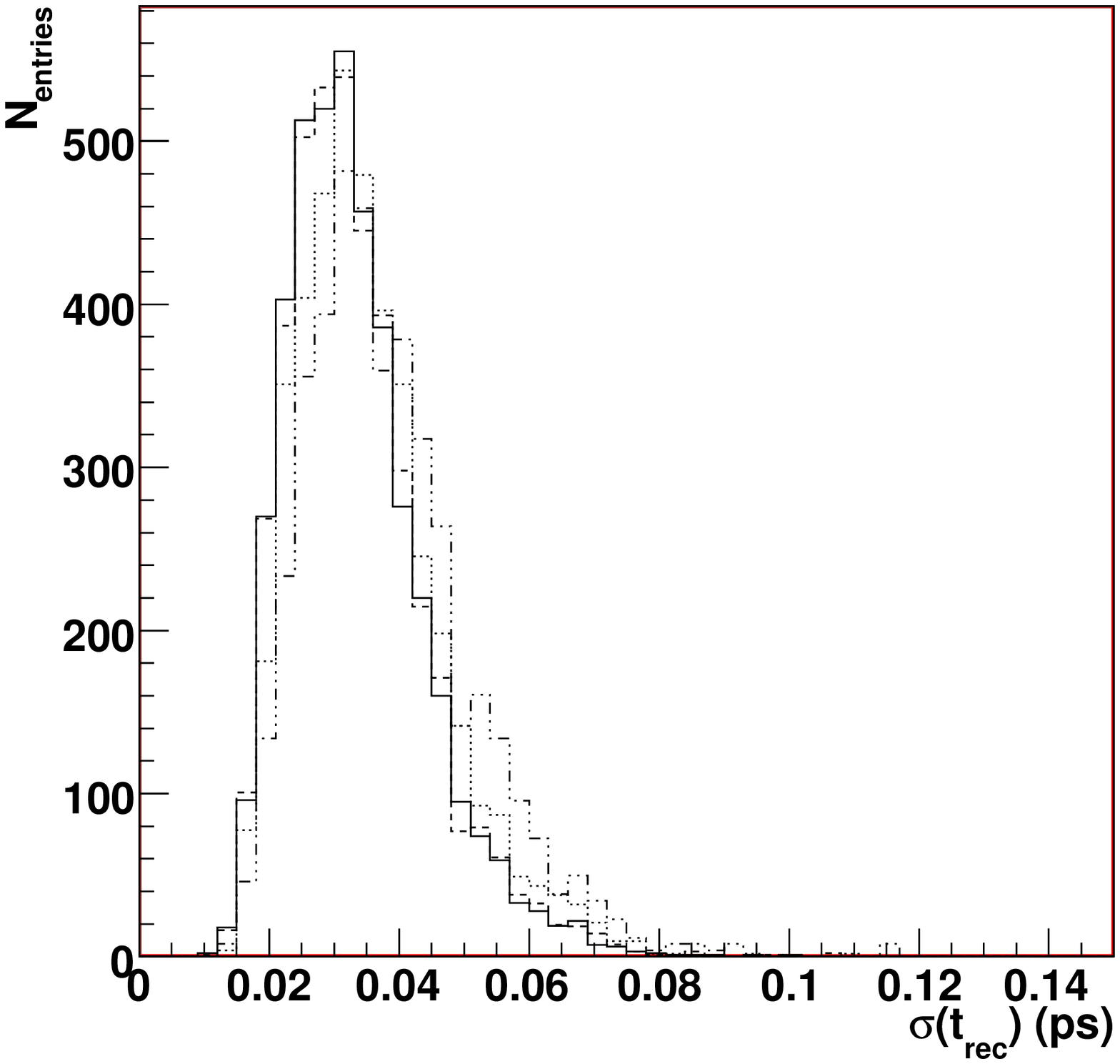}}}
\put(7.0,0.){\scalebox{0.32}{\includegraphics{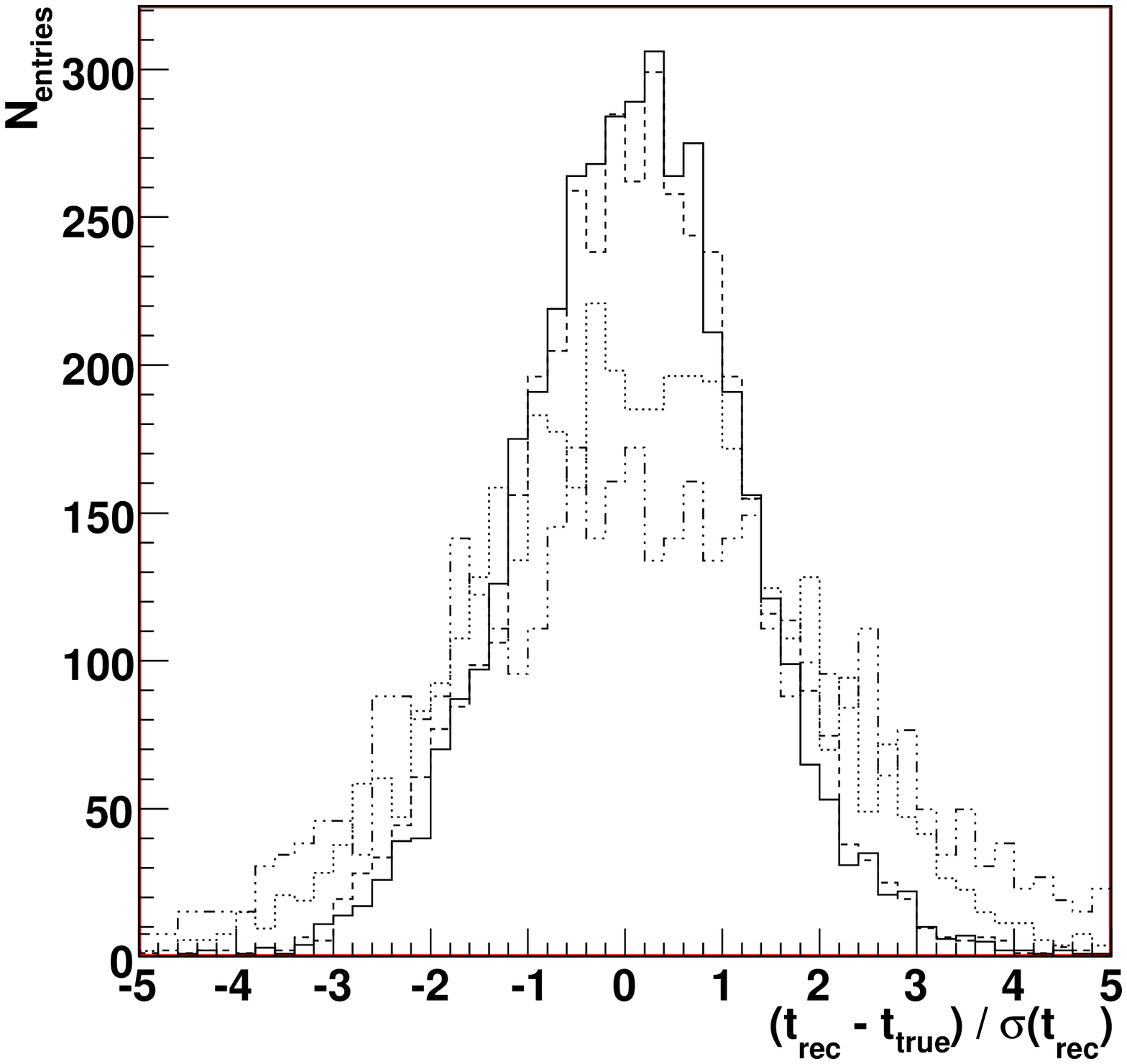}}}
\put(4.0,4.5){\small (a)}
\put(9.0,4.5){\small (b)}
\end{picture}
\end{center}
\caption{Effect of VELO misalignments on (a) the $B^0$ proper time error and
(b) on the pull distribution of the $B^0$ proper time.
The various line styles are as explained in Figure~\ref{fig:sel_velo}.}
\label{fig:tau_velo}
\vfill
\end{figure}

\begin{figure}[p]
\vfill
\begin{center}
\setlength{\unitlength}{1.0cm}
\begin{picture}(14.,18.5)
\put(0.0,12.6){\scalebox{0.32}{\includegraphics{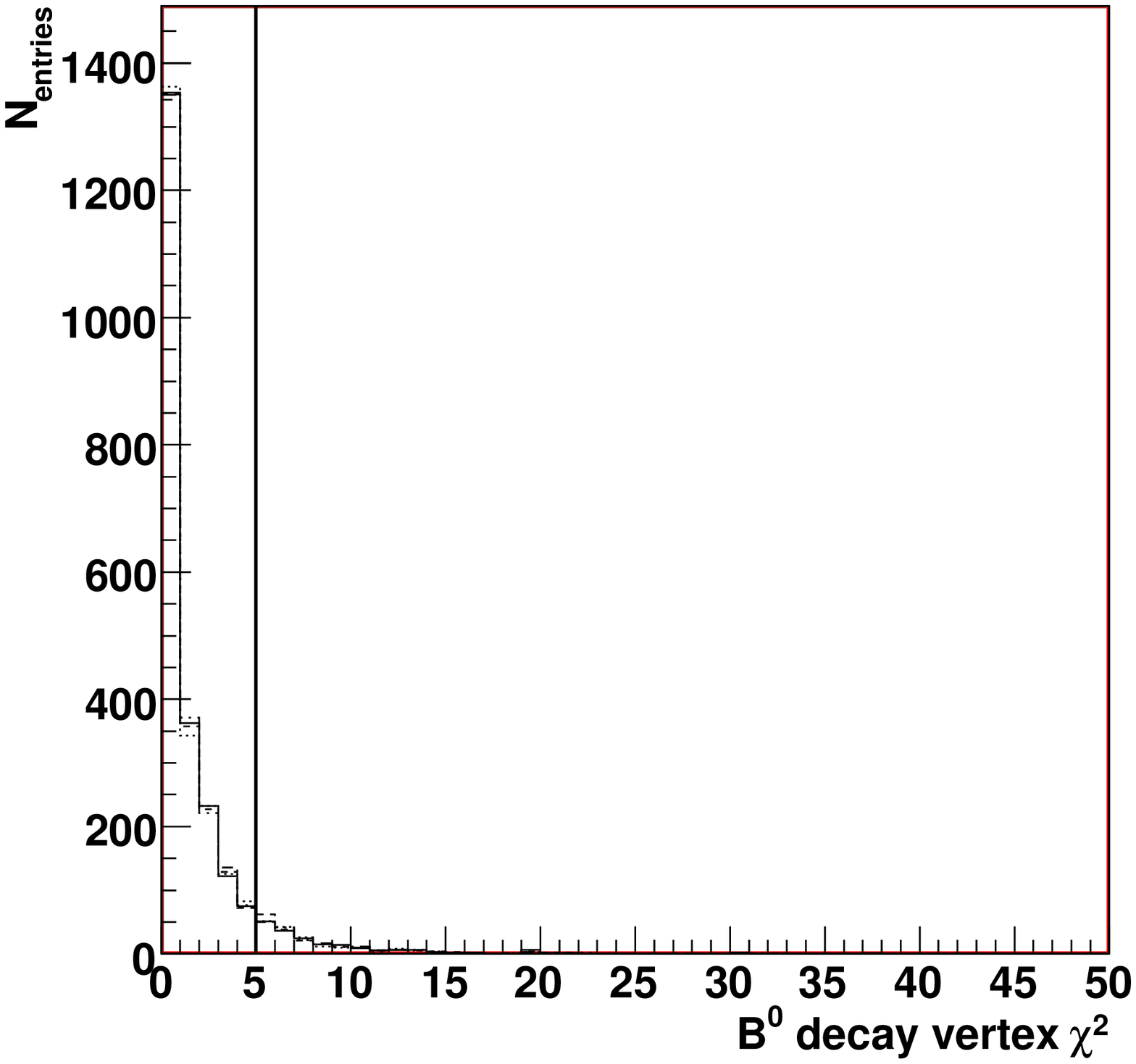}}}
\put(7.0,12.6){\scalebox{0.32}{\includegraphics{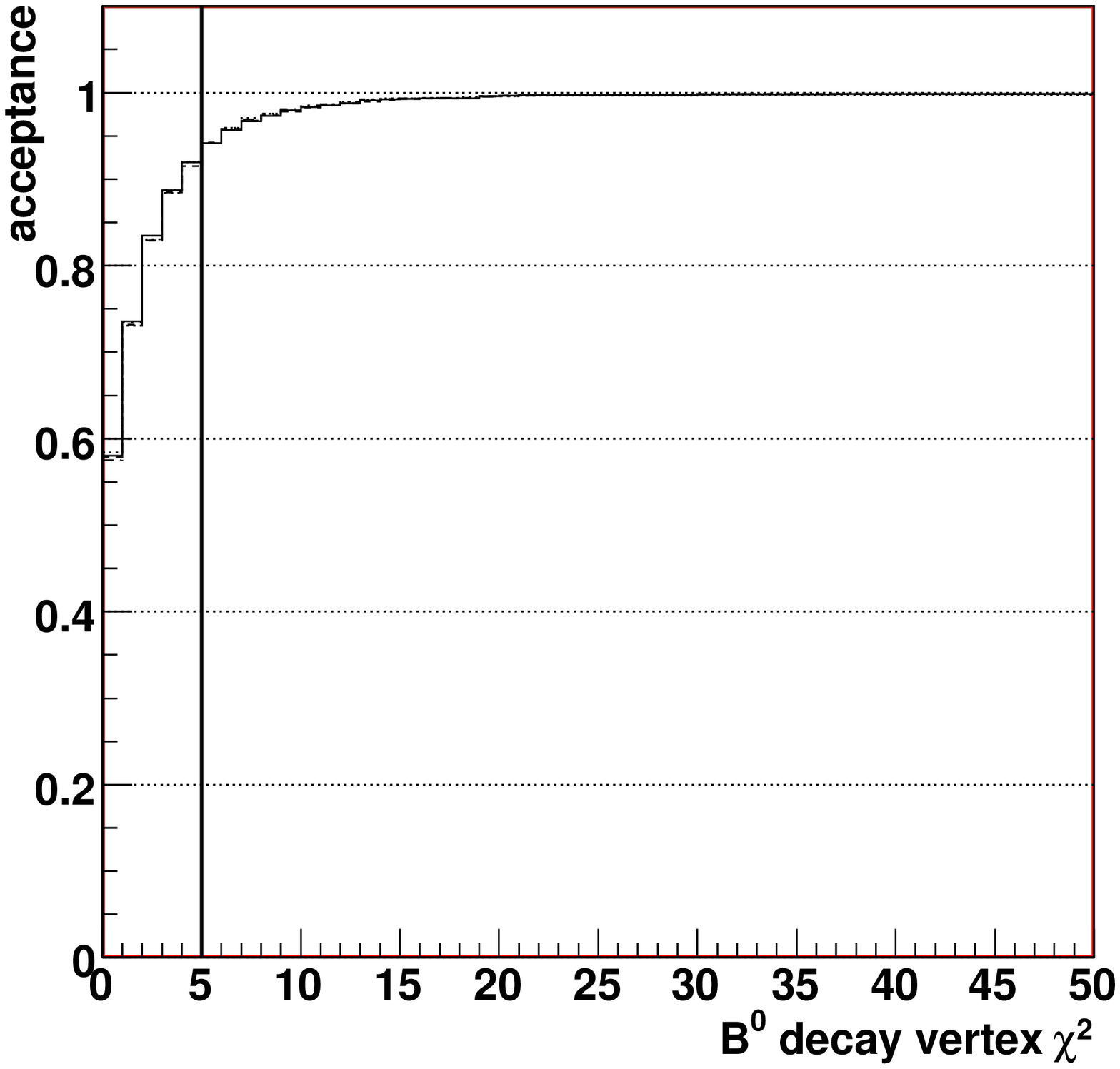}}}
\put(0.,6.3){\scalebox{0.32}{\includegraphics{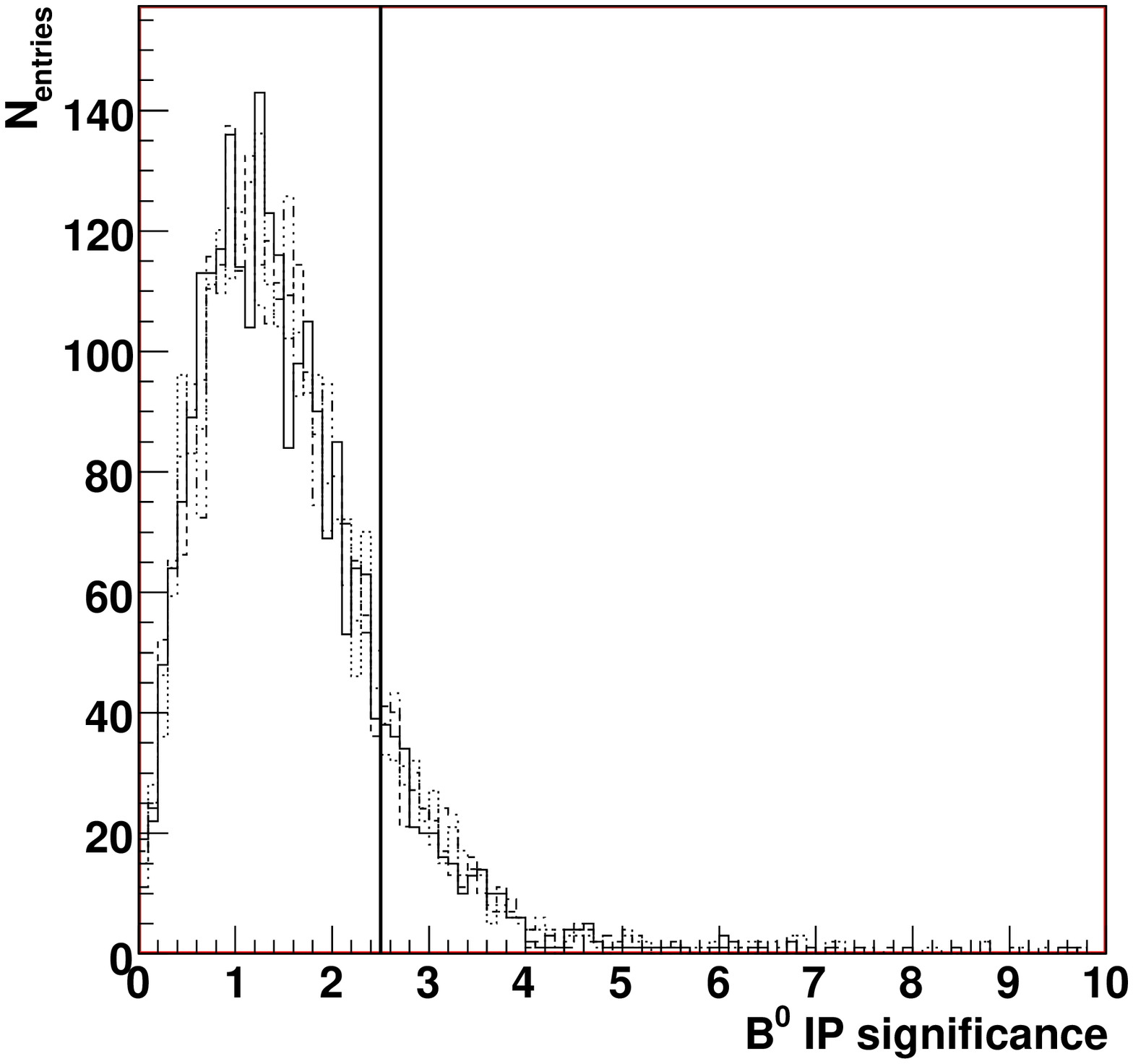}}}
\put(7.0,6.3){\scalebox{0.32}{\includegraphics{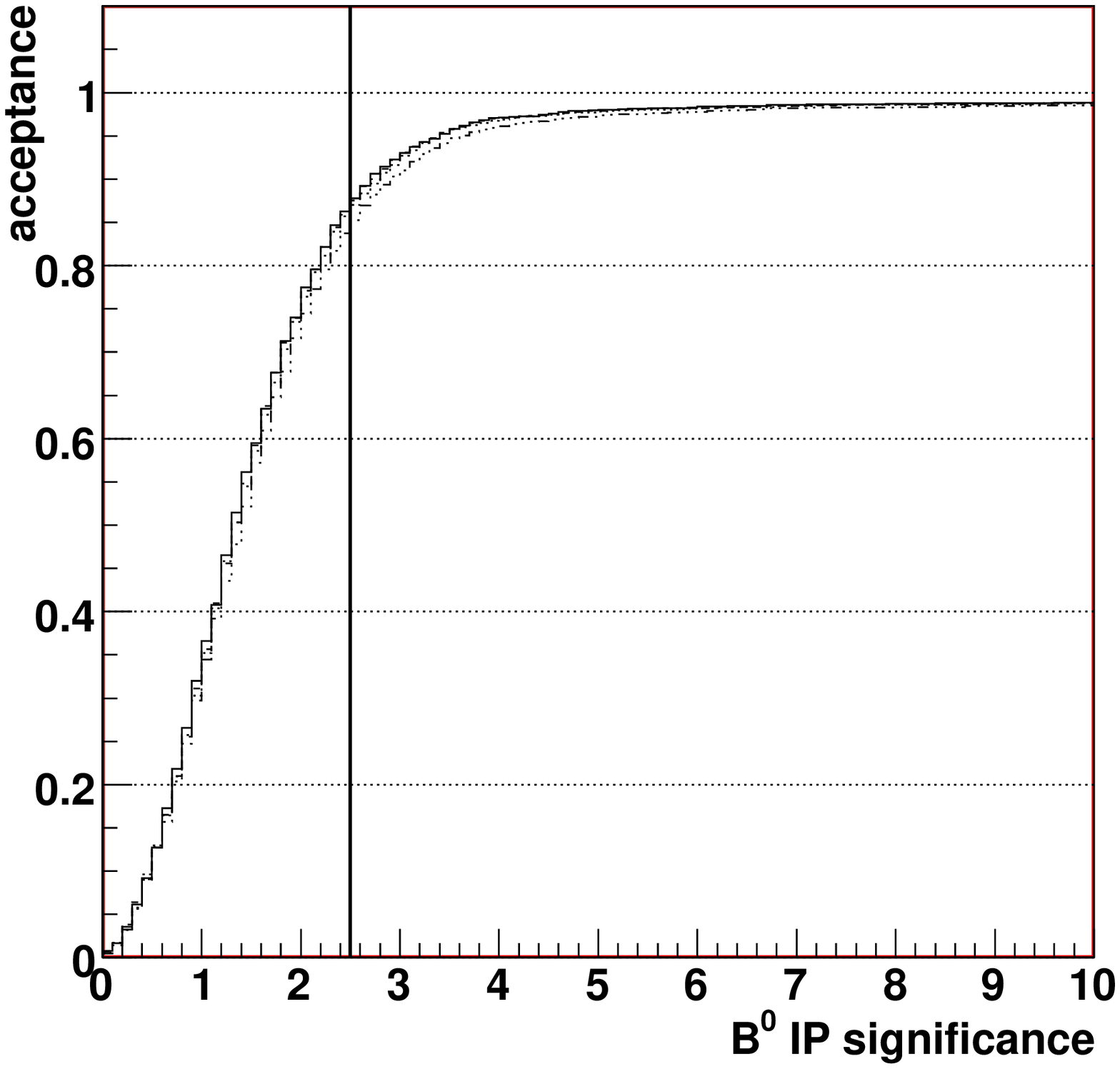}}}
\put(0.0,0.){\scalebox{0.32}{\includegraphics{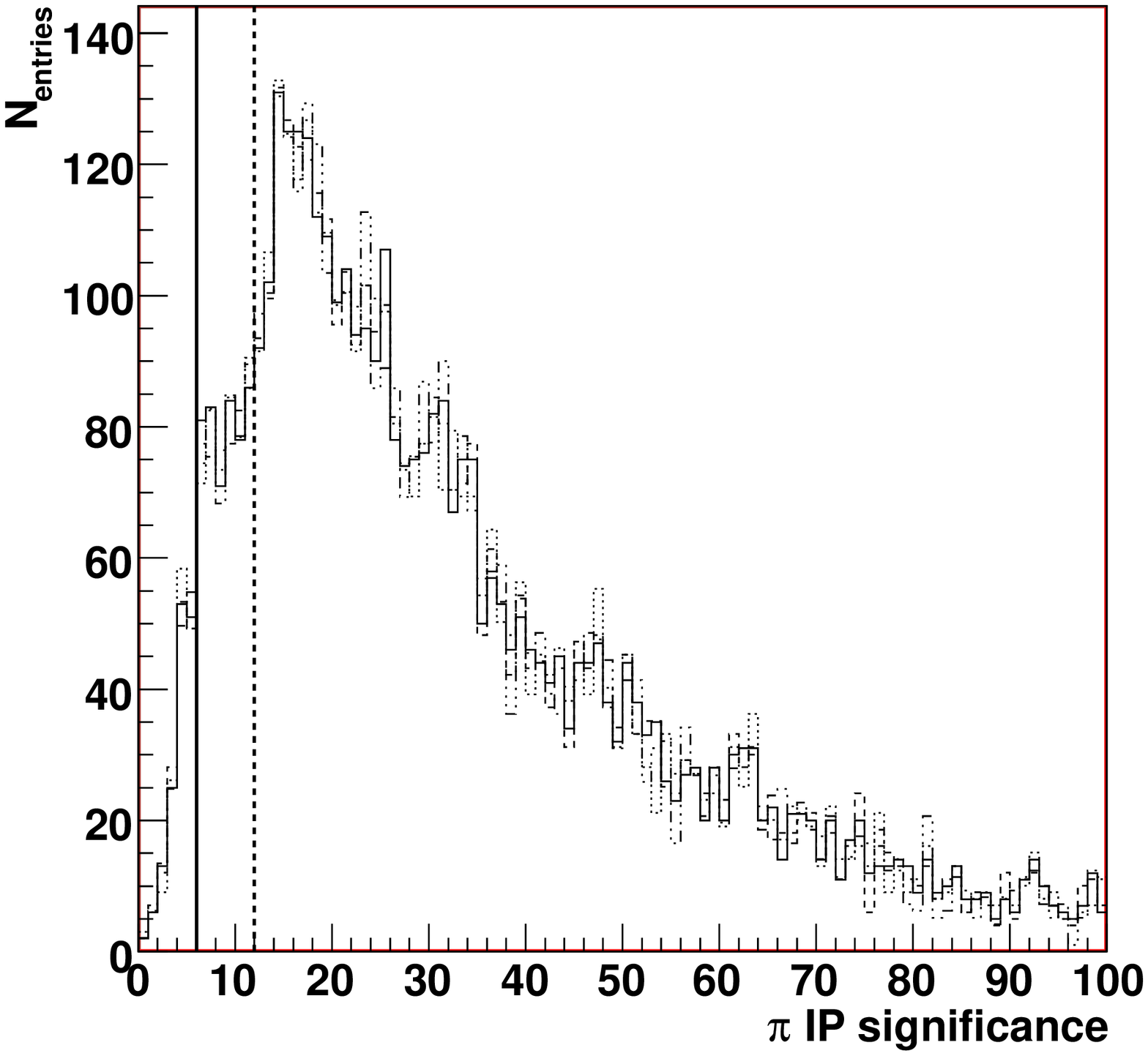}}}
\put(7.0,0.){\scalebox{0.32}{\includegraphics{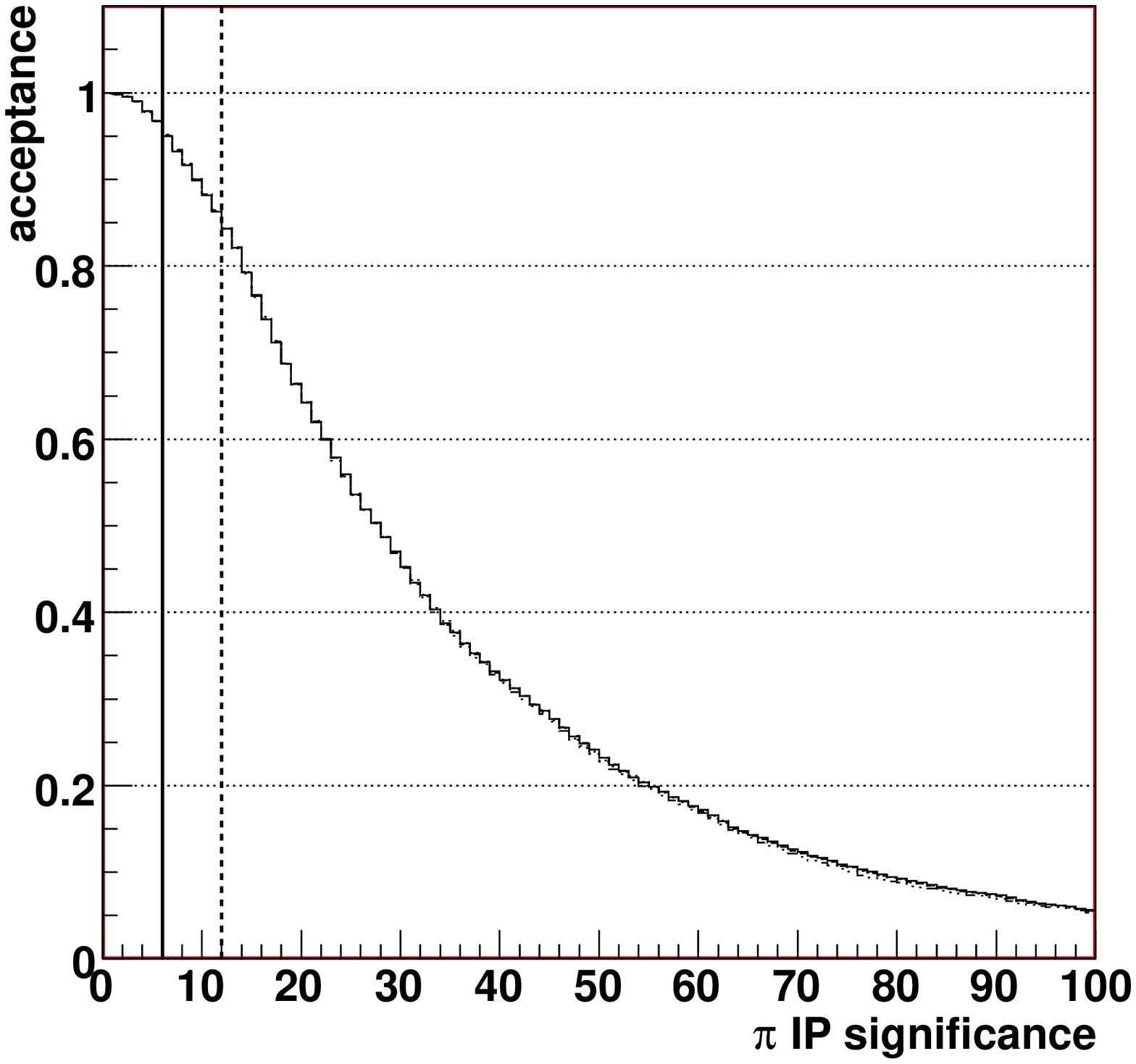}}}
\put(4.0,16.5){\small (a)}
\put(11.0,16.5){\small (b)}
\put(4.0,10.5){\small (c)}
\put(11.0,10.5){\small (d)}
\put(4.0,4.5){\small (e)}
\put(11.0,4.5){\small (f)}
\end{picture}
\end{center}
\caption{Effect of T-stations misalignments on the $B^0$ decay vertex $\chi^2$
and impact parameter significances for the $B^0$ candidate and its
daughter pions.
The right-hand-side distributions correspond to the integrated left-hand-side
distributions.
The various line styles are as explained in Figure~\ref{fig:sel_velo}.
The vertical cut lines are detailed in Section~\ref{sec:b2hh_velo_sel}.}
\label{fig:sel_ot}
\vfill
\end{figure}

\begin{figure}[p]
\vfill
\begin{center}
\setlength{\unitlength}{1.0cm}
\begin{picture}(14.,18.5)
\put(0.,12.6){\scalebox{0.32}{\includegraphics{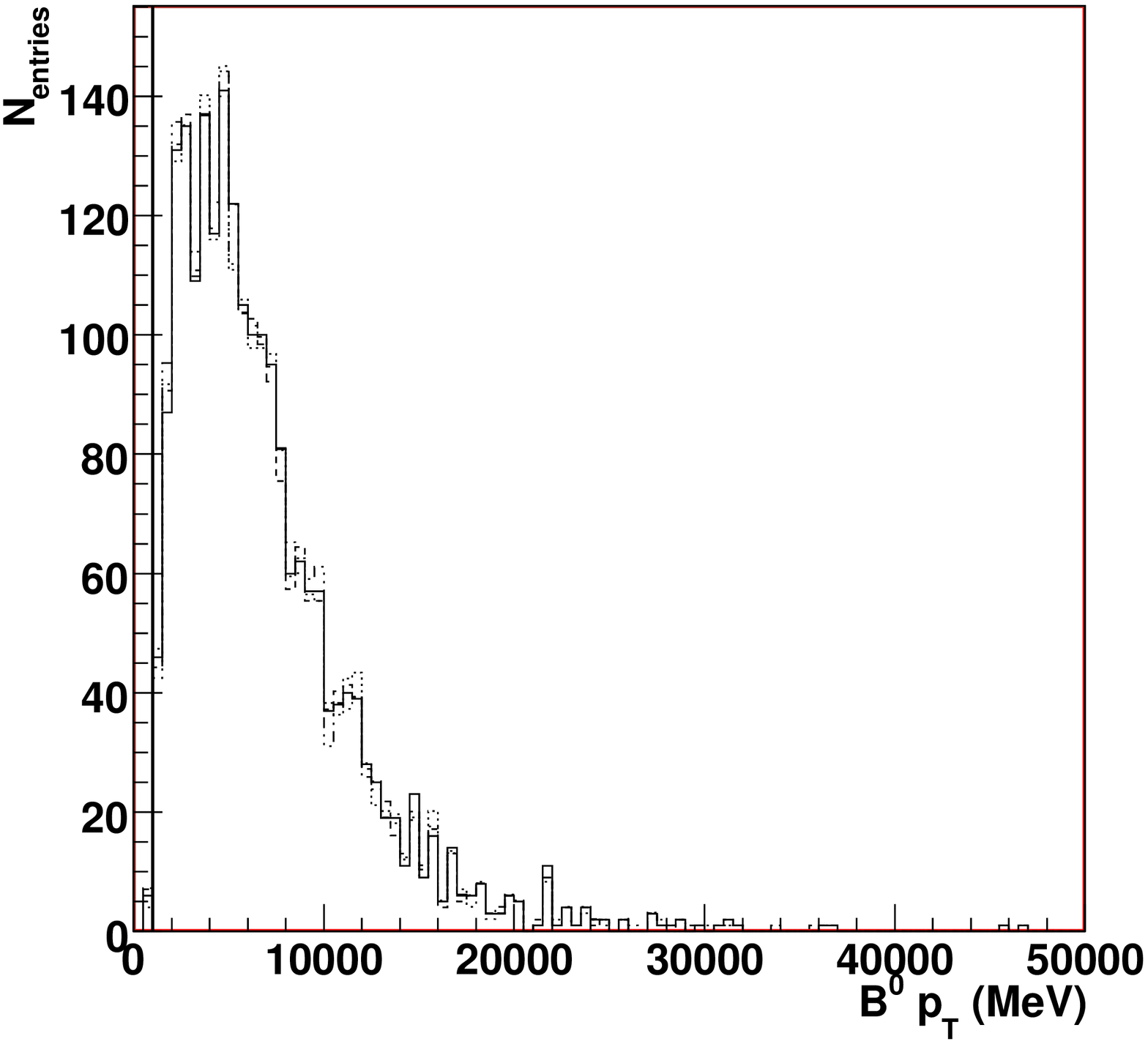}}}
\put(7.0,12.6){\scalebox{0.32}{\includegraphics{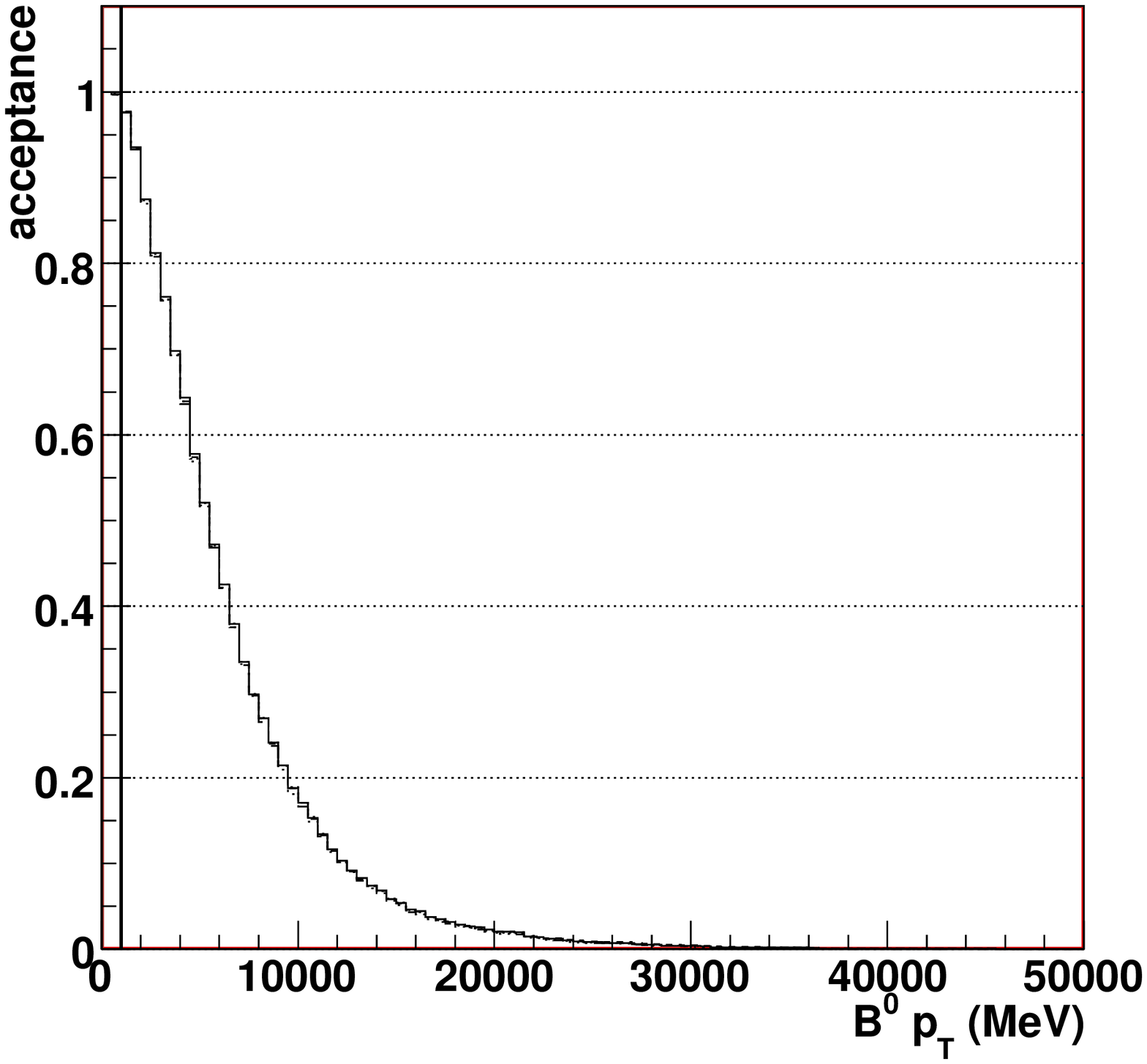}}}
\put(0.0,6.3){\scalebox{0.32}{\includegraphics{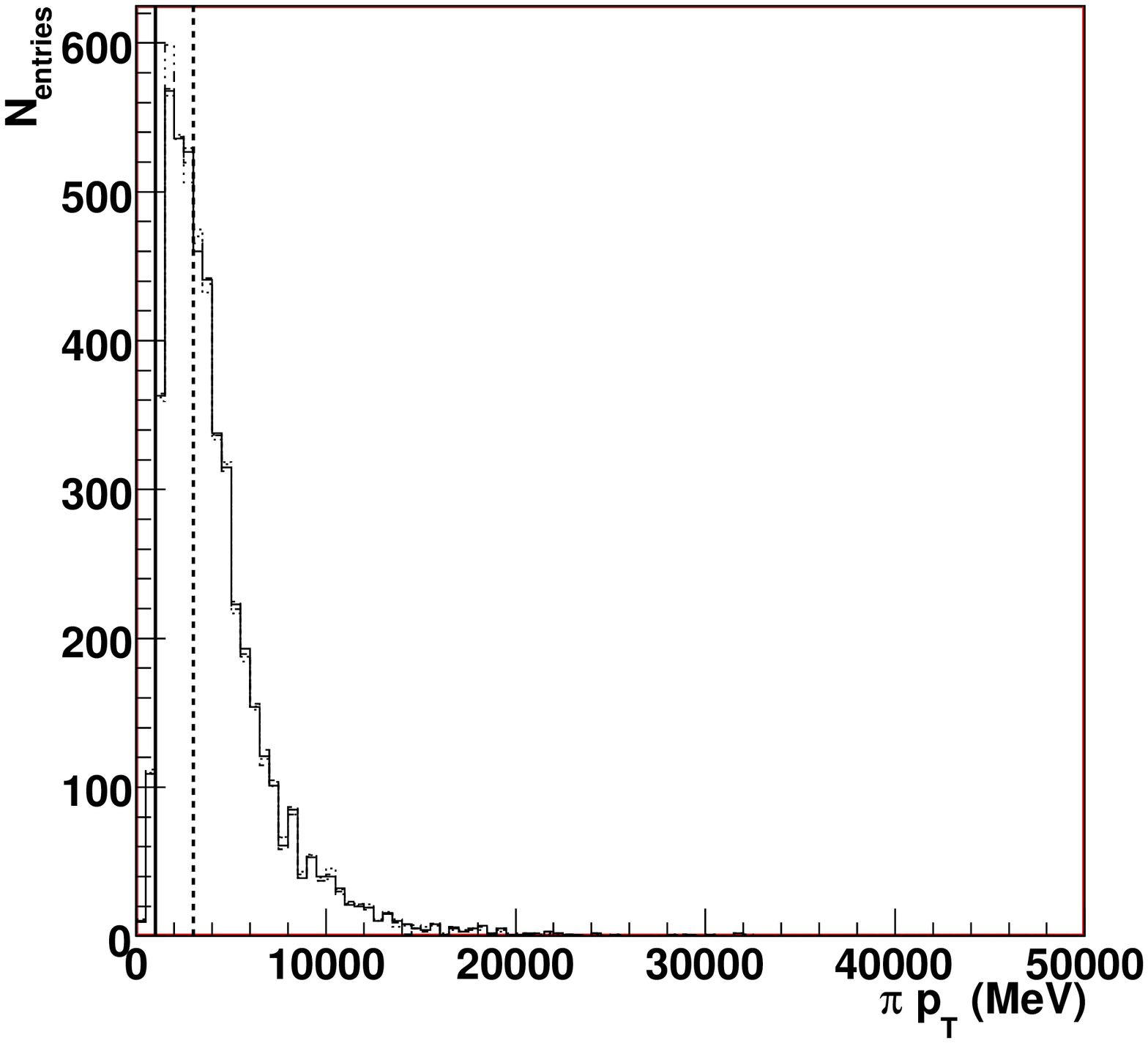}}}
\put(7.0,6.3){\scalebox{0.32}{\includegraphics{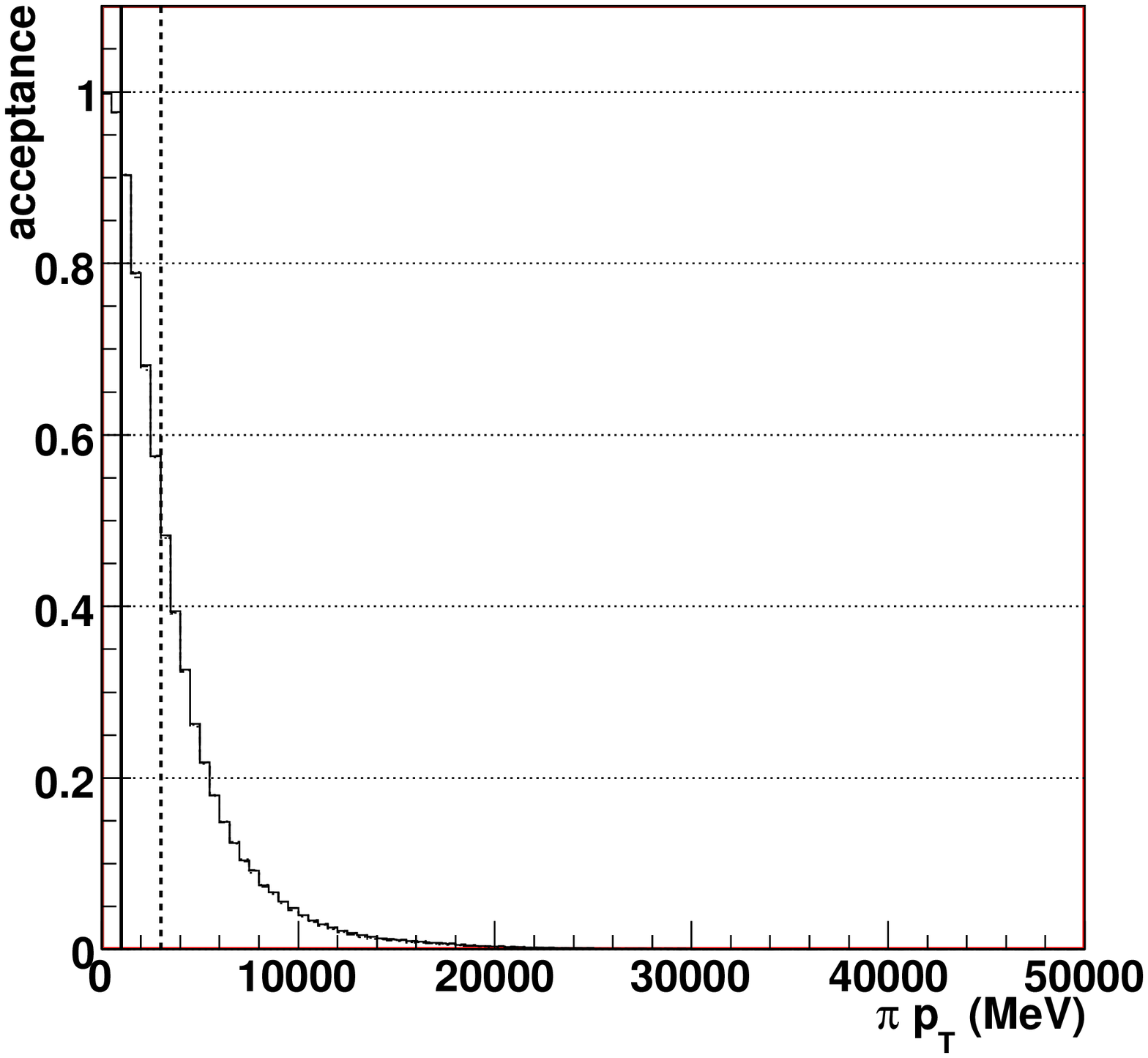}}}
\put(0.0,0.){\scalebox{0.32}{\includegraphics{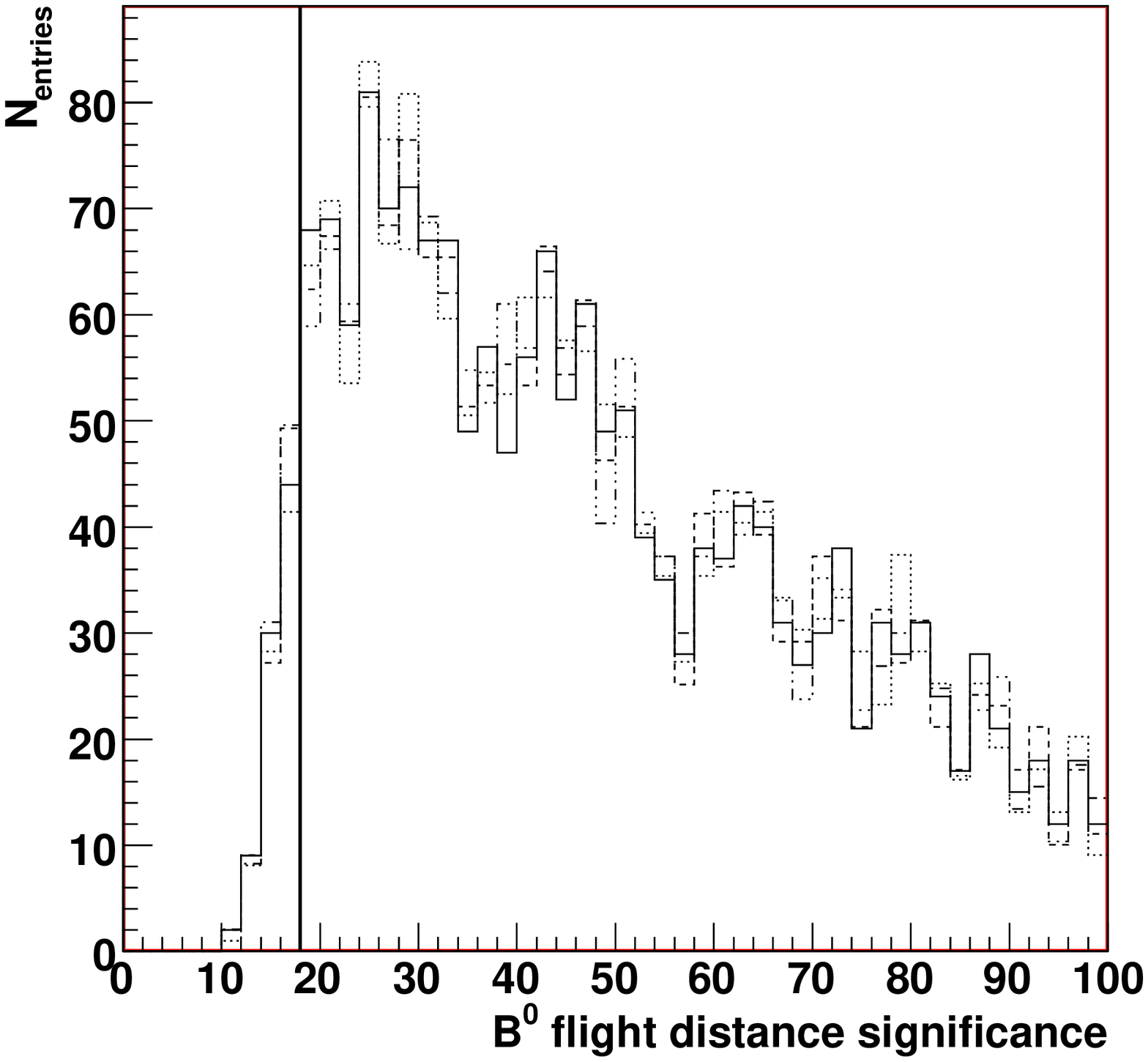}}}
\put(7.0,0.){\scalebox{0.32}{\includegraphics{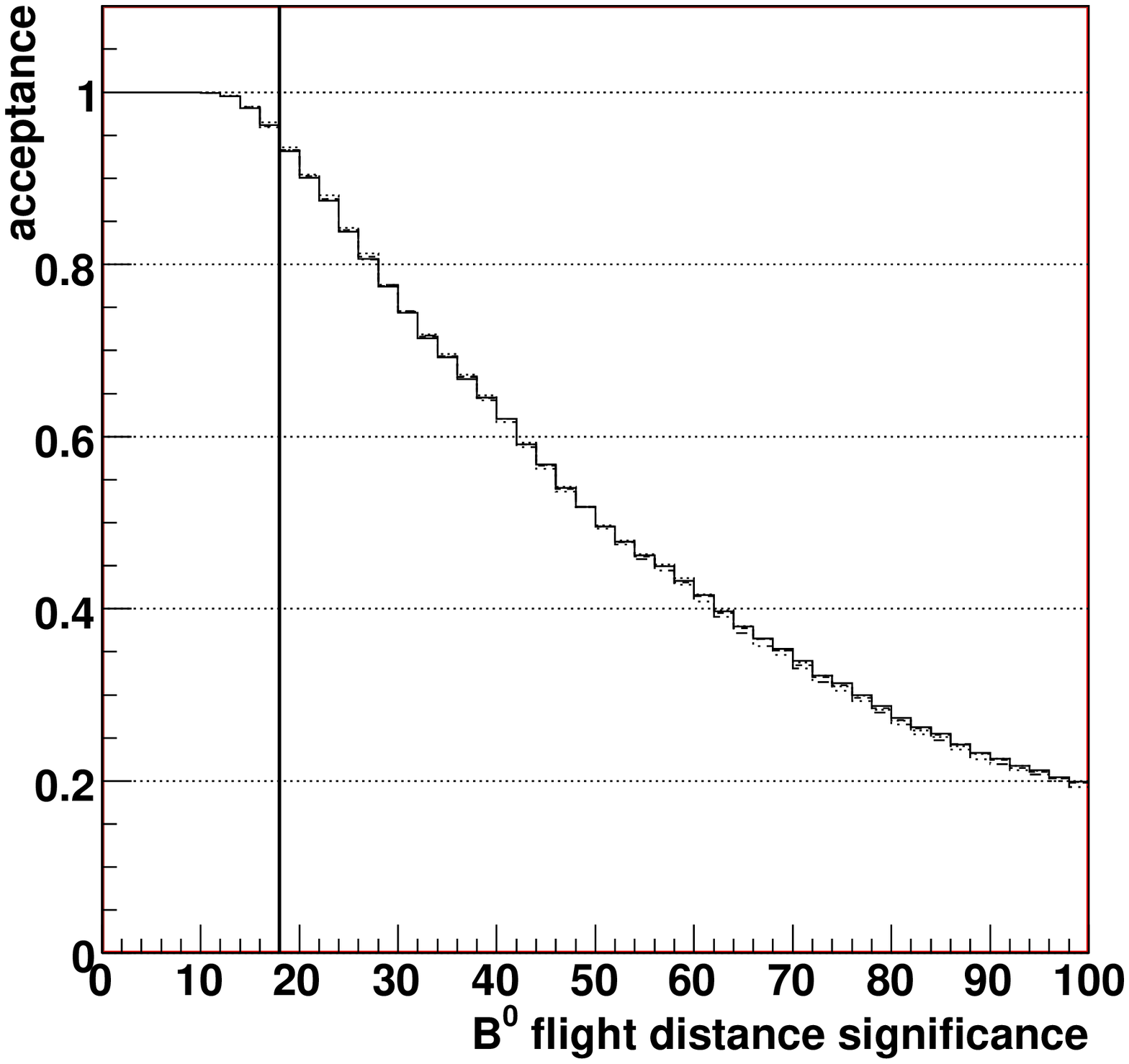}}}
\put(4.0,16.5){\small (a)}
\put(11.0,16.5){\small (b)}
\put(4.0,10.5){\small (c)}
\put(11.0,10.5){\small (d)}
\put(4.0,4.5){\small (e)}
\put(11.0,4.5){\small (f)}
\end{picture}
\end{center}
\caption{Effect of T-stations misalignments on the transverse momentum of the
$B^0$ and daughter pions, and $B^0$ flight distance significance.
The right-hand-side distributions correspond to the integrated left-hand-side
distributions.
The various line styles are as explained in Figure~\ref{fig:sel_velo}.
The vertical cut lines are detailed in Section~\ref{sec:b2hh_velo_sel}.}
\label{fig:sel_ot_2}
\vfill
\end{figure}
\begin{figure}[p]
\vfill
\begin{center}
\setlength{\unitlength}{1.0cm}
\begin{picture}(14.,13.)
\put(0.,6.5){\scalebox{0.32}{\includegraphics{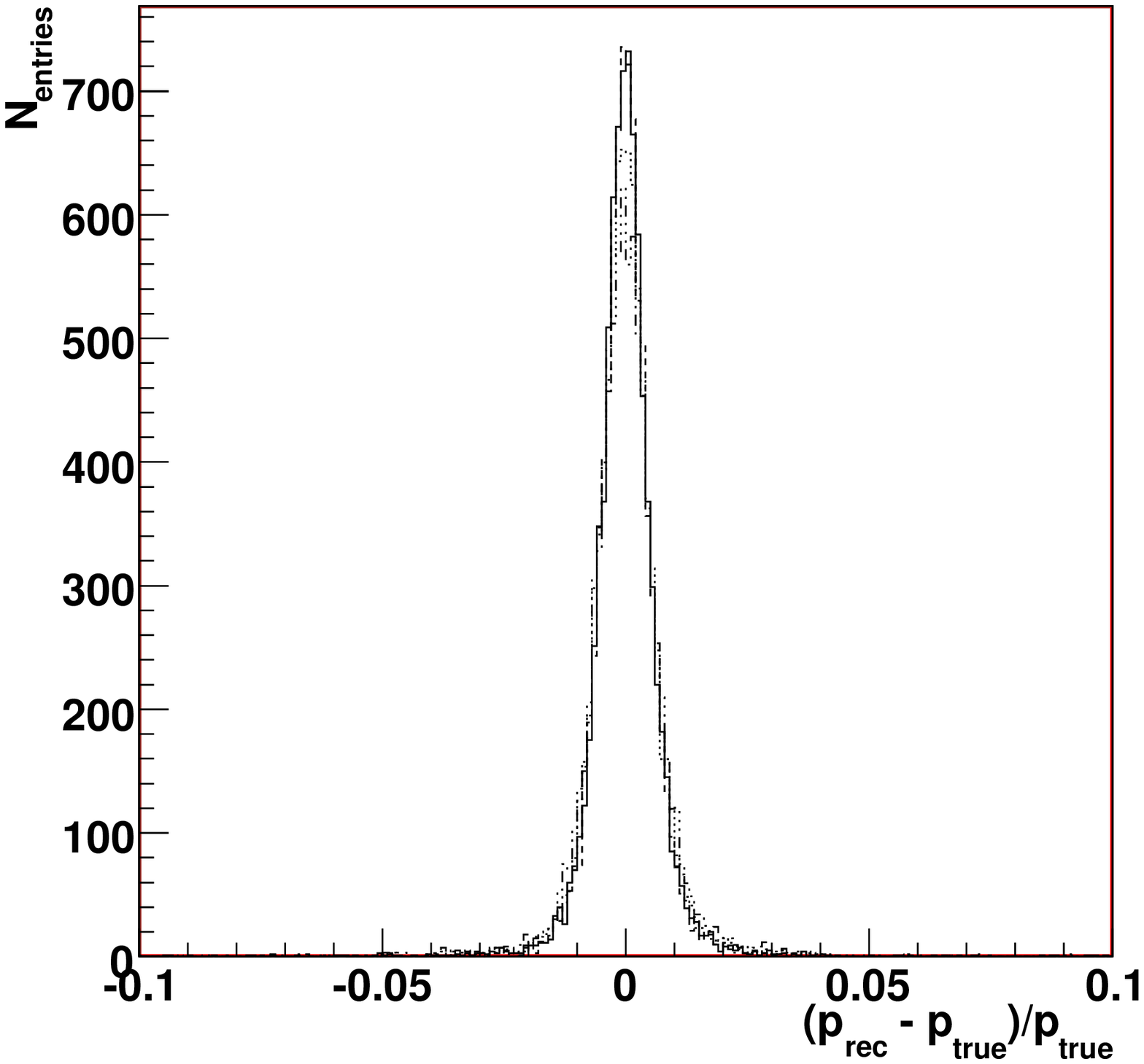}}}
\put(7.,6.5){\scalebox{0.32}{\includegraphics{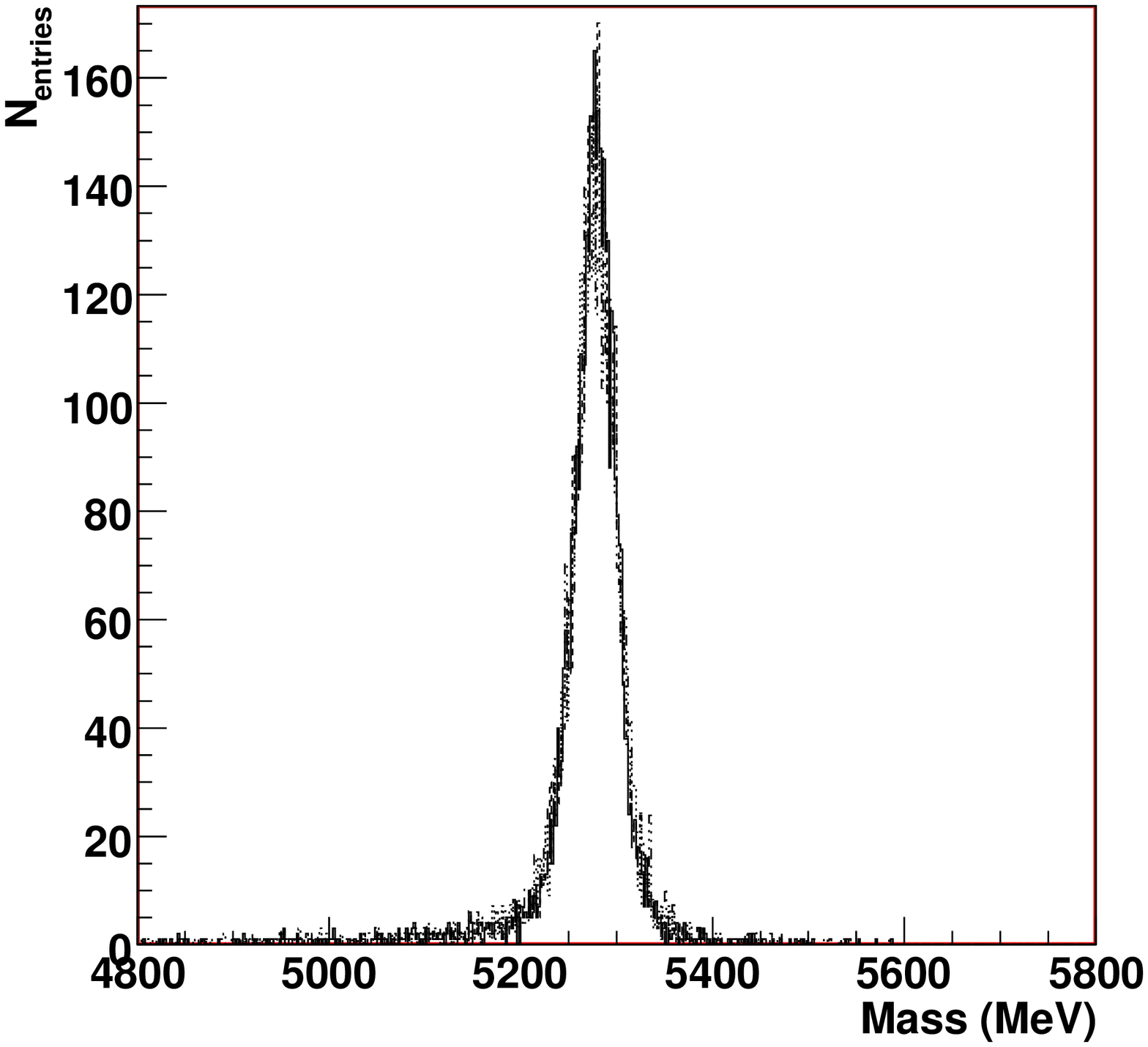}}}
\put(0.,0.){\scalebox{0.32}{\includegraphics{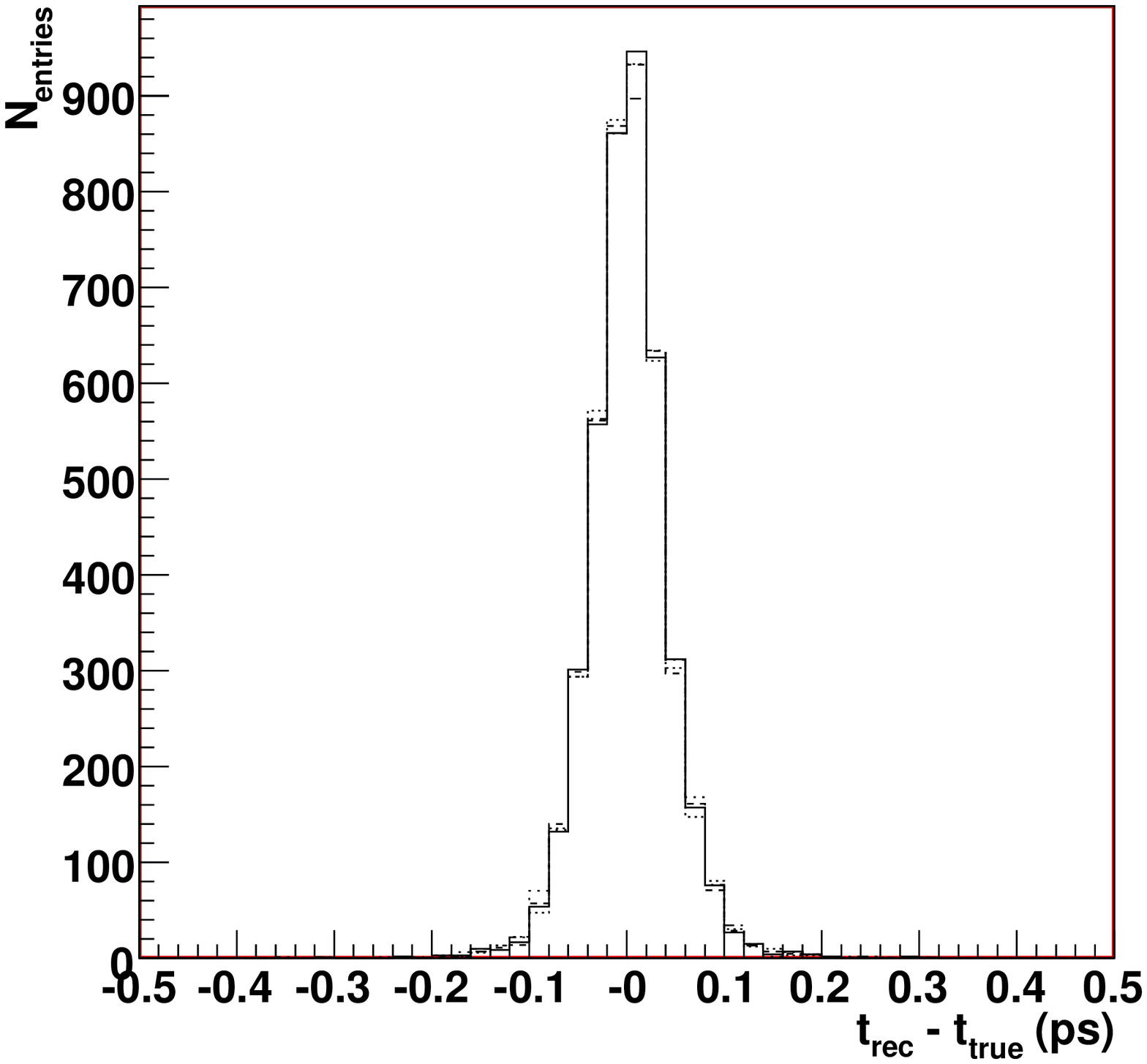}}}
\put(2.0,10.5){\small (a)}
\put(9.0,10.5){\small (b)}
\put(2.0,4.5){\small (c)}
\end{picture}
\end{center}
\caption{Effect of T-stations misalignments on the resolutions in (a) momentum
of the daughter pions, in (b) $B^0$ invariant mass and in (c) $B^0$ proper time.
The various line styles are as explained in Figure~\ref{fig:sel_velo}.}
\label{fig:resolution_ot}
\vfill
\end{figure}

\begin{figure}[p]
\vfill
\begin{center}
\setlength{\unitlength}{1.0cm}
\begin{picture}(14.,18.)
\put(0.0,0.){\scalebox{0.9}{\includegraphics[angle=90]{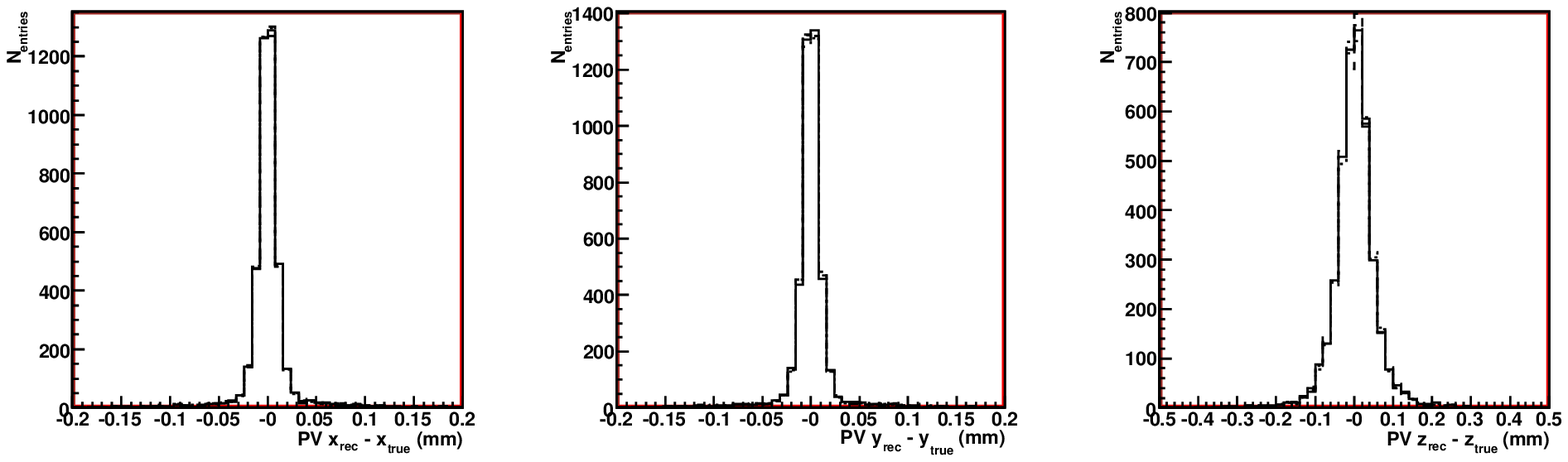}}}
\put(7.0,0.){\scalebox{0.9}{\includegraphics[angle=90]{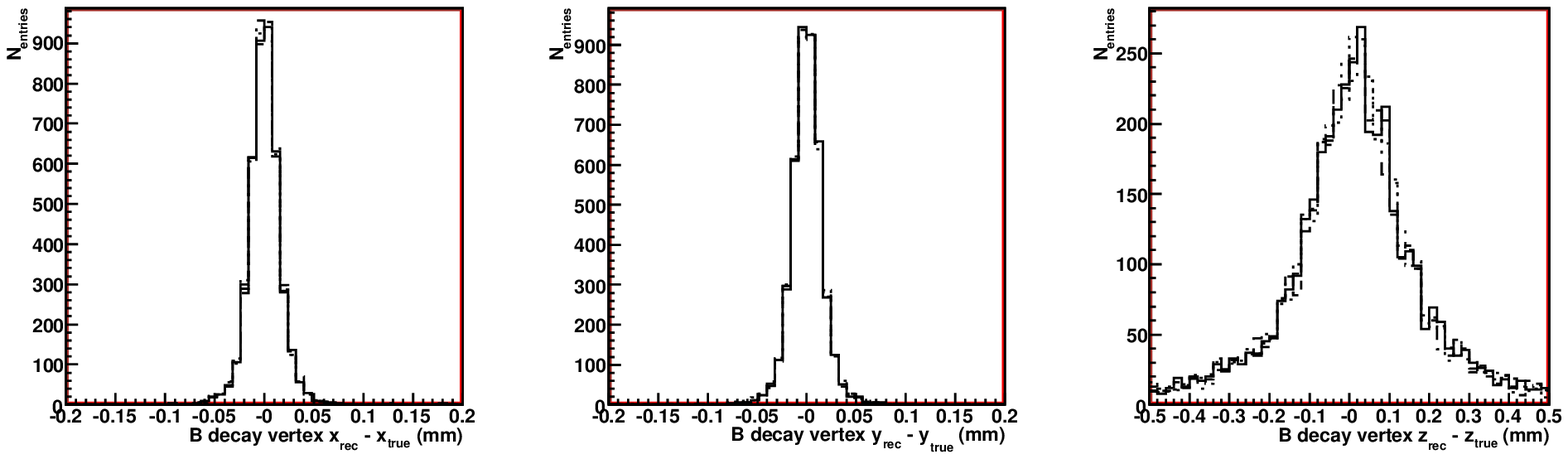}}}
\put(2.0,4.5){\small (a)}
\put(9.0,4.5){\small (b)}
\end{picture}
\end{center}
\caption{Effect of T-stations misalignments on the resolutions of the
(a) primary vertex and (b) the $B^0$ vertex.
The various line styles are as explained in Figure~\ref{fig:sel_velo}.}
\label{fig:resolution_ot_2}
\vfill
\end{figure}

\begin{figure}[p]
\vfill
\begin{center}
\setlength{\unitlength}{1.0cm}
\begin{picture}(14.,6.)
\put(0.,0.){\scalebox{0.32}{\includegraphics{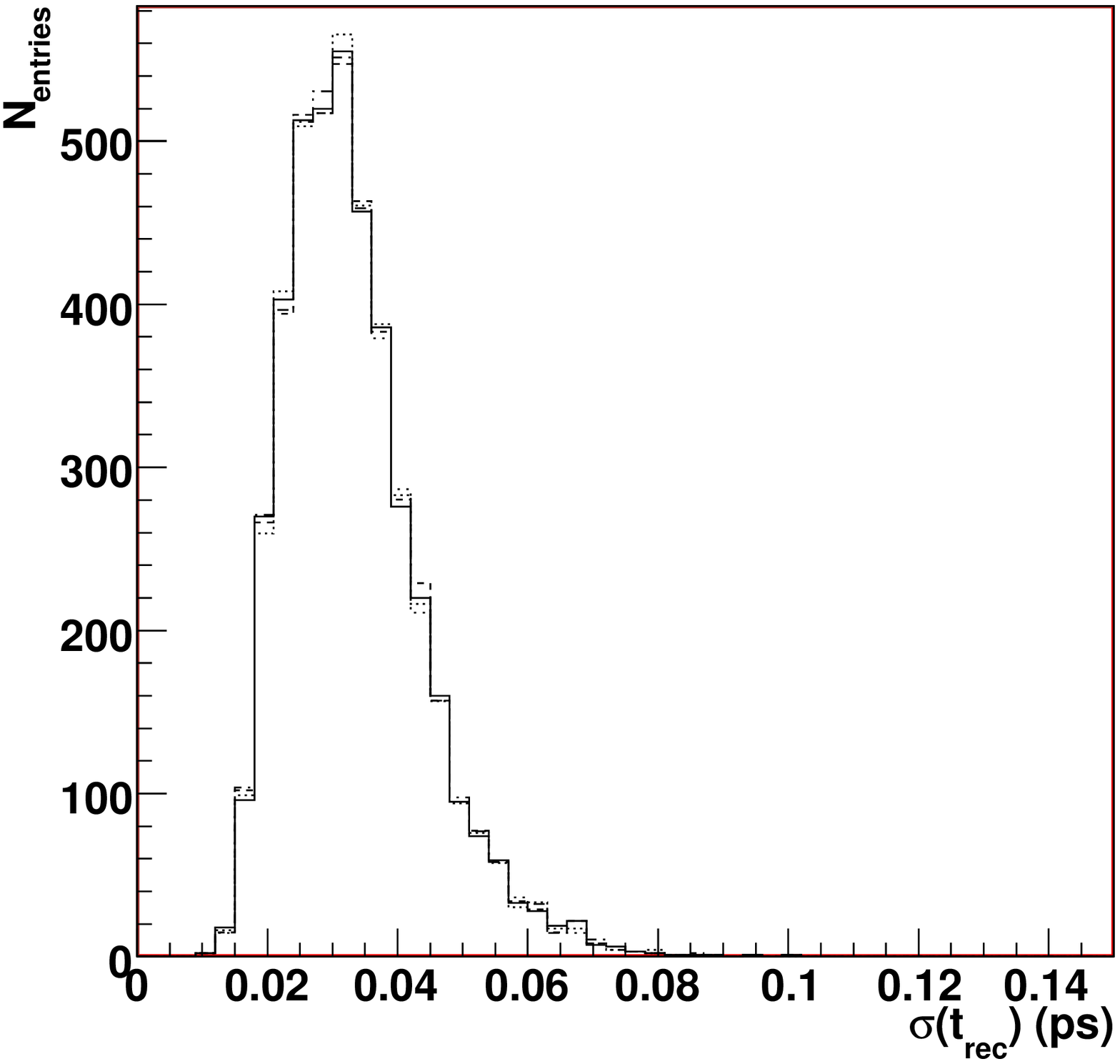}}}
\put(7.,0.){\scalebox{0.32}{\includegraphics{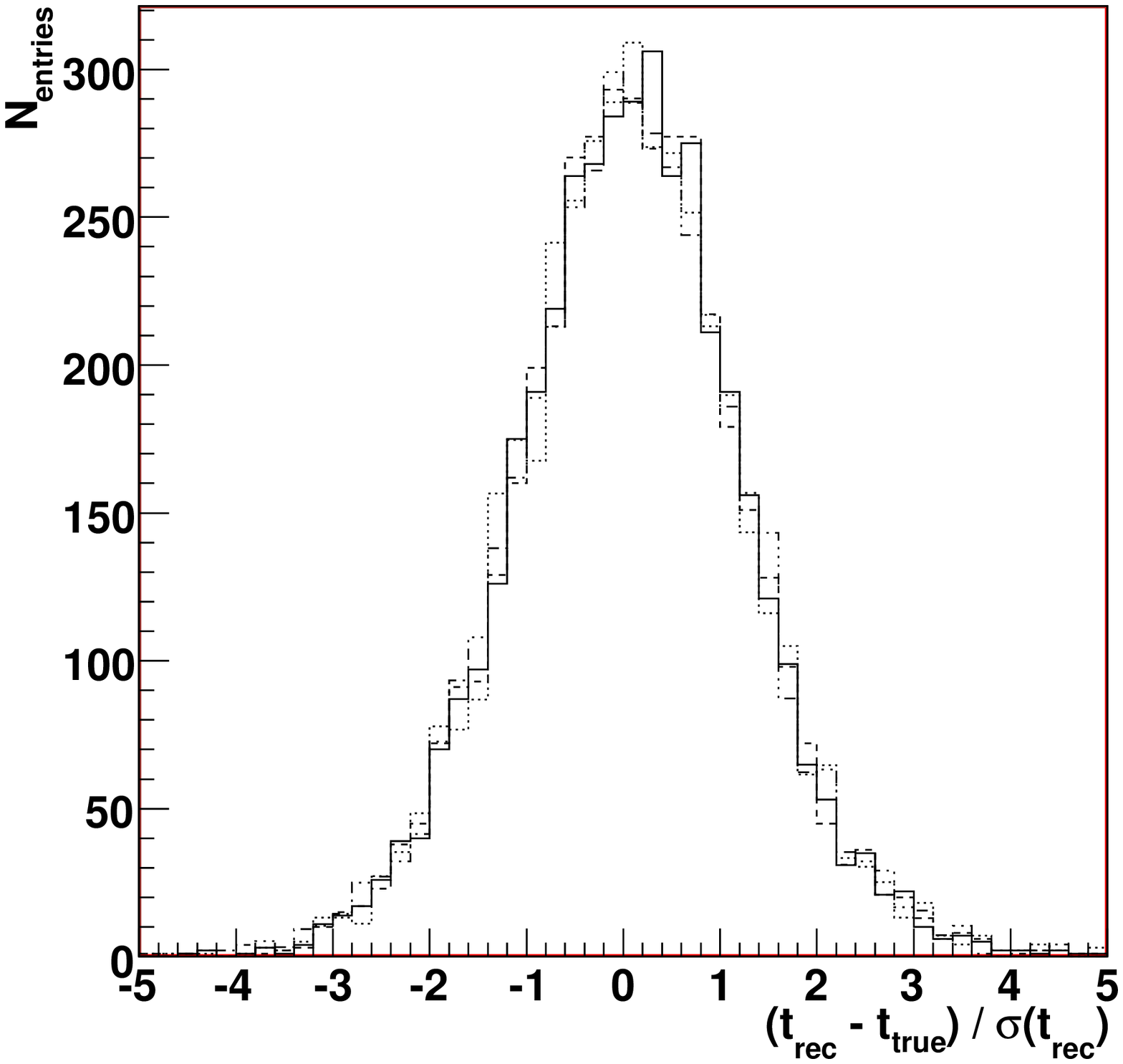}}}
\put(4.0,4.5){\small (a)}
\put(9.0,4.5){\small (b)}
\end{picture}
\end{center}
\caption{Effect of T-stations misalignments on (a) the $B^0$ proper time error
and (b) on the pull distribution of the $B^0$ proper time.
The various line styles are as explained in Figure~\ref{fig:sel_velo}.}
\label{fig:tau_ot}
\end{figure}

\begin{figure}[p]
\vfill
\begin{center}
\setlength{\unitlength}{1.0cm}
\begin{picture}(14.,13.)
\put(0.0,6.5){\scalebox{0.32}{\includegraphics{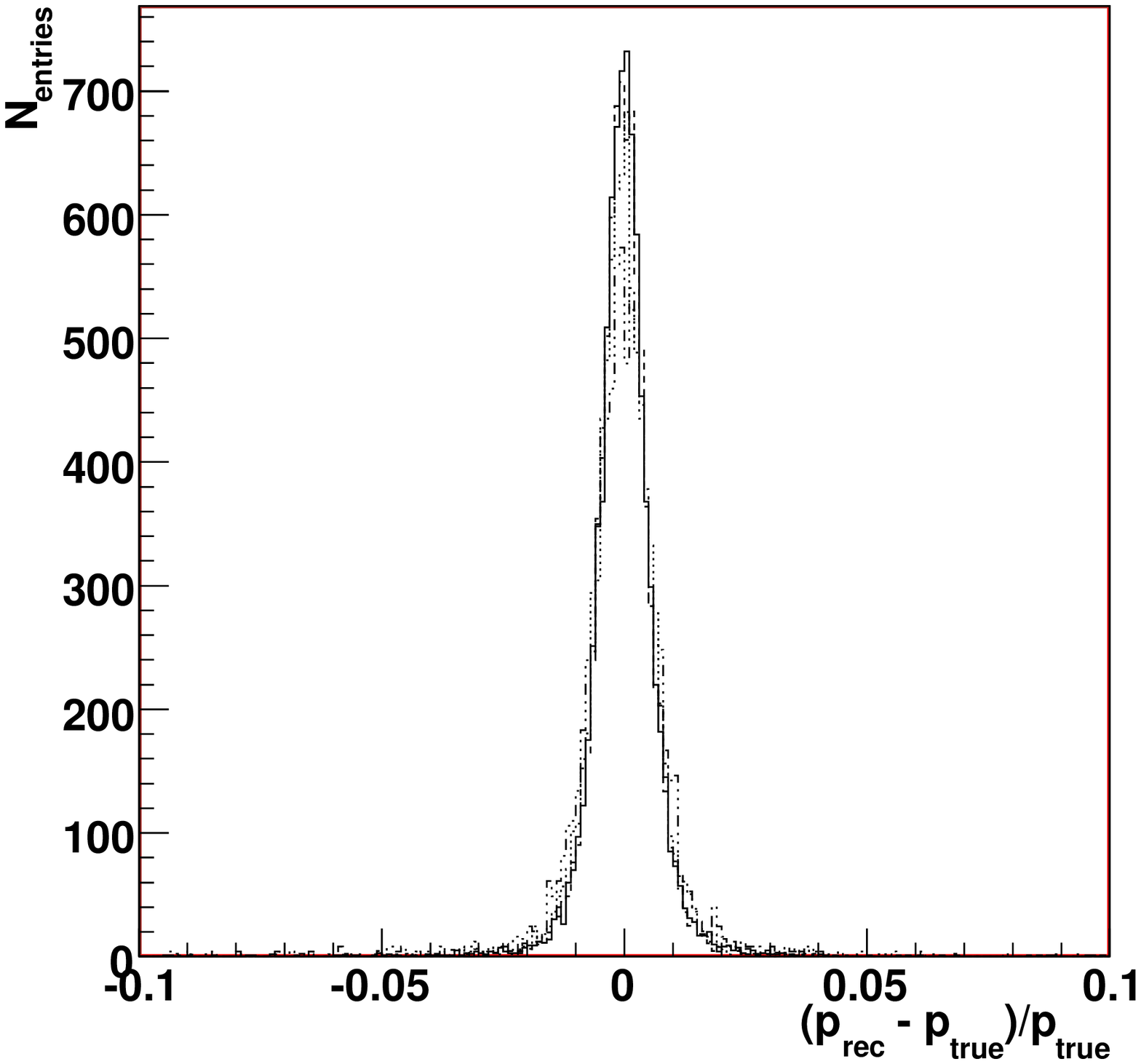}}}
\put(7.0,6.5){\scalebox{0.32}{\includegraphics{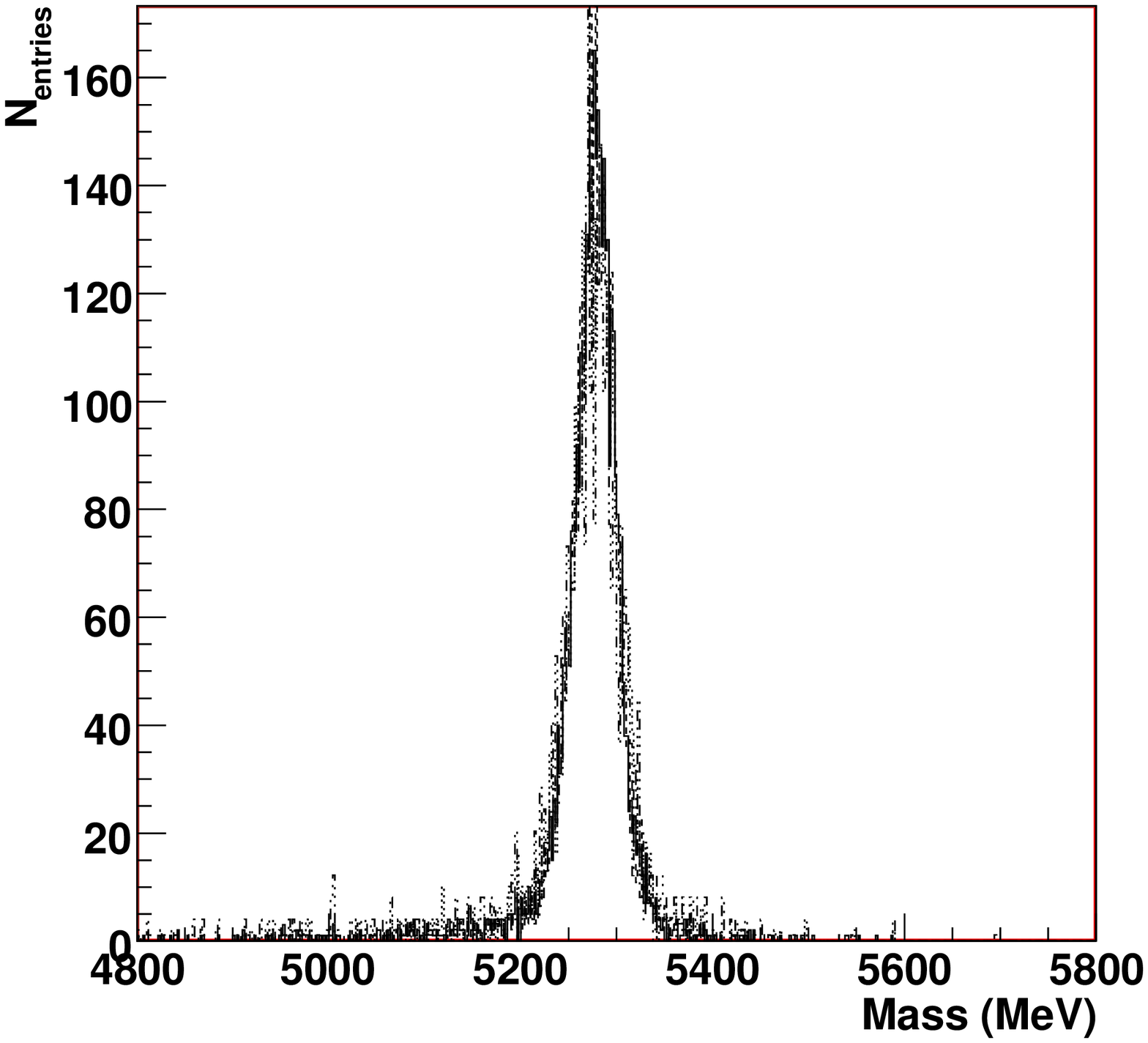}}}
\put(0.0,0.0){\scalebox{0.32}{\includegraphics{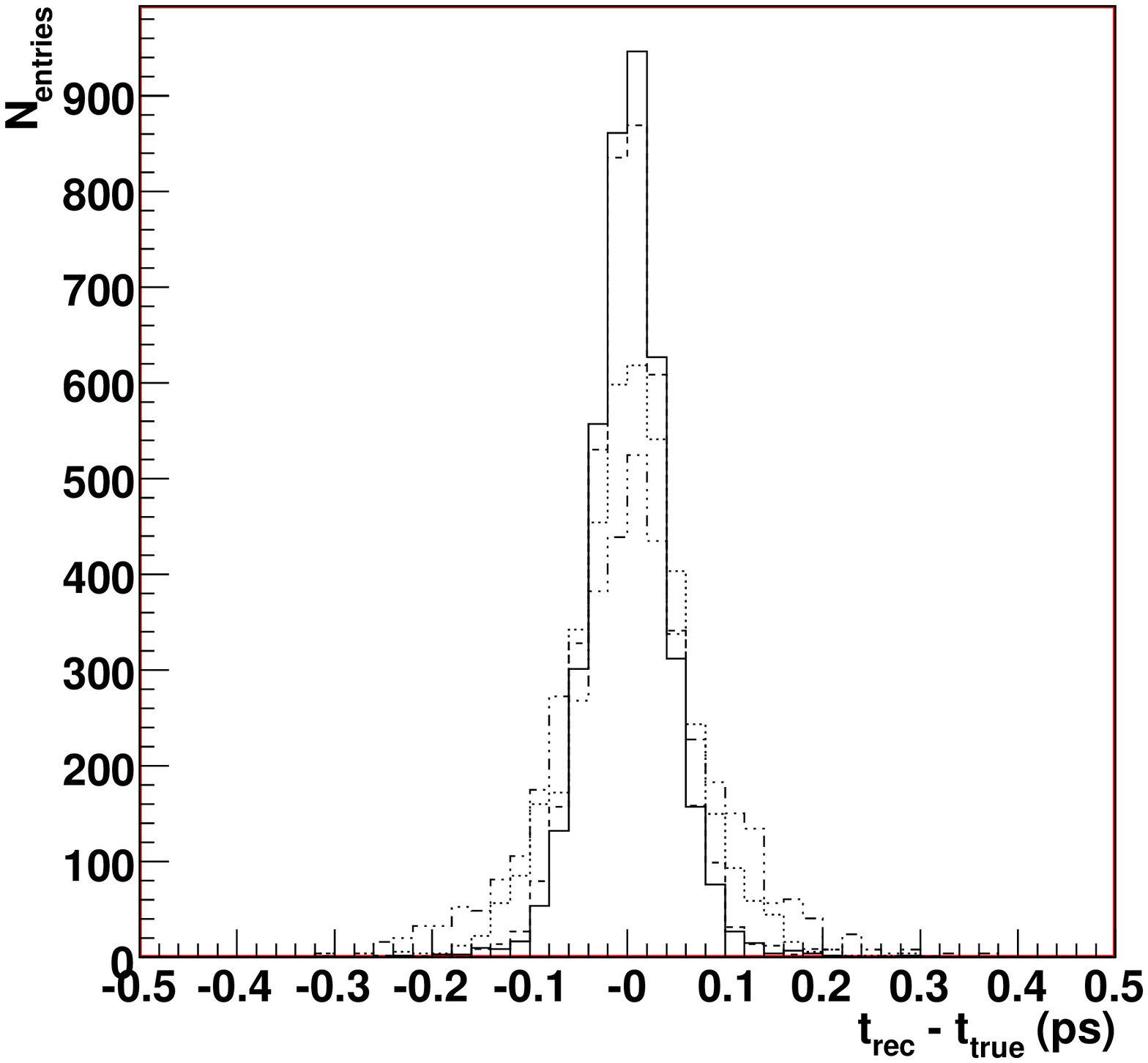}}}
\put(2.0,10.5){\small (a)}
\put(9.0,10.5){\small (b)}
\put(2.0,4.5){\small (c)}
\end{picture}
\end{center}
\caption{Effect of VELO and T-stations misalignments on the resolutions
in (a) momentum of the daughter pions, in (b) $B^0$ invariant mass and
in (c) $B^0$ proper time.
The various line styles are as explained in Figure~\ref{fig:sel_velo}.}
\label{fig:resolution_all}
\vfill
\end{figure}

\begin{figure}[p]
\vfill
\begin{center}
\setlength{\unitlength}{1.0cm}
\begin{picture}(14.,18.)
\put(0.0,0.){\scalebox{0.9}{\includegraphics[angle=90]{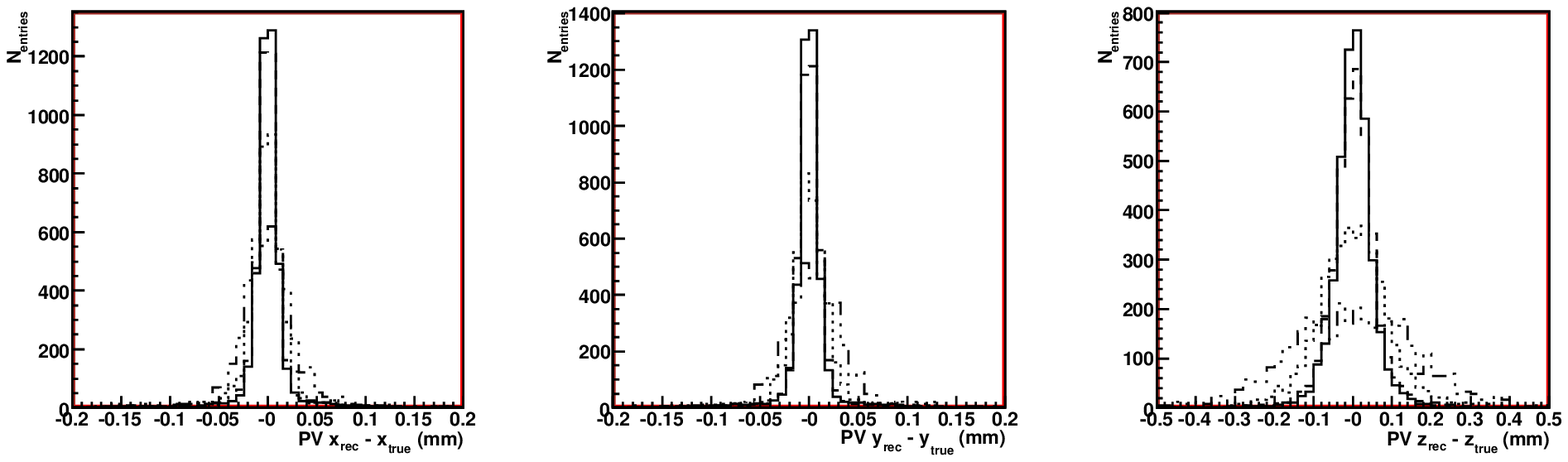}}}
\put(7.0,0.){\scalebox{0.9}{\includegraphics[angle=90]{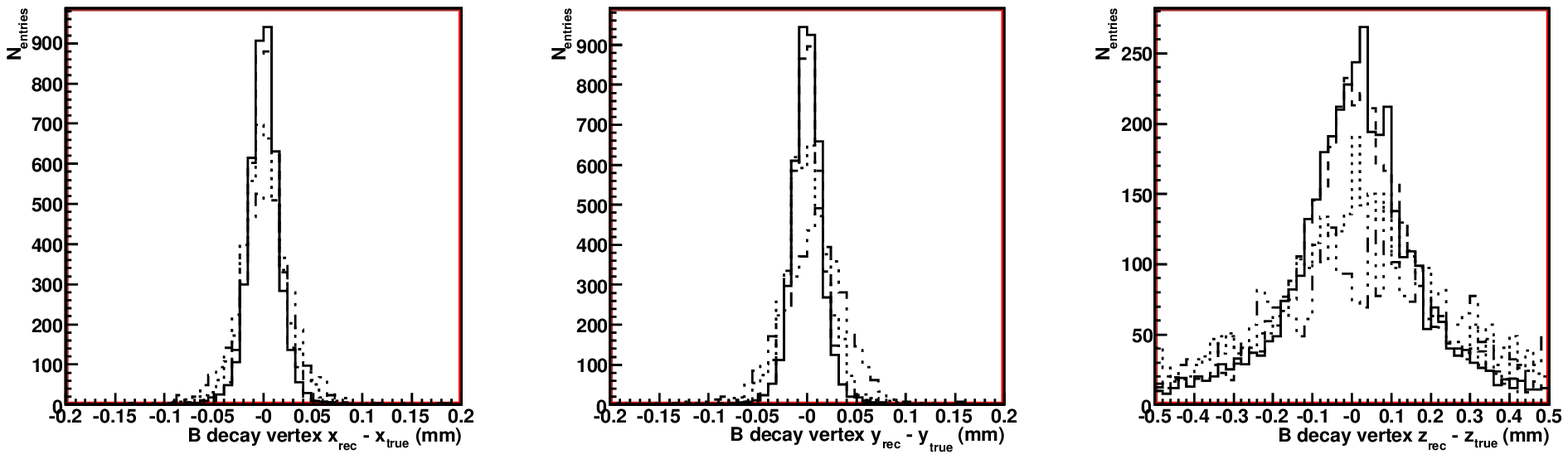}}}
\put(2.0,4.5){\small (a)}
\put(9.0,4.5){\small (b)}
\end{picture}
\end{center}
\caption{Effect of VELO and T-stations misalignments on the resolutions of the
(a) primary vertex and (b) the $B^0$ vertex.
The various line styles are as explained in Figure~\ref{fig:sel_velo}.}
\label{fig:resolution_all_2}
\vfill
\end{figure}

\begin{figure}[p]
\vfill
\begin{center}
\setlength{\unitlength}{1.0cm}
\begin{picture}(14.,6.)
\put(0.,0.){\scalebox{0.32}{\includegraphics{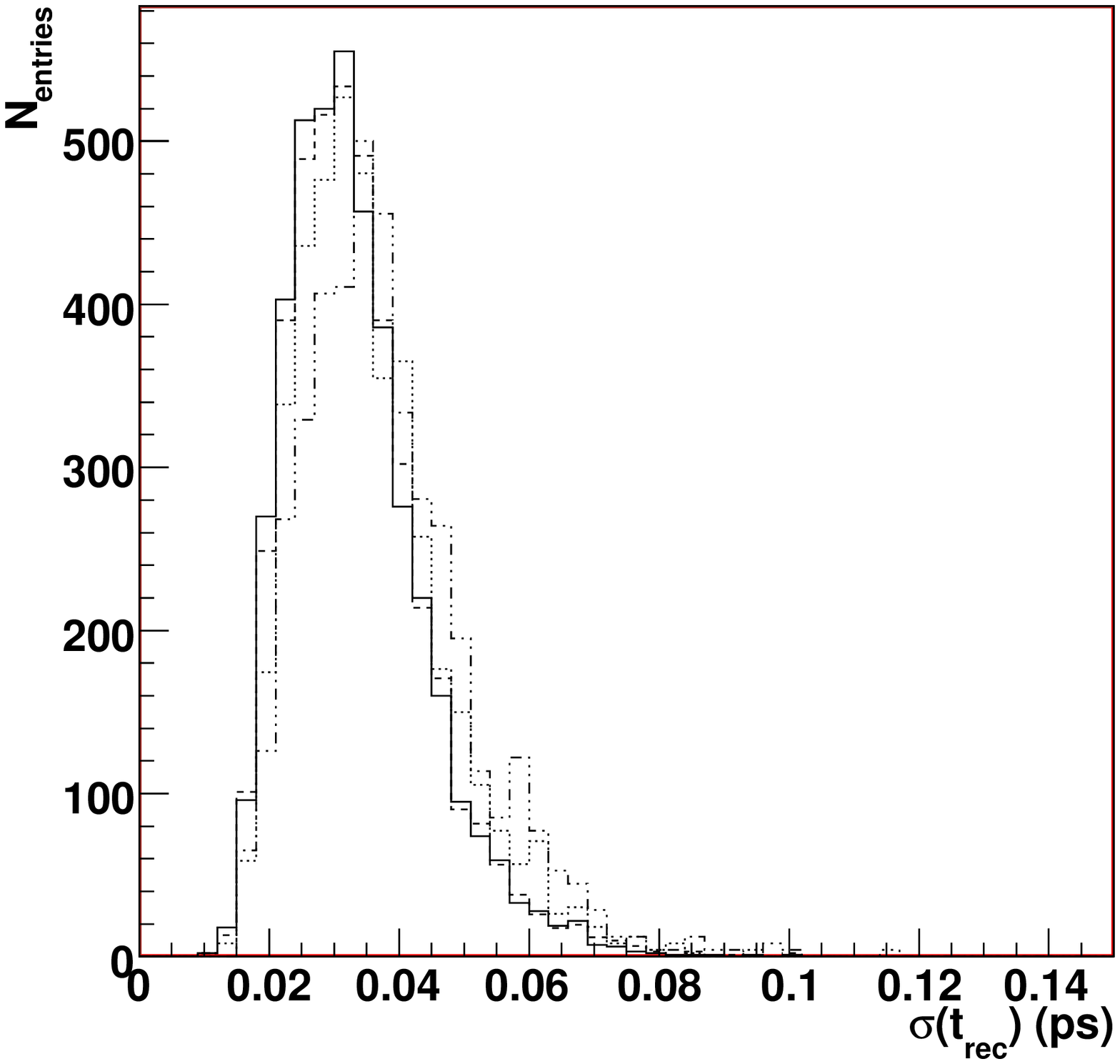}}}
\put(7.,0.){\scalebox{0.32}{\includegraphics{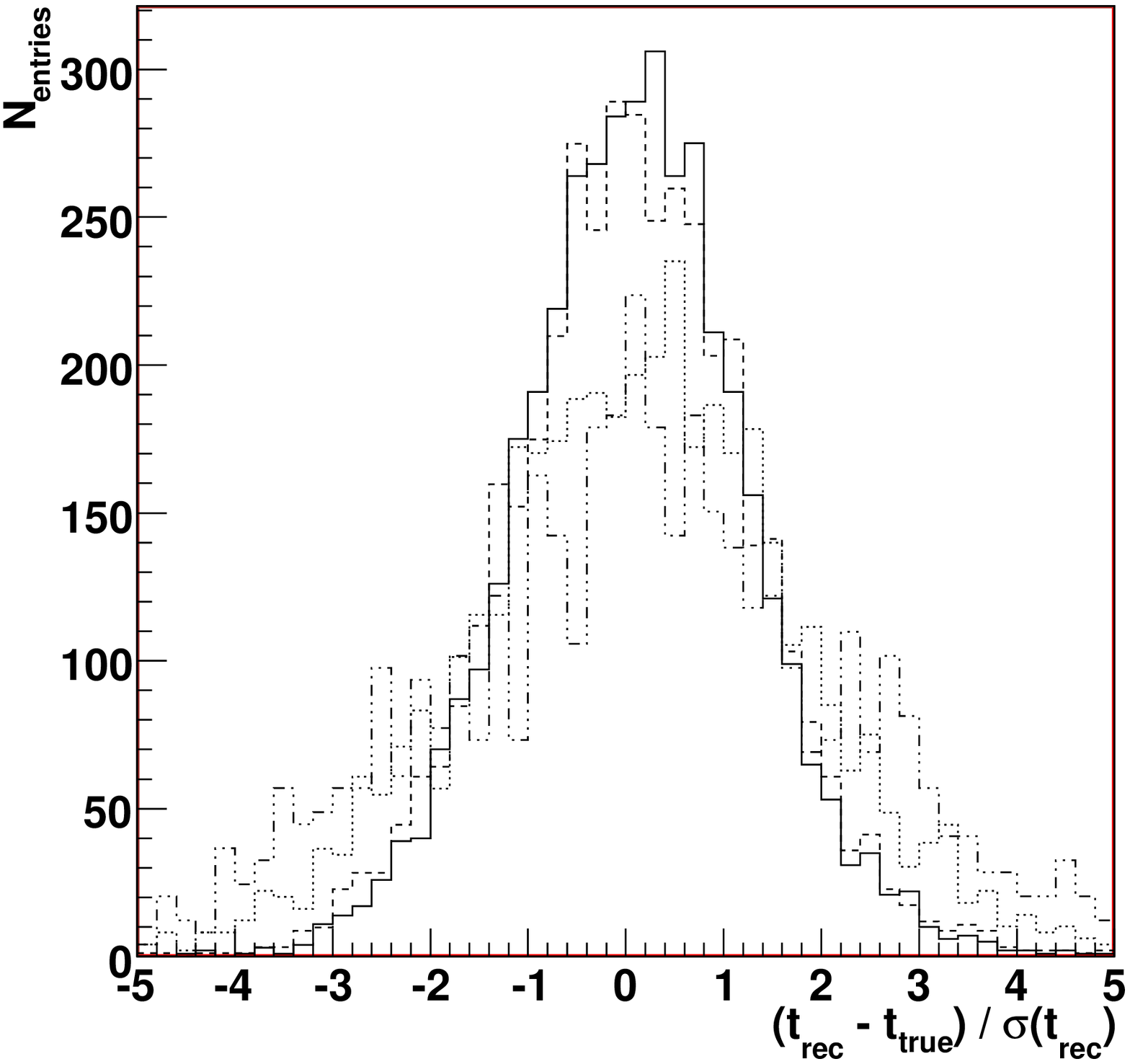}}}
\put(4.0,4.5){\small (a)}
\put(9.0,4.5){\small (b)}
\end{picture}
\end{center}
\caption{Effect of VELO and T-stations misalignments on (a) the $B^0$ proper time error and
(b) on the pull distribution of the $B^0$ proper time.
The various line styles are as explained in Figure~\ref{fig:sel_velo}.}
\label{fig:tau_all}
\vfill
\end{figure}

\begin{figure}[p]
\vfill
\begin{center}
\setlength{\unitlength}{1.0cm}
\begin{picture}(14.,13.)
\put(0.,6.5){\scalebox{0.32}{\includegraphics{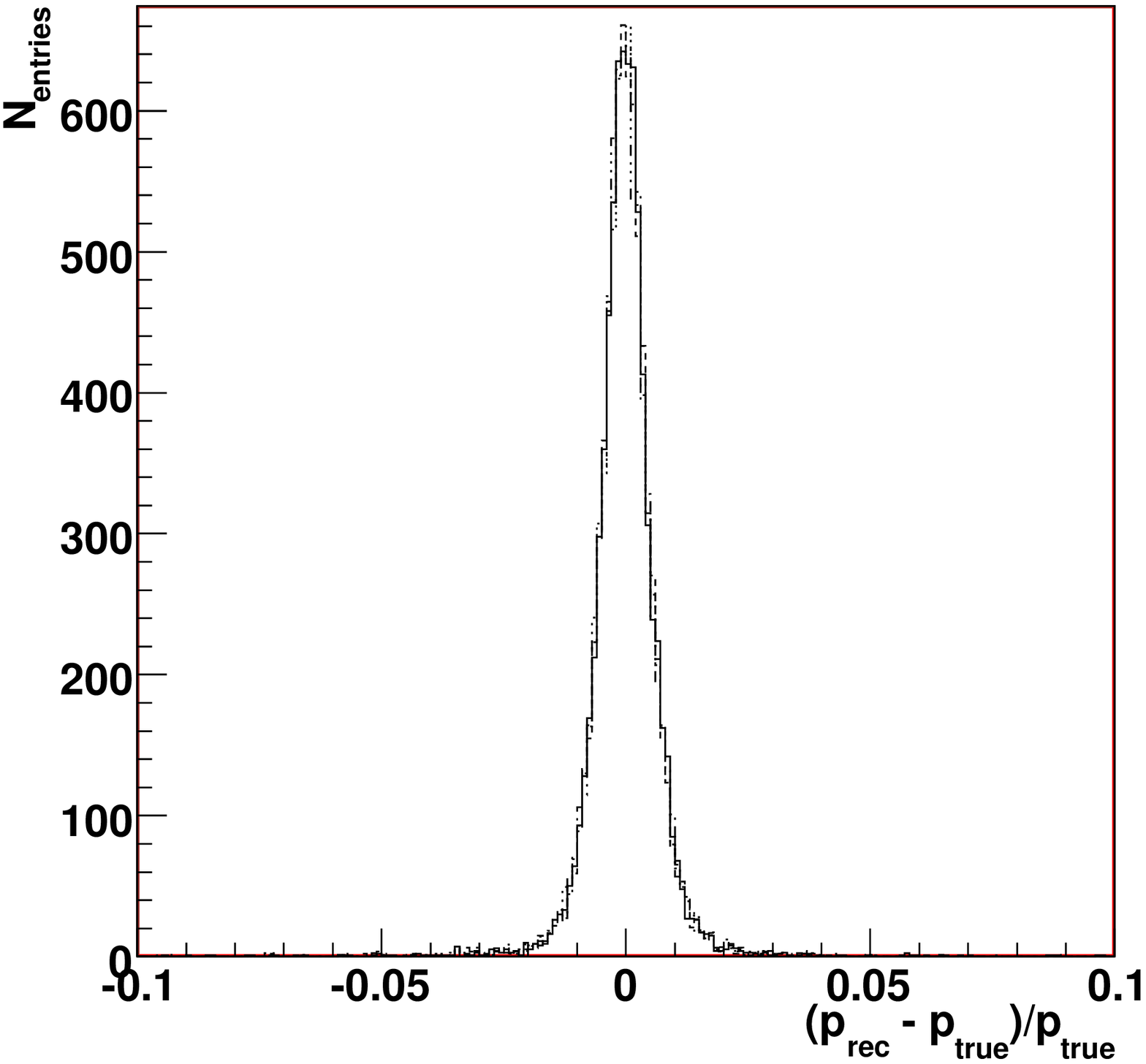}}}
\put(7.,6.5){\scalebox{0.32}{\includegraphics{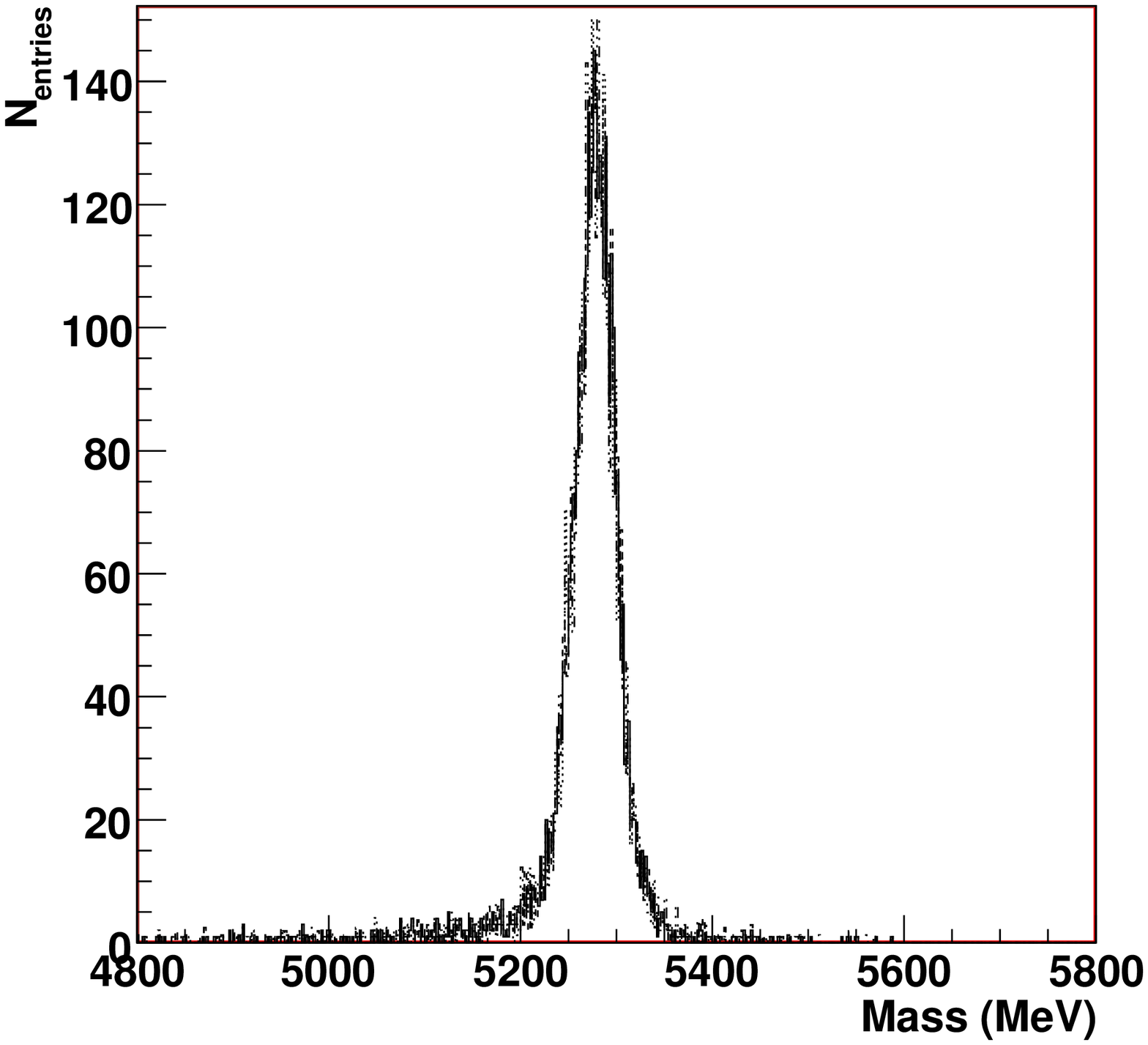}}}
\put(0.,0.){\scalebox{0.32}{\includegraphics{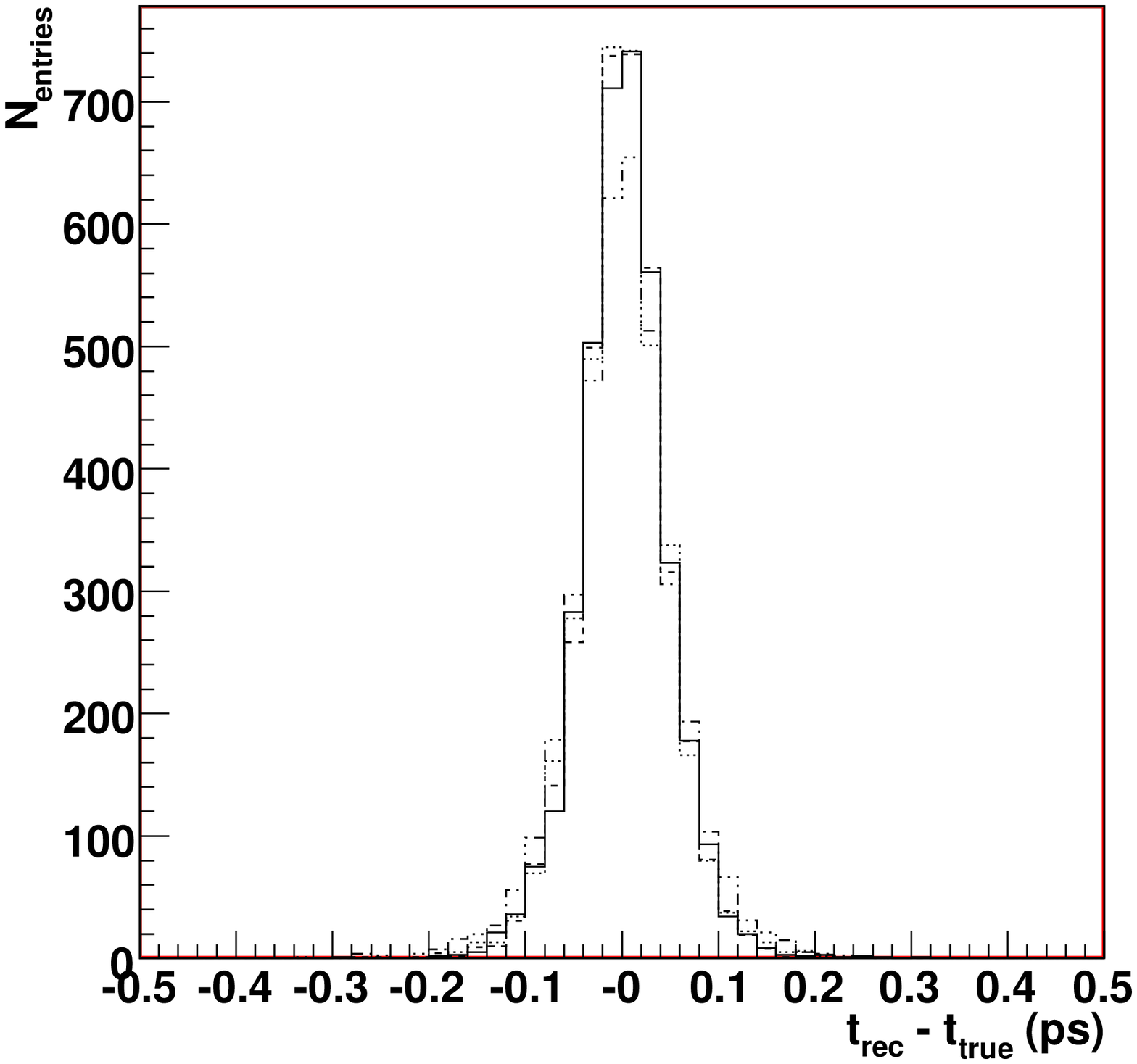}}}
\put(2.0,10.5){\small (a)}
\put(9.0,10.5){\small (b)}
\put(2.0,4.5){\small (c)}
\end{picture}
\end{center}
\caption{Effect of VELO $z$-scaling misalignments on the resolutions
in (a) momentum of the daughter pions, (b) $B^0$ invariant mass and in
(c) $B^0$ proper time.
The full line corresponds to a $z$-scale of $1.00003$; the dashed line
corresponds to a $z$-scale of $1.00010$; the dotted line corresponds to a
$z$-scale of $1.00033$ and the dot-dashed line corresponds to a $z$-scale of 
$1.00100$.}
\label{fig:res_z}
\vfill
\end{figure}

\begin{figure}[p]
\vfill
\begin{center}
\setlength{\unitlength}{1.0cm}
\begin{picture}(14.,18.)
\put(0.0,0.){\scalebox{0.9}{\includegraphics[angle=90]{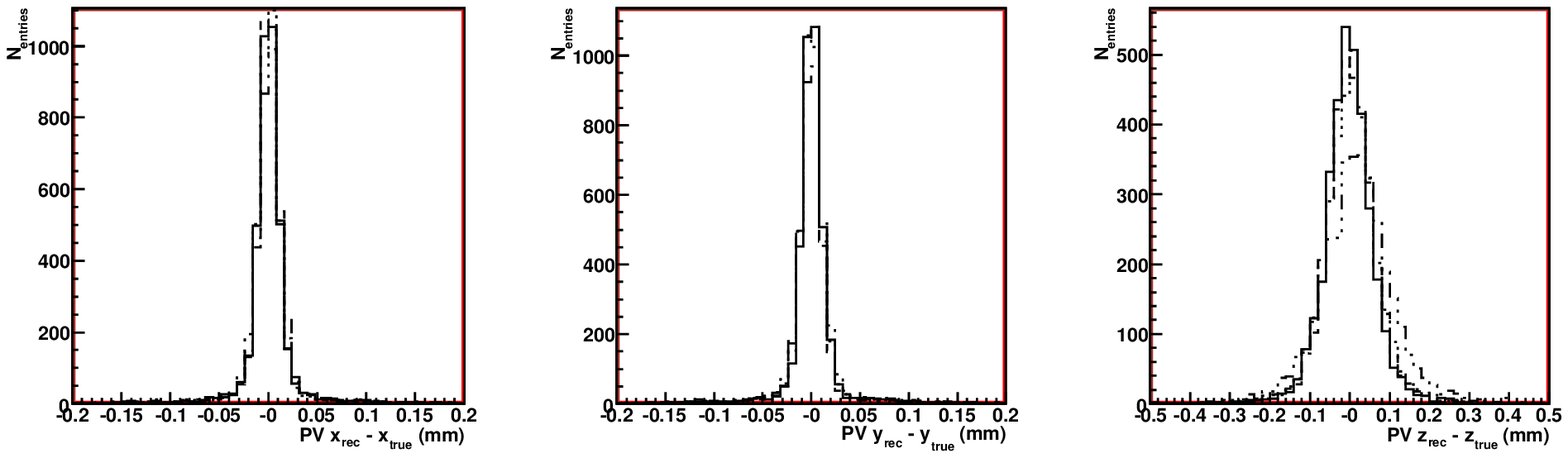}}}
\put(7.0,0.){\scalebox{0.9}{\includegraphics[angle=90]{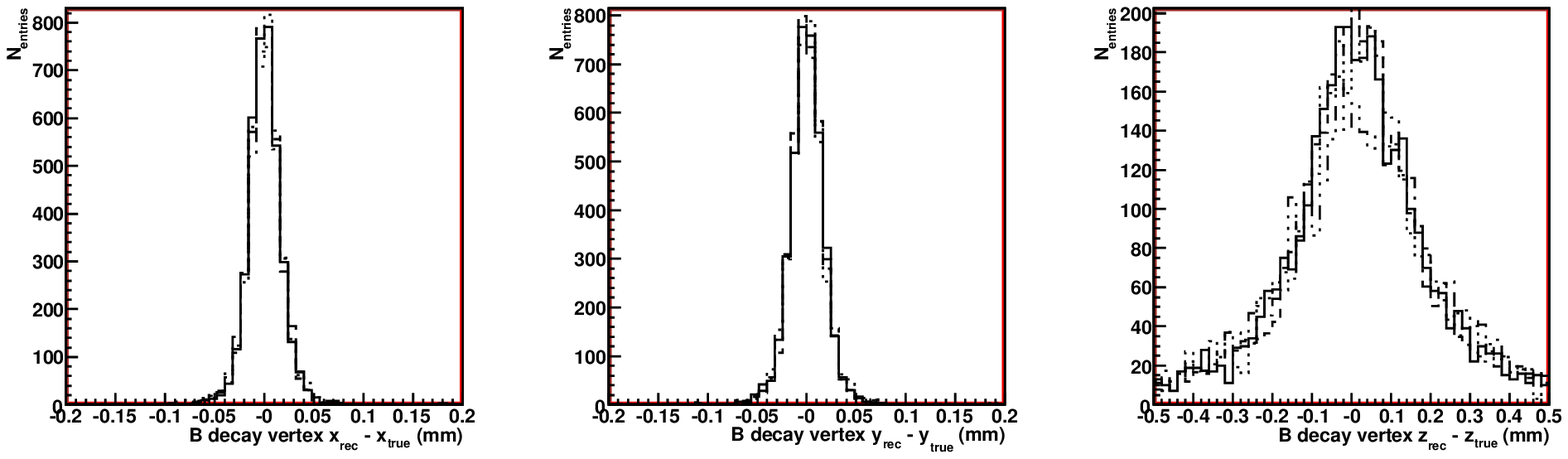}}}
\put(2.0,4.5){\small (a)}
\put(9.0,4.5){\small (b)}
\end{picture}
\end{center}
\caption{Effect of VELO $z$-scaling misalignments on the resolutions of the
(a) primary vertex and (b) the $B^0$ vertex.
The various line styles are as explained in Figure~\ref{fig:res_z}.}
\label{fig:res_z_2}
\vfill
\end{figure}

\begin{figure}[p]
\vfill
\begin{center}
\setlength{\unitlength}{1.0cm}
\begin{picture}(14.,6.)
\put(0.,0.){\scalebox{0.32}{\includegraphics{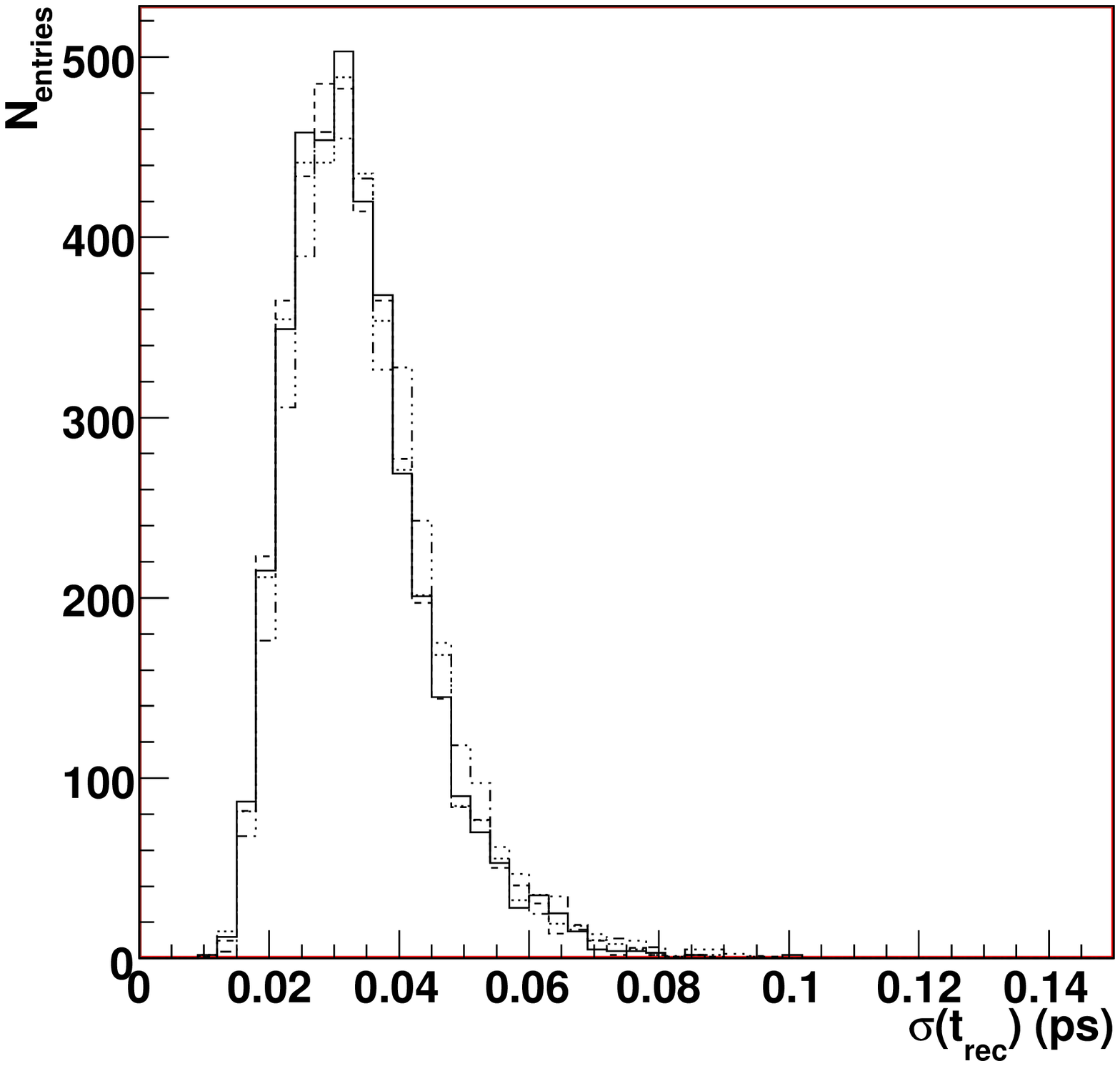}}}
\put(7.,0.){\scalebox{0.32}{\includegraphics{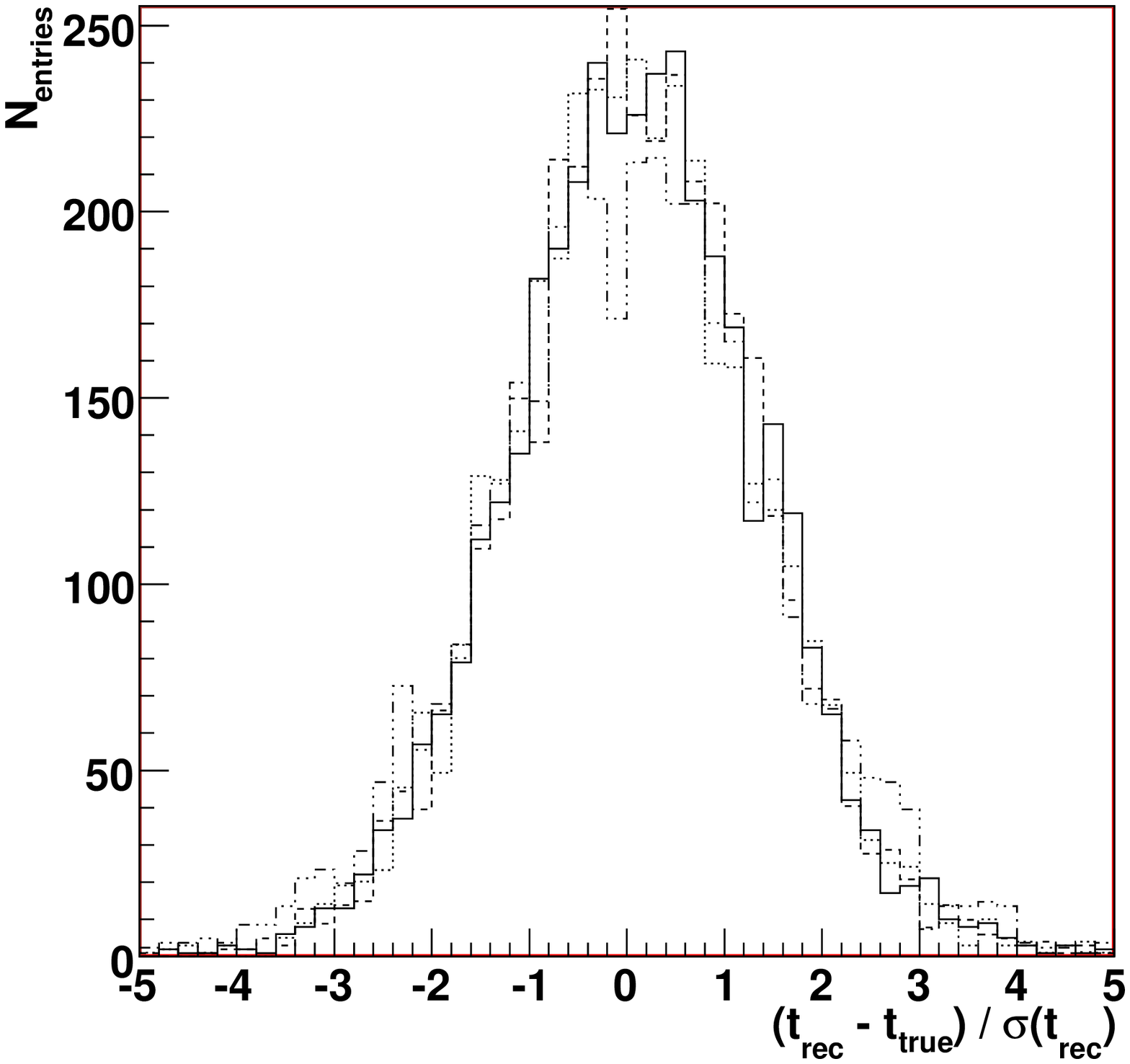}}}
\put(4.0,4.5){\small (a)}
\put(9.0,4.5){\small (b)}
\end{picture}
\end{center}
\caption{Effect of VELO $z$-scaling misalignments on (a) the $B^0$ proper time
error and (b) on the pull distribution of the $B^0$ proper time.
The various line styles are as explained in Figure~\ref{fig:res_z}.}
\label{fig:tau_z}
\vfill
\end{figure}

\clearpage



%
%
\end{document}